\documentclass[%
twocolumn,
letterpaper,
superscriptaddress,
showpacs,
preprintnumbers,
 amsmath,amssymb,
 aps,
 prd,
floatfix,
]{revtex4-1}

\usepackage{graphicx}
\usepackage{dcolumn}
\usepackage{bm}

\usepackage{fix-cm}
\usepackage{epsfig,float}
\usepackage{mathptmx}
\usepackage{latexsym}
\usepackage[colorlinks,citecolor=blue,urlcolor=blue,linkcolor=blue]{hyperref}

\usepackage[utf8]{inputenc}
\usepackage[T1]{fontenc}
\usepackage{graphicx}
\usepackage{longtable}
\usepackage{float}
\usepackage{wrapfig}
\usepackage{rotating}
\usepackage[normalem]{ulem}
\usepackage{amsmath}
\usepackage{amsfonts}

\usepackage{textcomp}
\usepackage{marvosym}
\usepackage{wasysym}
\usepackage{amssymb}
\usepackage[usenames,dvipsnames,svgnames,table]{xcolor}

\usepackage{booktabs}
\usepackage{etoolbox}

\newcommand{\pbarp}{\ensuremath{\bar{p}p}}
\newcommand{\pbar}{\ensuremath{\bar{p}}}
\newcommand{\mompbar}{\ensuremath{p_{\bar{p}}}}

\newcommand{\jpsi}{\ensuremath{J/\psi}}
\newcommand{\gstar}{\ensuremath{\gamma^*}}
\newcommand{\piz}{\ensuremath{\pi^0}}
\newcommand{\epem}{\ensuremath{e^+e^-}}
\newcommand{\pipm}{\ensuremath{\pi^+\pi^-}}
\newcommand{\pip}{\ensuremath{\pi^+}}
\newcommand{\pim}{\ensuremath{\pi^-}}

\newcommand{\jpsitoepem}{\ensuremath{\jpsi\to\epem}}

\newcommand{\jpsipiz}{\ensuremath{\jpsi\piz}}
\newcommand{\jpsipizpiz}{\ensuremath{\jpsi\piz\piz}}

\newcommand{\pipmpiz}{\ensuremath{\pipm\piz}}
\newcommand{\pipmpizpiz}{\ensuremath{\pipm\piz\piz}}
\newcommand{\pipmpipmpiz}{\ensuremath{\pipm\pipm\piz}}

\newcommand{\epempiz}{\ensuremath{\epem\piz}}
\newcommand{\epempizpiz}{\ensuremath{\epem\piz\piz}}
\newcommand{\gstarpiz}{\ensuremath{\gstar\piz}}

\newcommand{\sigrxnshort}{\ensuremath{\pbarp\to\jpsipiz}}
\newcommand{\sigrxn}{\ensuremath{\sigrxnshort\to\epempiz}}
\newcommand{\sigrxnepem}{\ensuremath{\pbarp\to\gstarpiz\to\epempiz}}

\newcommand{\bgrxn}{\ensuremath{\pbarp\to\pipmpiz}}

\newcommand{\jpsitopbarp}{\ensuremath{\jpsi\to\pbarp}}

\newcommand{\bgtwopizshort}{\ensuremath{\pbarp\to\jpsipizpiz}}
\newcommand{\bgtwopiz}{\ensuremath{\bgtwopizshort\to\epempizpiz}}

\newcommand{\mjpsiwin}{2.8~$<$ \allowbreak$M_{\epem}$ \allowbreak$<$~3.3~GeV/$c^{2}$}
\newcommand{\fitwin}{2.5~$<$ \allowbreak$M_{\epem}$ \allowbreak$<$~3.8~GeV/$c^{2}$}

\newcommand{\detlist}{\ensuremath{\mathrm{EMC},\mathrm{DIRC},\mathrm{DISC},\mathrm{STT},\mathrm{MVD}}}
\newcommand{\splist}{\ensuremath{e^{\pm},\ \mu^{\pm},\ \pi^{\pm},\ K^{\pm},\ p^{\pm}}}

\newcommand{\prob}{\ensuremath{\mathcal{P}}}

\newcommand{\probeidcomb}{\ensuremath{\prob^{ID}_{comb}(e^{\pm})}}

\newcommand{\fbi}{fb$^{-1}$}

\newcommand{\tgamma}{\ensuremath{\gamma\gamma}}
\newcommand{\sigh}{\ensuremath{2\gamma\epem}}
\newcommand{\bgh}{\ensuremath{4\gamma\epem}}

\newcommand{\rowa}{\ensuremath{\varepsilon_{\sigrxnshort}}}
\newcommand{\rowb}{\ensuremath{\mathcal{R}_{\jpsipizpiz}}}
\newcommand{\rowc}{\ensuremath{\mathcal{R}_{\pipmpiz}}}
\newcommand{\rowd}{\ensuremath{\mathcal{C}_{\jpsipizpiz}}}
\newcommand{\rowe}{\ensuremath{\mathcal{C}_{\pipmpiz}}}
\newcommand{\rowf}{\ensuremath{\mathcal{P}_{comb}}}
\newcommand{\rowg}{S/B}

\newcommand{\epth}{\ensuremath{\theta^{e^+}_{\jpsi}}}
\newcommand{\epcth}{\ensuremath{\cos(\epth)}}
\newcommand{\opcthsq}[1]{\ensuremath{1+#1\cos^2(\epth)}}

\newcommand{\bmom}[1]{\ifstrequal{#1}{0}{5.5}{\ifstrequal{#1}{1}{8.0}{\ifstrequal{#1}{2}{12.0}{}}}}
\newcommand{\bmomu}[1]{\ifstrequal{#1}{0}{5.5~GeV/$c$}{\ifstrequal{#1}{1}{8.0~GeV/$c$}{\ifstrequal{#1}{2}{12.0~GeV/$c$}{}}}}
\newcommand{\plfm}[1]{{(#1)}}

\newcommand{\PANDA}{\texorpdfstring{$\overline{\mbox{\sf P}}${\sf ANDA}}{Panda}}
\newcommand{\PANDAbf}{\texorpdfstring{$\overline{\mbox{\sf \bf P}}${\sf \bf ANDA}}{Panda}}

\begin{document}

%\preprint{APS/123-QED}

\title{Feasibility study for the measurement of $\pi N$ TDAs at
  \PANDAbf\ in $\sigrxnshort$}

%%%% BEGIN AUTOHRS %%%%
% PANDA Collaboration - authorlist
% 2016-10-01 03:15:01
\author{B.~Singh}
\affiliation{Aligarth Muslim University, Physics Department,{ \bf Aligarth}, India}
\author{W.~Erni}
\author{B.~Krusche}
\author{M.~Steinacher}
\author{N.~Walford}
\affiliation{Universität Basel,{ \bf Basel}, Switzerland}
\author{H.~Liu}
\author{Z.~Liu}
\author{B.~Liu}
\author{X.~Shen}
\author{C.~Wang}
\author{J.~Zhao}
\affiliation{Institute of High Energy Physics, Chinese Academy of Sciences,{ \bf Beijing}, China}
% ...
\collaboration{$\overline{\mbox{\sf P}}${\sf ANDA} Collaboration}
\noaffiliation
\author{\vskip -0.3cm and K.~M.~Semenov-Tian-Shansky~$^{50}$}

\date{\today}

\begin{abstract}
  The exclusive charmonium production process in $\pbarp$ annihilation
  with an associated $\piz$ meson $\sigrxnshort$ is studied in the
  framework of QCD collinear factorization. The feasibility of
  measuring this reaction through the $\jpsitoepem$ decay channel with
  the \PANDA\ (AntiProton ANnihilation at DArmstadt) experiment is
  investigated. Simulations on signal reconstruction efficiency as
  well as the background rejection from various sources including the
  $\bgrxn$ and $\bgtwopizshort$ reactions are performed with
  PandaRoot, the simulation and analysis software framework of the
  \PANDA\ experiment. It is shown that the measurement can be done at
  \PANDA\ with significant constraining power under the assumption of
  an integrated luminosity attainable in four to five months of data
  taking at the maximum design luminosity.
\end{abstract}


\maketitle

\section{\label{sec:intro}Introduction}

Understanding of the hadronic structure in terms of the fundamental
degrees of freedom of QCD is one of the fascinating questions of the
present day physics. Lepton beam initiated reactions, allowing to
resolve individual quarks and gluons inside hadrons, proved to be a
handy tool for this issue. The factorization property established for
several classes of hard (semi-)inclusive and exclusive processes
allows to separate the short distance dominated stage of interaction
and the universal non-perturbative hadronic matrix elements. Some of
the matrix elements which have been the subject of significant
interest include the Parton Distribution Functions
(PDFs)~\cite{Collins:1989gx}, Generalized Parton Distributions
(GPDs)~\cite{Guidal:2013rya,Mueller:2014hsa}, Transverse Momentum
Dependent Parton Distribution Functions (TMD
PDFs)~\cite{Barone:2010zz}, (Generalized) Distribution Amplitudes
((G)DAs)~\cite{Braun:1999te} and Transition Distribution Amplitudes
(TDAs)~\cite{Frankfurt:1999fp,Pire:2005ax} encoding valuable
information on the hadron constituents.

Alongside with the study of lepton beam induced reactions, one can get
access to the same non-perturbative functions in a complementary way
by considering the cross conjugated channels of the corresponding
reactions. For example, proton-antiproton annihilation into a lepton
pair and a photon (or a meson) can be seen as the cross conjugated
counterpart of the leptoproduction of photons (or mesons) off protons,
and provides access to nucleon GPDs and/or nucleon-to-photon
(nucleon-to-meson) TDAs.

Such investigations have been hindered up to now by the limitations of
antiproton beam luminosities. However, very significant results on the
electromagnetic form factors in the time-like region using the $p
\bar{p} \to e^+ e^-$ reaction were obtained by the E835 experiment at
FNAL (Fermilab National Accelerator
Laboratory)~\cite{Andreotti:2003bt}. But inclusive lepton pair
production and hard exclusive channels still remain unexplored.

This situation will be largely improved in the next decade with the
availability of the high intensity antiproton beam at FAIR (Facility
for Antiproton and Ion Research) with momentum up to $15$~GeV/$c$. The
\PANDA\ experiment~\cite{PandaPhysBook09c} will dedicate an important
part of its physics program to the investigation of the nucleon
structure in antiproton-proton annihilation reactions. It includes the
detailed study of the time-like electromagnetic nucleon form factors
employing both the $e^+e^-$ and $\mu^+\mu^-$ production channels in a
broad kinematic range.  It is also planned to access PDFs through the
Drell-Yan mechanism, by measuring inclusive $e^+e^-$ production and
GPDs considering $\gamma^* \gamma$ and $\gamma^* \pi^0$ exclusive
channels at large production angles.  Finally, the reactions $\bar{p}p
\to \gamma^* M \to e^+e^- M $ and $\bar{p}p \to J/\psi M \to e^+e^-
M$, where $M$ stands for a light meson $M=\{\pi^0, \, \eta, \rho^0, \,
\omega, \, \ldots\}$, are proposed to study nucleon-to-meson TDAs.

Nucleon-to-meson (and particularly nucleon-to-pion) TDAs were
introduced as a further generalization of the concepts of both GPDs
and nucleon light-cone wave functions (DAs). They describe partonic
correlations inside nucleons and allow to access the non-minimal Fock
components of the nucleon light-cone wave function with additional
quark-antiquark pair seen as a light meson. Therefore, in particular,
$\pi N$ TDAs provide information on the nucleon's pion cloud.

Nucleon-to-pion TDAs arise in the collinear factorized description of
several hard exclusive reactions such as backward electroproduction of
pions off nucleons~\cite{Lansberg:2007ec,Lansberg:2011aa}, which can
be studied at JLab~\cite{Kubarovskiy:2012yz} and COMPASS in the
space-like regime, while \PANDA\ will provide access to the same
non-perturbative functions in the time-like regime
\cite{Lansberg:2007se,Lansberg:2012ha}.

On the theory side, the possibility to study nucleon-to-meson TDAs is
provided by the collinear factorization theorem similar to the well
known collinear factorization theorem for hard meson electroproduction
\cite{Collins:1996fb}, giving rise to the description in terms of
GPDs. However, the collinear factorization theorem for the TDA case
has never been proven explicitly. Therefore, one of the important
experimental tasks is to look for experimental evidence of the
validity of the factorized description of the corresponding reactions
in terms of nucleon-to-meson TDAs.  This can be done either by
verifying the appropriate scaling behavior or by checking the angular
dependence of the produced lepton pair specific for the dominant
reaction mechanism. Bringing trustworthy evidence for the validity of
the factorized description of a new class of hard exclusive reaction
will, by itself, represent a major experimental achievement of
\PANDA\@.

Recently, a detailed study of the access to $\pi N$ TDAs in the
reaction $\bar{p}p \to \gamma^* \pi^0 \to e^+e^- \pi^0 $ following the
cross section estimates of Ref.~\cite{Lansberg:2007se} with \PANDA\
has been presented in Ref.~\cite{Singh:2014pfv}.  The investigation of
the reaction $\bar{p}p \to J/\psi \pi^0 \to e^+e^- \pi^0 $ constitutes
a natural complement to the latter study. The resonant case presents
the noticeable advantage of a larger cross section, and a cleaner
signal selection due to the resonant $e^+e^-$ production. The
simultaneous measurement of both resonant and non-resonant channels
provides constraints to the $\pi N$ TDAs in different kinematic
ranges, and allows to test the universality of $\pi N$ TDAs. While the
non-resonant $\bar{p}p \to \gamma^* \pi^0 \to e^+e^- \pi^0 $ has never
been measured, some scarce data exist for $\bar{p}p \to J/\psi \pi^0
\to e^+e^-
\pi^0$~\cite{Armstrong:1992ae,Joffe:2004ce,Andreotti:2005vu}, which
can be used to constrain the predictions. Production of a $\jpsi$ with
an associated $\pi^0$ in $\bar{p} p $ collisions has indeed been
investigated in the past by the E760 experiment at FNAL, since it
constitutes a background in the search for charmonium states via their
decay into $\jpsipiz$. An important part of the \PANDA\ program will
also focus on such studies, as described in
Ref.~\cite{Lundborg:2005am}. This brings additional motivation for the
detailed measurements of the $\bar{p}p \to J/\psi \pi^0 \to e^+e^-
\pi^0$ reaction. From the theory point of view, access to the $\pi N$
TDA in the $J/\psi$ production channel is also more favorable, since
one can take advantage of the known $\jpsitopbarp$ decay
width~\cite{Agashe:2014kda} in order to reduce ambiguities related to
the choice of the phenomenological parametrization for the relevant
nucleon DA~\cite{Pire:2013jva}.

The aim of the present study is therefore to explore the feasibility
of the measurement of the reaction $\bar{p}p \to J/\psi \pi^0 \to
e^+e^- \pi^0 $ with \PANDA\ at different incident momenta of the
antiproton beam, based on the cross section estimates of
Ref.~\cite{Pire:2013jva}. The paper is organized as follows:
Section~\ref{sec:pnd_exp} outlines the design of the \PANDA\
experimental setup with a focus on the most relevant components to the
analysis. Section~\ref{sec:sig_prop} covers the properties of the
$\bar{p}p \to J/\psi \pi^0 \to e^+e^- \pi^0 $ reaction which
constitutes the signal. In Section~\ref{sec:bg_prop}, the different
background contributions are discussed. Section~\ref{sec:simul} is
devoted to the description of the simulation and analysis
procedure. In Section~\ref{sec:results}, the expected precision on
differential cross section measurements is presented.

\section{\label{sec:pnd_exp}\PANDAbf\ Experimental Setup}

\subsection{\label{sec:fair}FAIR Accelerator Complex}

The FAIR accelerator complex which is under construction to extend the
existing GSI (Gesellschaft f\"ur Schwerionenforschung) facilities in
Darmstadt, Germany, will provide beam for four experimental pillars,
one of which is the \PANDA\ experiment dedicated to hadronic
physics. FAIR will use the existing SIS18 synchrotron as an injection
ring into a new larger synchrotron SIS100. The SIS100 ring will
generate an intense pulsed beam of protons with energies reaching up
to 29~GeV that can be directed at an antiproton production
target. Time averaged production rates in the range of $5.6\times10^6$
to $10^7~\bar{p}~s^{-1}$ are expected. Antiprotons are collected and
phase space cooled in the CR (Collector Ring), then transferred to the
RESR (Recycled Experimental Storage Ring) accumulator, and then
injected into the HESR (High Energy Storage Ring), equipped with
stochastic and electron cooling, where they will be used by the
\PANDA\ experiment in a fixed target setup. This full setup is
designed to provide beams with up to $10^{11}$ antiprotons, and peak
instantaneous luminosities reaching up to
$2\times10^{32}$~cm$^{-2}$s$^{-1}$. Such a scenario will allow the
accumulation of an integrated luminosity of 2~\fbi\ in about five
months, which will be used as the basis for all results shown in this
analysis. However, the more likely scenario currently is a staged
construction with a reduced setup at the start of operations without
the RESR until the full design can be realized. In this case, the HESR
will be used as an accumulator in addition to its original task of
cooling the beam and storing it for experiments with internal
targets. This will result in a luminosity that is about a factor ten
lower than the full design goal during the initial phases of operating
FAIR\@.

\subsection{\label{sec:pnd_setup}The \PANDAbf\ Detector}

\begin{figure*}
  \includegraphics[width=\textwidth]{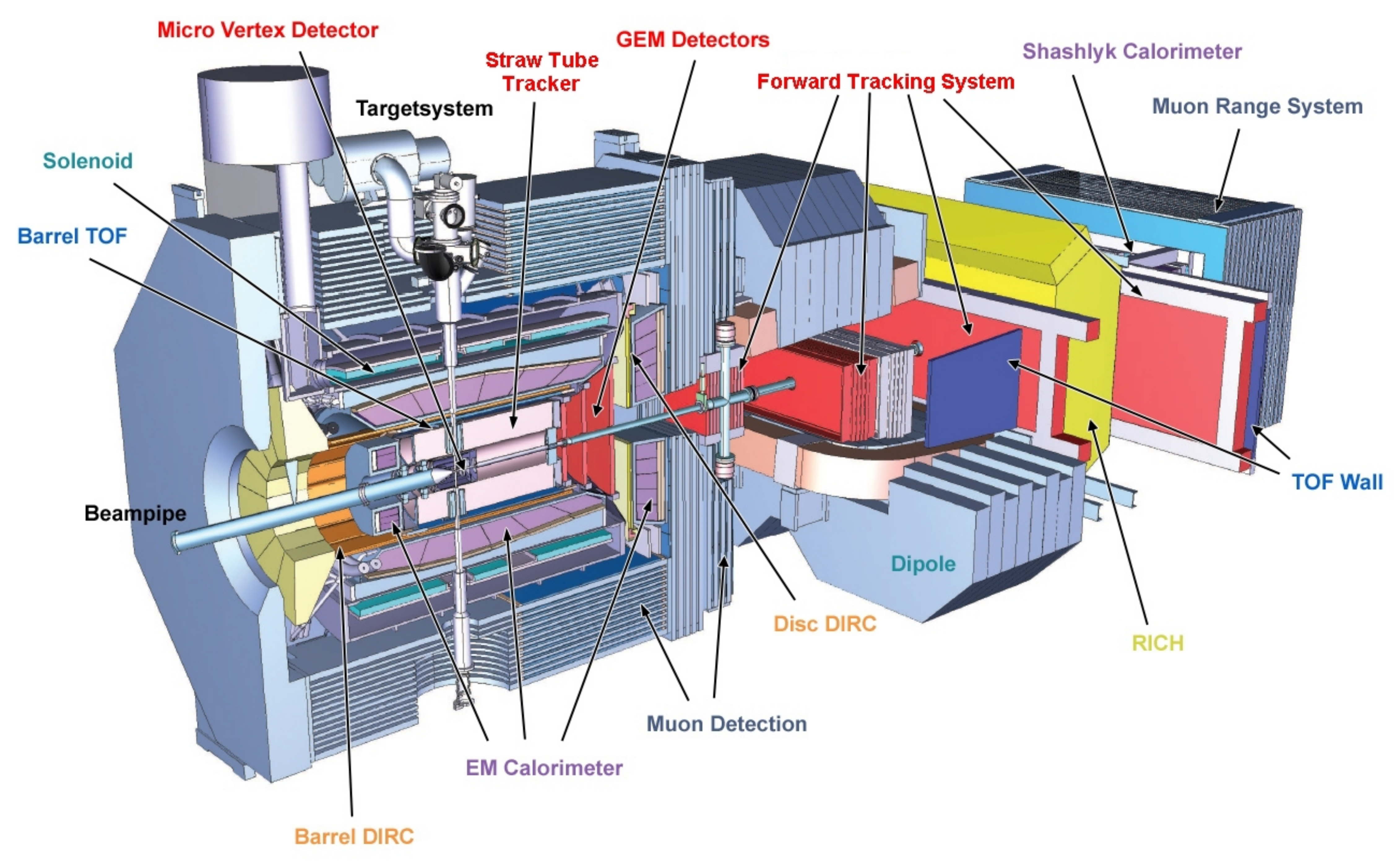}
  \caption{\label{fig:pnd_setup} The proposed \PANDA\ experimental
    setup~\cite{PandaPhysBook09c}.}
\end{figure*}

The proposed \PANDA\ detector is depicted in
Fig.~\ref{fig:pnd_setup}. The discussion here will focus on the
subsystems that are particularly relevant for the presented
analysis. The \PANDA\ detector consists of the {\emph{target
    spectrometer}} surrounding the target area and the {\emph{forward
    spectrometer}} designed to detect particles in the forward
rapidity region. The target spectrometer is divided into a barrel
region with polar angle reach from 22$^\circ$ to 145$^\circ$, and an
endcap region that covers polar angles below 22$^\circ$, down to
10$^\circ$ in the horizontal plane and 5$^\circ$ in the vertical
plane. Particles with polar angles below the endcap coverage are
detected by the forward spectrometer. In addition, \PANDA\ will be
equipped with a Luminosity Monitor Detector (LMD) at very forward
angles, built for precise determination of both absolute and relative
time integrated luminosities.

Two technologies are being developed to provide a hydrogen target with
sufficient density that allows to reach the design luminosity within
the restricted space available~\cite{PandaPhysBook09c}. A target
thickness of $4\times10^{15}$ hydrogen atoms per cm$^2$ is required to
achieve a peak luminosity of $2\times10^{32}$~cm$^{-2}$s$^{-1}$
assuming 10$^{11}$ stored antiprotons in the HESR\@. The Cluster-Jet
Target system operates by pumping pressurized cold hydrogen gas into
vacuum through a Laval-type nozzle, leading to a condensation of
hydrogen molecules into a narrow jet of hydrogen clusters with each
cluster containing $10^3$~--~$10^6$ hydrogen molecules. The main
advantages of this setup are the homogeneous density profile and the
ability to focus the antiproton beam at the highest areal density
point. The Pellet Target system creates a stream of frozen hydrogen
micro-spheres (pellets) of diameter 25~--~40~$\mathrm{\mu}$m, that
cross the antiproton beam perpendicularly. The Pellet Target will be
equipped with an optical tracking system that can determine the vertex
position of individual events with high precision.

The innermost part of the barrel region is occupied by charged
particle tracking detectors, which in turn are surrounded by particle
identification detectors, followed by a solenoid magnet that generates
a nearly uniform 2~T field pointing in the direction of the beam. The
innermost layers of tracking are provided by the Micro Vertex Detector
(MVD)~\cite{PandaMVD12c}, based on silicon pixel detectors for the
innermost two layers, and a double sided strip detectors for the
remaining two layers. The Straw Tube Tracker (STT)~\cite{PandaSTT12c}
is constructed from aluminized mylar tubes with gold-plated tungsten
anode wires running along the axis. The tubes operate with an active
gas mixture composed of argon and CO$_2$ held at a pressure of 2~bar
allowing them to be mechanically self-supporting. The STT adds only
about X/X$_0\approx~1.2\%$ to the total radiation length of the
tracking system on top of the $\approx~10\%$ expected from the
MVD\@. The MVD and the STT also measure ionization energy loss by
charged particles in their layers. A truncated mean of the specific
energy loss of the tracks measured by the MVD and STT layers is used
to estimate the energy loss $dE/dx$ for each track.

The barrel region tracking subsystems are immediately surrounded by
various dedicated particle identification (PID) detectors. The
innermost PID detector that is used in this analysis is the barrel
DIRC~\cite{Hoek20149} (Detection of Internally Reflected Cherenkov
light), where particles are identified by the size of the Cherenkov
opening angle. The DIRC is followed by the barrel Electromagnetic
Calorimeter (EMC)~\cite{PandaEMC08c}, constructed from lead tungstate
(PbWO$_4$) doped crystals, operated at a temperature of -25$^\circ$C
to optimize the light yield. Photons from each crystal in the barrel
EMC are detected by a pair of APDs (Avalanche Photodiodes). The EMC
constitutes the most powerful detector for the identification of
electrons through the momentum-energy correlation, particularly at
momenta higher than $\approx$~1~GeV/$c$. The DIRC, together with the
MVD and STT $dE/dx$ measurements, ensure coverage at lower momenta
where the EMC electron identification (EID) capacity is weaker.

The design of the barrel spectrometer of \PANDA\ also includes a Muon
Range System (MRS) surrounding the Solenoid for the identification of
muons as well as Time of Flight (TOF) detectors between the tracking
layers and the DIRC for general PID\@. The MRS and TOF are not used in
the analysis presented here.

For the endcap section of the target spectrometer, tracking points are
provided by the STT as well as a set of four disc shaped MVD layers,
and three chambers of Gas Electron Multiplier (GEM) trackers. PID is
performed using information from the endcap EMC and the endcap
DIRC~\cite{Merle201496} (also called Disc DIRC in reference to its
geometrical shape). Apart from its location, and angular coverage, the
Disc DIRC operates using the same basic principle as the barrel
DIRC\@. The endcap EMC is instrumented using the same PbWO$_4$
crystals as the barrel EMC, however photon detection is performed
using APDs only for the outer lying crystals. The crystals close to
the beam axis are readout by VPTs (Vacuum Phototriodes) because of the
stringent requirements on radiation hardness there.

The design of \PANDA\ also provides for coverage at angles below those
of the target spectrometer ($<5^\circ$), through the Forward Tracking
System (FTS), a Ring Imaging Cerenkov (RICH) detector system and a
Shashlyk calorimeter. Charged particles traversing the forward
tracking system are subject to a field integral of 2~Tm generated by a
dipole magnet, allowing for momentum determination. For this analysis,
the forward Shashlyk calorimeter was not included in the
simulations. As a result, our efficiency prediction is underestimated
for events whose kinematics leads to charged particles requiring
energy measurement for identification in the extremely forward
direction. With the full \PANDA\ setup, the performance for such
events will be better than that reported in this paper.

Precise determination of integrated luminosity is a critically
important ingredient for the whole \PANDA\ physics program. The LMD is
a detector that has been designed to provide both absolute and
relative time integrated luminosity measurements with 5\% and 1\%
systematic uncertainty, respectively~\cite{Panda:LumiTdr}. The LMD
will rely on the determination of the differential cross section of
antiproton -- proton elastic scattering into a polar angle (laboratory
reference frame) range of 3.5~--~8~mrad and full azimuth to achieve
this goal. The LMD tracks antiprotons using four planes of HV-MAPS
(High Voltage Monolithic Active Pixel Sensors) tracking stations, a
setup chosen to fulfil the constraints of high spatial resolution and
low material budget. The LMD will be placed 10.5~m downstream from the
interaction point within a vacuum sealed enclosure to reduce
systematic uncertainties from multiple scattering.

\subsection{\label{sec:trk_and_pid}Simulation and Analysis Software Environment}

The simulation and analysis software framework called
PandaRoot~\cite{Bertini:PandaRoot,Spataro:PandaRoot} is used for the
feasibility study described here. PandaRoot is a collection of tools
used for the simulation of the transport of particles through a
GEANT4~\cite{Agostinelli:2002hh} implementation of the \PANDA\
detector geometry, as well as detailed response simulation and
digitization of hits in the various detector elements that takes into
account electronic noise. Software for the reconstruction of tracks
based on the simulated tracking detector hit points is implemented in
PandaRoot, as well as the association of reconstructed tracks to
signal in outer PID detectors. A PID probability is assigned to each
track based on the response in all the outer detectors, using a
simulation of five possible particle species: $\splist$. Clusters in
the electromagnetic calorimeter that are not associated to a
reconstructed track are designated as neutral candidates, and used for
the reconstruction of photons. The simulation studies are done in the
$\epem$ decay channel for the $\jpsi$ due to much higher EID
efficiency as compared to muon identification efficiency, at a fixed
pion rejection probability. This is particularly pertinent for
$\sigrxnshort$, due to the very high pionic background event rates, as
will be discussed in Section~\ref{sec:bg_prop}.

The tracking points from the tracking detectors are used for pattern
recognition to find charged particle tracks. Points that are found to
belong to the same track are in a first step fitted to a simplified
helix for an initial estimate of the momentum, which is then used as a
starting point for an iterative Kalman Filter procedure relying on the
GEANE~\cite{GeaneNote} track follower. The output from the Kalman
Filter is a more refined estimate of the momentum of tracks that takes
into account multiple scattering as well as changes in curvature due
to energy loss in the detector material.

Since the Kalman Filter does not take into account non-Gaussian
alterations of track parameters, it can not correctly handle changes
of track momentum through Bremsstrahlung energy loss. This is
particularly pernicious for electrons that can lose on average up to
10\% of their total energy through photon emission. To correct this
effect event by event, a procedure was developed~\cite{Ma:2014iza}.
For each track this algorithm looks for potential Bremsstrahlung
photon candidates in the EMC, and adds that energy to the track. This
method was demonstrated to work over a wide range of momenta and
angles including those relevant for this analysis.

\section{\label{sec:sig_prop}Theoretical Description of the Signal Channel}

The feasibility study presented here is carried out at three values of
the square of the center-of-mass (c.m.) energy $s$: $12.3$~GeV$^2$,
$16.9$~GeV$^2$ and $24.3$~GeV$^2$. The first value is chosen to
coincide with existing data from E835 for
$\sigrxnshort$~\cite{Armstrong:1992ae,Joffe:2004ce,Andreotti:2005vu}.
The remaining two values are chosen at the incident $\pbar$ momenta of
$8$~GeV/$c$ and $12$~GeV/$c$, respectively, to explore the kinematic
zone between the first point and the maximum available $\bar{p}$
momentum at FAIR of $15$ GeV/$c$.

\subsection{Kinematics}

In order to present the cross section estimates within the description
based on the $\pi N$ TDAs employed for our feasibility study, we would
like to review briefly the kinematics of the signal reaction:
\begin{eqnarray}
\label{eq:reac}
 N (p_N) \;+ \bar N (p_{\bar N}) \; \to  J/\psi(p_{\psi})\;+\; \pi(p_{\pi}),
\end{eqnarray}
\noindent
making special emphasis on the kinematic quantities employed in the
collinear factorization approach. The natural hard scale for the
reaction~(\ref{eq:reac}) is introduced by the c.m.\@ energy squared
$s={(p_N+p_{\bar{N}})}^2$ and the charmonium mass squared
$M^2_\psi$. The collinear factorized description is supposed to be
valid in the two distinct kinematic regimes, corresponding to the
generalized Bjorken limit (large $s$ and $Q^2 \equiv M^2_\psi$ for a
given $s/Q^2$ ratio and small cross channel momentum transfer
squared):
\begin{itemize}
\item the near-forward kinematics $|t| \equiv
  |{(p_\pi-p_{\bar{N}})}^2| \ll s,\ M^2_\psi$; it corresponds to the
  pion moving almost in the direction of the initial antinucleon in
  the $N \bar{N}$ center-of-mass system (CMS);
\item the near-backward kinematics $|u| \equiv |{(p_\pi-p_{N})}^2| \ll
  s,\ M^2_\psi $ corresponding to the pion moving almost in the
  direction of the initial nucleon in the $N \bar{N}$ CMS\@.
\end{itemize}

Due to the charge-conjugation invariance of the strong interaction
there exists a perfect symmetry between the near-forward and
near-backward kinematic regimes of the reaction~(\ref{eq:reac}). These
two regimes can be considered in exactly the same way, and the
amplitude of the reaction within the $u$-channel factorization regime
can be obtained from that within the $t$-channel factorization regime,
with the help of the obvious change of the kinematic variables (see
Eq.~(\ref{eq:tu_change}) below). In the $N \bar{N}$ CMS these two
regions look perfectly symmetric. However, we note that \PANDA\
operates with the antibaryon at beam momentum and the baryon at rest
in the lab frame. Consequently, the symmetry between the near-forward
and near-backward kinematics is not seen immediately in the \PANDA\
detector. Moreover, this introduces acceptance differences between the
two regimes which will be explored in Section~6 as a function of the
incident $p_{\bar N}$ momentum.

For definiteness below, we consider the near-forward ($t$-channel)
kinematic regime. The detailed account of the relevant kinematic
quantities is presented in Appendix~A of
Ref.~\cite{Lansberg:2012ha}. It is convenient to choose the $z$-axis
along the nucleon-antinucleon colliding path, selecting the direction
of the antinucleon as the positive direction. Introducing the
light-cone vectors $p^t$ and $n^2$ ($2 p^t \cdot n^t=1)$, one can
perform the Sudakov decomposition of the particle momenta. Neglecting
the small mass corrections (assuming $m_\pi=0$ and $m_N \ll \sqrt{s}$) we
obtain:
\begin{eqnarray}
  && p_{\bar{N}}=(1+\xi^t)p^t; \ \ \ p_N=\frac{s}{1+\xi}n^t; \nonumber \\
  && p_\pi=(1-\xi^t)p^t; \ \ \ p_\psi=2 \xi^t p^t+ \frac{s}{1+\xi}n^t,
\end{eqnarray}
\noindent
where $\xi^t$ stands for the $t$-channel skewness variable, which
characterizes the $t$-channel longitudinal momentum transfer:
\begin{eqnarray}
\xi^t \equiv - \frac{(p_\pi-p_{\bar{N}}) \cdot
  n^t}{(p_\pi+p_{\bar{N}}) \cdot n^t}.
\label{eq:skewdness}
\end{eqnarray}

We also introduce the transverse (with respect to the selected
$z$-axis) $t$-channel momentum transfer squared
${(\Delta_T^t)}^2$. This quantity can be expressed in terms of the
skewness variable $\xi$ from Eq.~(\ref{eq:skewdness}) and the
$t$-channel momentum transfer squared ${(\Delta^t)}^2 \equiv t$ as:
\begin{eqnarray}
  {{(\Delta_T^t)}^2}= \frac{1-\xi^t}{1+\xi^t}
  \left({(\Delta^t)}^2- 2\xi \left[\frac{m_N^2}{1+\xi^t}
  -\frac{m_\pi^2}{1-\xi^t} \right] \right).
\label{eq:Delta_T_expr}
\end{eqnarray}

For ${(\Delta_T^t)}^2=0$ the momentum transfer is purely longitudinal
and hence the pion is produced exactly in the forward direction. This
corresponds to the maximal possible value of the momentum transfer
squared:
\begin{eqnarray}
  \Delta^2_{\max} \equiv \frac{2 \xi^t \left( {m_N^2} (\xi^t
  -1)+m_\pi ^2 (\xi^t +1)\right)}{{\xi^t} ^2-1}.
\label{eq:Dmax}
\end{eqnarray}

The $t$ -channel skewness variable can be expressed through the
reaction invariants as:
\begin{eqnarray}
  \xi^t \simeq \frac{M_\psi^2-t-m_N^2}{2s-M_\psi^2+t-3 m_N^2}.
\label{eq:xi_more_acc}
\end{eqnarray}

In the present study, following Ref.~\cite{Pire:2013jva} we neglect
all $t/s$ and $m_N^2/s$ corrections, and employ the simple
expression for the skewness variable:
\begin{eqnarray}
  \xi^t \equiv - \frac{(p_\pi-p_{\bar{N}}) \cdot
  n^t}{(p_\pi+p_{\bar{N}}) \cdot n^t} \simeq \frac{M^2_\psi}{2s
  -M^2_\psi}.
\label{eq:xi_const}
\end{eqnarray}

In order to apply the same formalism for the $u$-channel
(near-backward) kinematic regime, it suffices to perform the following
variable transformations in the relevant formula:
\begin{eqnarray}
  && p_N \to p_{\bar{N}}; \ \ \ p_{\bar{N}} \to
     p_{{N}}; \nonumber \\
  && \Delta^t \equiv (p_\pi-p_{\bar{N}})
     \to \Delta^u \equiv (p_\pi-p_N); \nonumber \\
  && t \to
     u; \ \ \ \xi^t \to \xi^u.
\label{eq:tu_change}
\end{eqnarray}

Therefore, in what follows we omit the superscript referring to the
kinematic regime for the kinematic variables.

The generalized Bjorken limit, in which the validity of the collinear
factorized description of the reaction~(\ref{eq:reac}) is assumed, is
defined by the requirement $\Delta^2 \ll s, \, Q^2 \equiv M_\psi^2$.
There is no explicit theoretical means to specify quantitatively the
condition $|\Delta^2| \ll s, \, Q^2$. However, the common practice
coming from studies of the similar reactions suggests
$|\Delta^2|<1$~GeV$^2$ can be taken as a reasonable estimate. For the
fixed value of the skewness parameter this condition can be translated
into the corresponding kinematic cut for $\Delta_T^2$ or,
equivalently, to cuts in the pion scattering angle in the CMS and
(after the appropriate boost transformation) in the lab frame. This
allows to specify the span of the ``forward'' and ``backward'' cones
in which the collinear factorized description is supposed to be valid
for the reaction~(\ref{eq:reac}).

However, once such a kinematic cut has been implemented, one has to be
prudent. Indeed, the kinematic formulas derived in the Appendix~A of
Ref.~\cite{Lansberg:2012ha} represent the approximation, which is
valid in the generalized Bjorken limit $s, \, Q^2 \to \infty$ while
its validity for a given kinematic setup is not necessarily
ensured. Certainly it is useless to employ the approximate kinematic
formulas in the region where the approximation does not provide a
satisfactory description of the kinematic quantities.

Therefore, it is instructive to compare the approximate result for
${(\Delta_T^t)}^2$ given by Eq.~(\ref{eq:Delta_T_expr}) with $\xi^t$
given by Eq.~(\ref{eq:xi_more_acc}) and by the less accurate
expression from Eq.~(\ref{eq:xi_const}) with the general result
obtained with the help of the exact kinematic relation for the CMS
pion scattering angle $\theta_\pi^*$:
\begin{eqnarray}
  {(\Delta_T^t)}^2 \big|_{\rm exact} = -|\vec{p}_\pi|^2(1-\cos^2
  \theta_\pi^*).
\label{eq:Delta_T_exact}
\end{eqnarray}

Here,
$$
|\vec{p}_\pi| = \frac{1}{2\sqrt{s}} \Lambda(s,\ M_\psi^2,\ m_\pi^2)
$$
denotes the CMS momentum of the produced pion, where:
$$
\Lambda(x,y,z) = \sqrt{x^2+y^2+z^2-2xy-2xz-2yz}
$$
is the usual Mandelstam function, and the $s$-channel CMS scattering
angle is expressed as:
\begin{eqnarray}
  \cos \theta_\pi^*=-
  \frac{s(2m_N^2+m_\pi^2+M_\psi^2-s-2t)}{\Lambda(s,\ m_N^2,\ m_N^2)\Lambda(s,\ M_\psi^2,\ m_\pi^2)}.
\end{eqnarray}

\begin{figure}[hbpt]
  \includegraphics[width=\columnwidth]{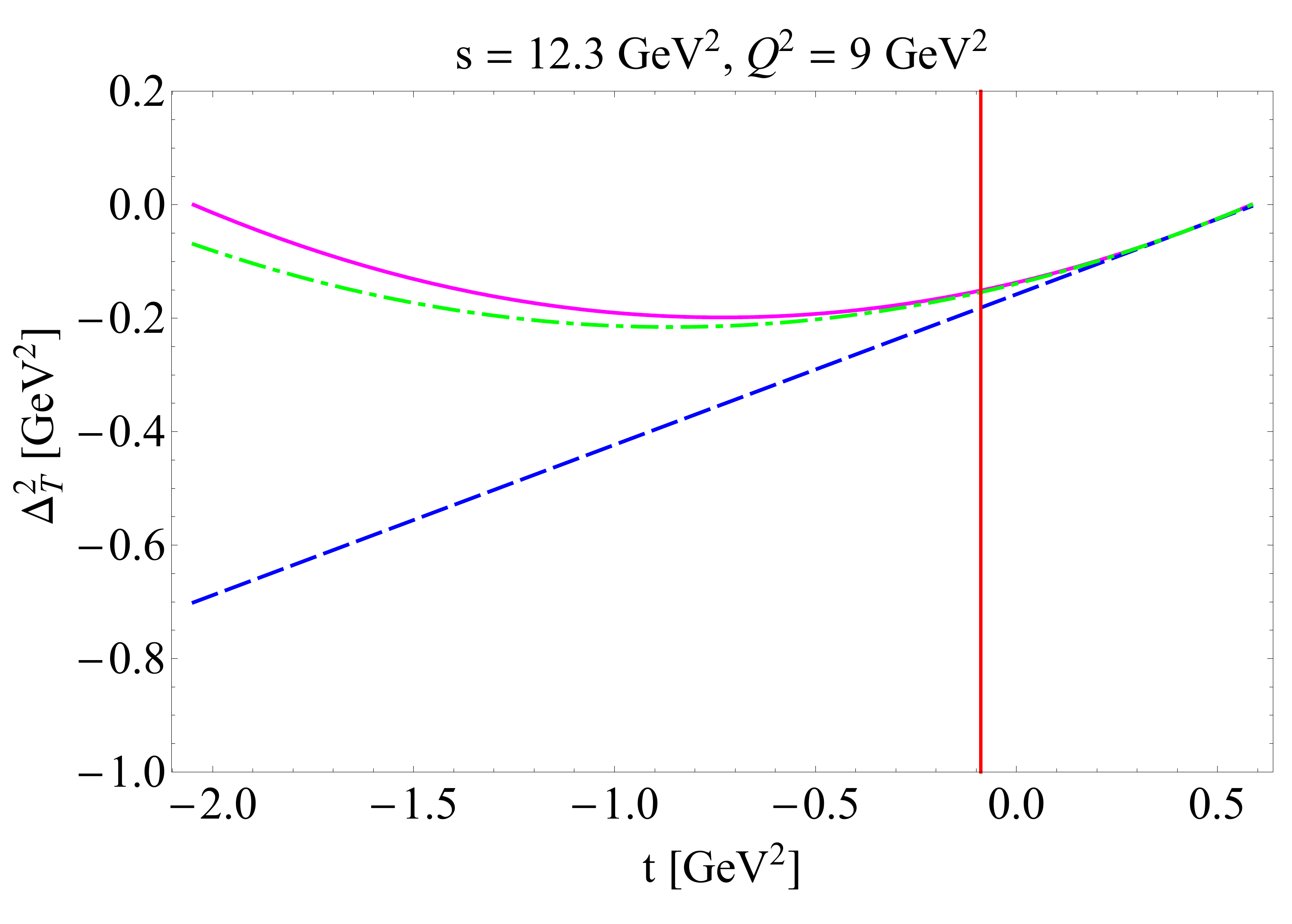}\\
  \includegraphics[width=\columnwidth]{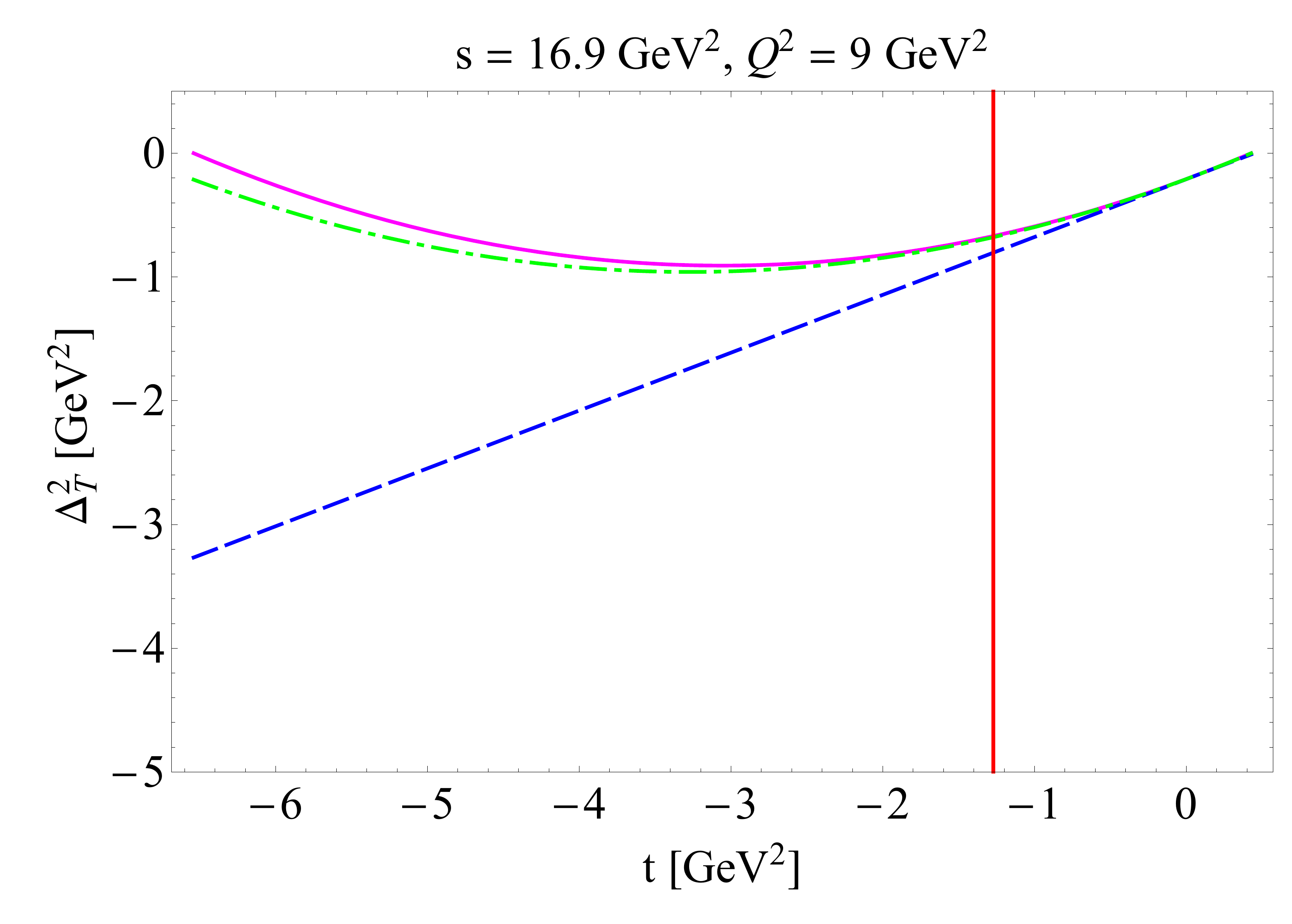}\\
  \includegraphics[width=\columnwidth]{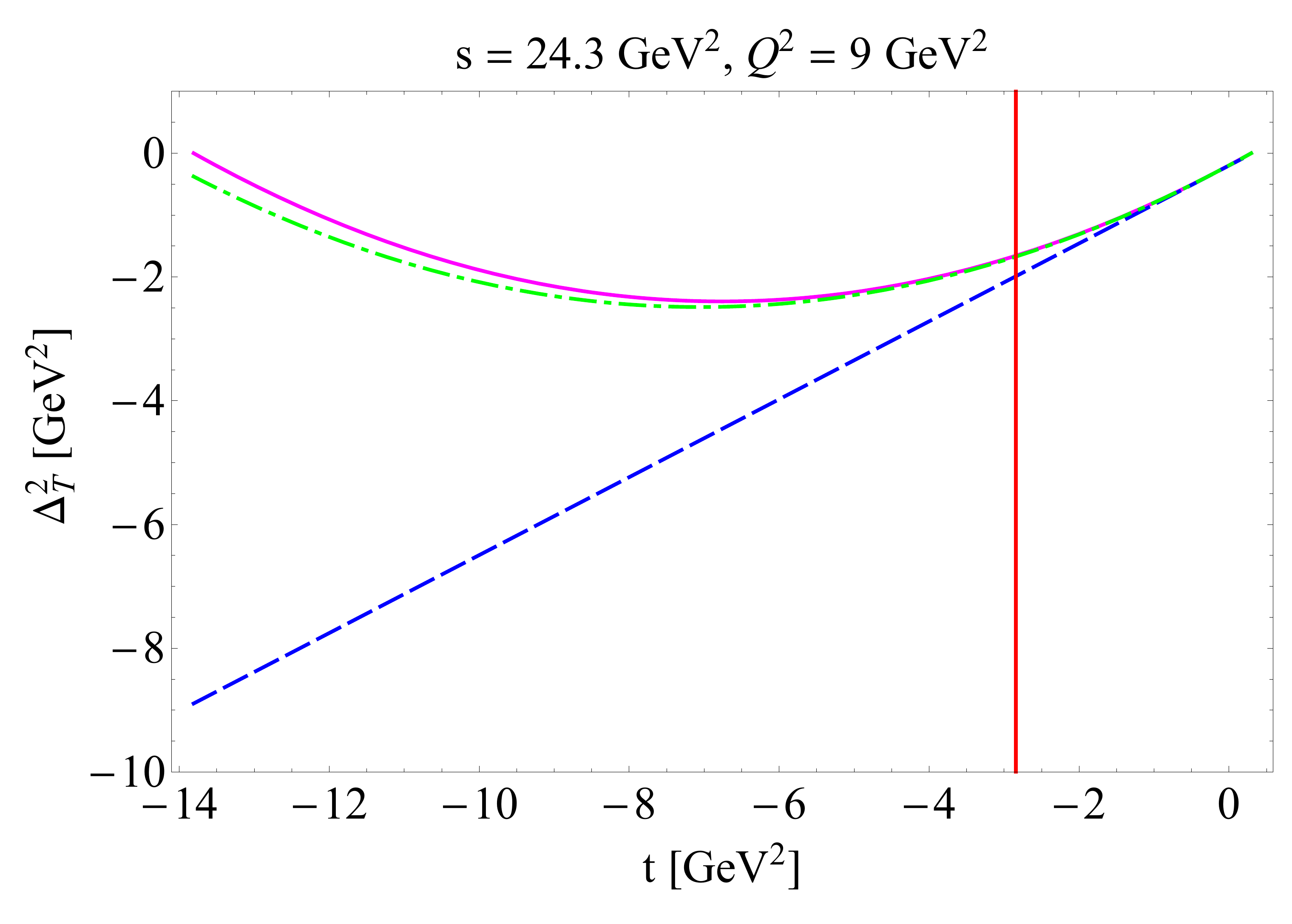}
  \caption{\label{fig:xi_approx}\small Transverse momentum transfer
    squared ${(\Delta_T^t)}^2$ as the function of $t={(\Delta^t)}^2$
    computed from the exact formula in Eq.~(\ref{eq:Delta_T_exact})
    \plfm{solid line} compared to the approximate result of
    Eq.~(\ref{eq:Delta_T_expr}) with $\xi$ expressed by
    Eq.~(\ref{eq:xi_more_acc}) \plfm{dash-dotted line} and approximate
    expression for $\xi$ in Eq.~(\ref{eq:xi_const}) \plfm{dashed
      line}. The functions are drawn for an incident $\bar{p}$
    momentum of \bmomu{0} \plfm{top}, \bmomu{1} \plfm{middle},
    \bmomu{2} \plfm{bottom}. The minimum value of $t$ for which these
    approximations are valid is indicated by the solid vertical
    lines.}
\end{figure}

Figure~\ref{fig:xi_approx} shows the comparison of ${(\Delta_T^t)}^2$
computed using the approximate formula from
Eq.~(\ref{eq:Delta_T_expr}) with the exact result from
Eq.~(\ref{eq:Delta_T_exact}) for the three selected values of
$s$. We plot ${(\Delta_T^t)}^2$ as a function of $\Delta^2$ for
$\Delta^2 \le \Delta^2_{\max}$, where $\Delta^2_{\max}$ is the maximal
kinematically accessible value of $\Delta^2$, corresponding to
$\Delta_T=0$ ({\it i.e.}, the pion produced exactly in the forward
direction). The validity limit of our kinematic approximation is shown
in Fig.~\ref{fig:xi_approx} by the solid vertical lines. It is
calculated by imposing the maximum allowable deviation of 20\% on the
value of ${(\Delta_T^t)}^2$ from the exact result from
Eq.~(\ref{eq:Delta_T_exact}).  The final validity range
$\Delta_{\min}^2$ is determined by picking the more conservative limit
of the kinematic approximation and the standard constraint $|\Delta^2|
\le 1$~GeV$^2$, ensuring the smallness of the $|\Delta^2|$ comparing
to $s$ and $Q^2=M_\psi^2$ in the generalized Bjorken limit.

Table~\ref{tab:validity} summarizes the valid ranges of $\Delta^2$ in
which we are going to apply the factorized description of the
reaction~(\ref{eq:reac}) for the three selected energies considered
here. The lower limit comes from the applicability of collinear
factorization (Bjorken limit) for the lowest beam momentum
(5.5~GeV/$c$) and from the shortcoming of the approximation employed
for the kinematic quantities (kinematic constraints) for the higher
two beam momenta (8 and 12~GeV/$c$). The last two columns show how
these limits translate to the polar angle of the $\piz$ in the lab
frame, $\theta^{\piz}_{lab}$, namely the maximum (minimum) valid
$\theta^{\piz}_{lab}$ in the near-forward (near-backward) validity
range. At the other end, the minimum (maximum) valid polar angle in
the near-forward (near-backward) validity range is $0^{\circ}$
($180^{\circ}$) for all energies.

\begin{table}[hbpt]
  \caption{\label{tab:validity} The kinematic range in which we assume
    the validity of the factorized description of the signal channel
    in terms of $\pi N$~TDAs and nucleon DAs for the three values of
    the incident $\bar{p}$ momentum employed in the present study. The
    last two columns give the limits in terms of the polar angle of the
    $\piz$ in the lab frame for the near-forward and near-backward
    regimes.}
  \begin{ruledtabular}
    %\begin{tabular}{|c|c|c|c|c|c|}
    \begin{tabular}{cccccc}
      %\toprule
      $s$ & $\mompbar$ & $\Delta^2_{\min}$ & $\Delta^2_{\max}$ & $\theta^{\piz}_{lab,\max}$ & $\theta^{\piz}_{lab,\min}$\\
      (GeV$^2$) & (GeV/$c$) & (GeV$^2$) & (GeV$^2$) & Fwd. & Bwd. \\
      \colrule
      %\midrule
      12.3 & \bmom{0} & -0.092 & 0.59 & 23.2$^\circ$ & 44.6$^\circ$ \\
      16.9 & \bmom{1} & -1.0 & 0.43 & 15.0$^\circ$ & 48.1$^\circ$ \\
      24.3 & \bmom{2} & -1.0 & 0.3 & 7.4$^\circ$ & 62.3$^\circ$ \\
      %\bottomrule
    \end{tabular}
  \end{ruledtabular}
\end{table}

\subsection{Cross Section Estimates within the Collinear Factorization Approach}

\begin{figure}[hbpt]
\includegraphics[width=0.92\columnwidth]{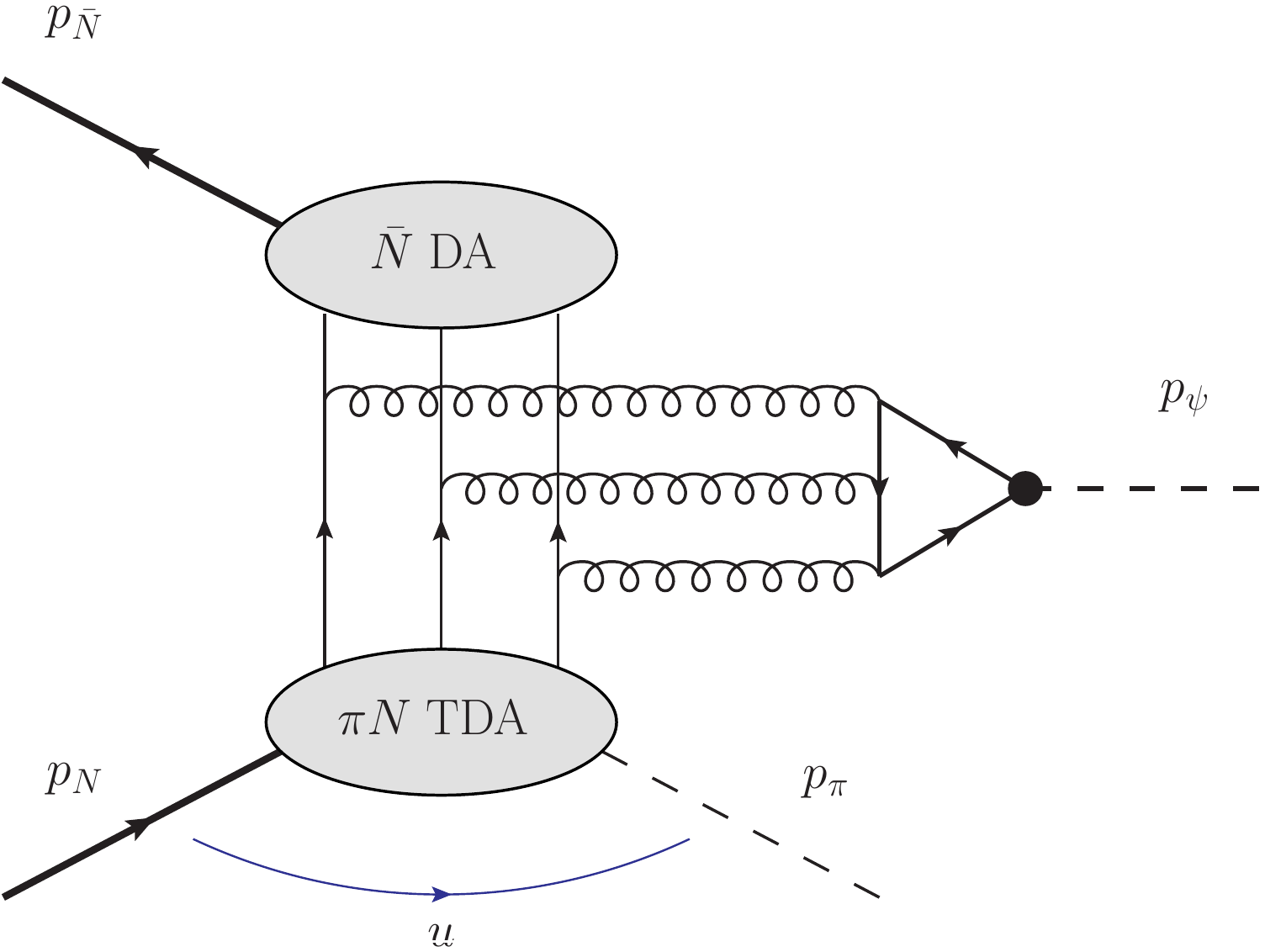}
\includegraphics[width=0.92\columnwidth]{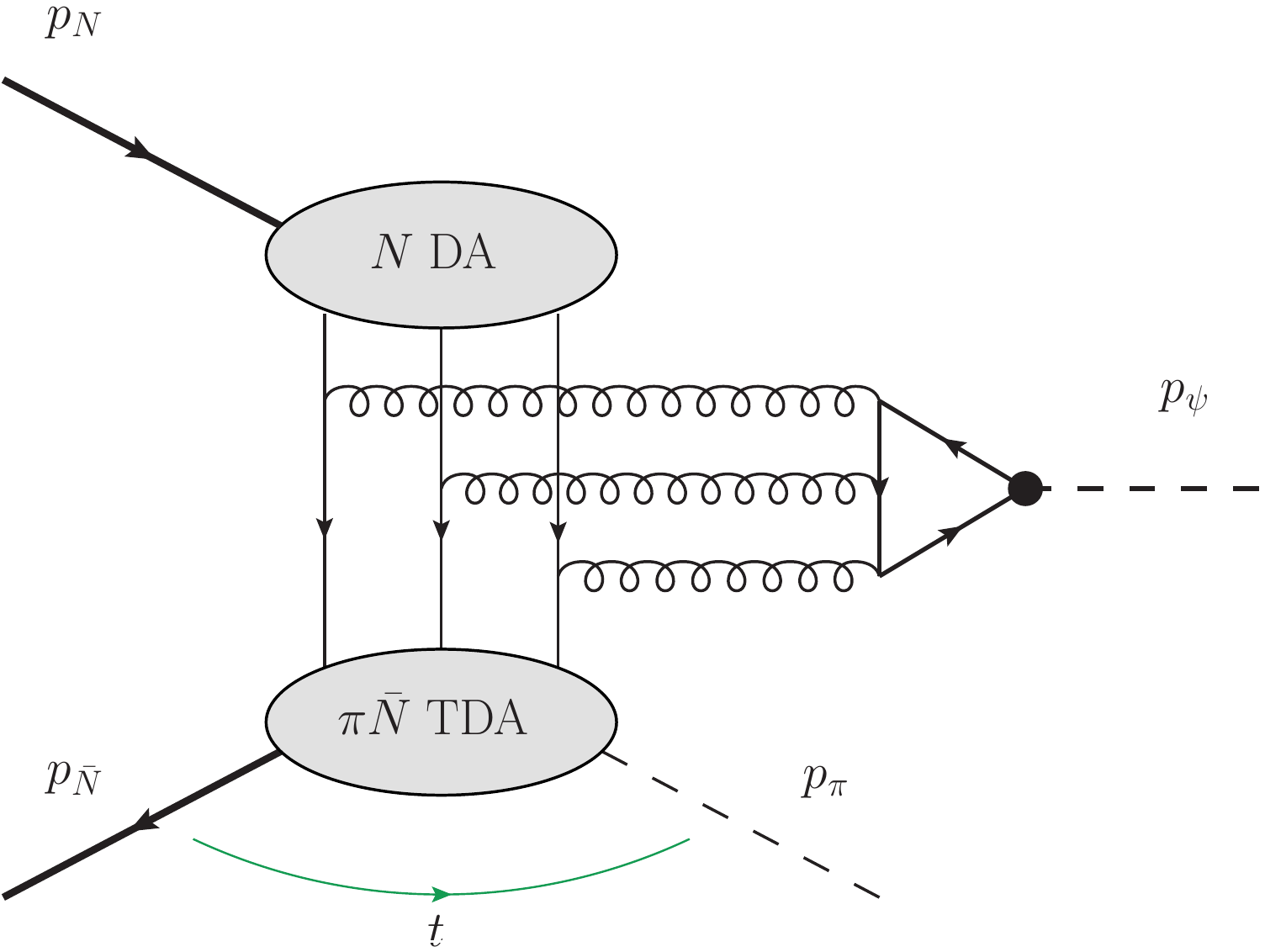}
\caption{\label{fig:factorization}\small Collinear factorization of
  the annihilation process
  $\bar{N}(p_{\bar{N}})N(p_N)\to\jpsi(p_{\psi})\pi(p_{\pi})$.  Top
  panel: near-backward kinematics ($u \sim 0$). Bottom panel:
  near-forward kinematics ($t \sim 0$). $\bar{N}(N)$ DA stands for the
  distribution amplitude of antinucleon (nucleon); $\pi N (\pi
  \bar{N})$ TDA stands for the transition distribution amplitude from
  a nucleon (antinucleon) to a pion. Figures reproduced from
  Ref.~\cite{Pire:2013jva}.}
\end{figure}

The calculation of the $N+\bar{N}\to\jpsi+\pi$ cross section within
the collinear factorization approach follows the same main steps as
those in the calculations of the \jpsitopbarp\ decay width in the
perturbative QCD (pQCD) approach
\cite{Brodsky:1981kj,Chernyak:1983ej,Chernyak:1987nv}. The small and
large distance dynamics is factorized, and the corresponding amplitude
is presented as the convolution of the hard part, computed in the
pQCD, with the hadronic matrix elements of the QCD light-cone
operators ($\pi N$ TDAs and nucleon DAs) encoding the long distance
dynamics (see Fig.~\ref{fig:factorization}). The hard scale, which
justifies the validity of the perturbative description of the hard
subprocess, is provided by the mass of heavy quarkonium $M_\psi \simeq
2m_c \simeq \bar{M} = 3$~GeV.

Within the approximation to order leading twist-three, only the
transverse polarization of the $J/\psi$ is relevant, and the cross
section with the suggested reaction mechanism for either the
near-forward or the near-backward kinematic regimes
reads~\cite{Pire:2013jva}:
\begin{eqnarray}
  \frac{d \sigma}{d \Delta^2}= \frac{1}{16 \pi
  \Lambda^2(s,\ m_N^2,\ m_N^2) } |\overline{\mathcal{M}_{T}}|^2,
\label{eq:CS_def_delta2}
\end{eqnarray}
\noindent
where the squared matrix element $|\overline{\mathcal{M}_{T}}|^2$ is
expressed as:
\begin{equation}
  |\overline{\mathcal{M}_{T}}|^2 = \frac{1}{4} |\mathcal{C}|^2
  \frac{2(1+\xi)}{\xi {\bar{M}}^8} \left( |\mathcal{I}(\xi,\ \Delta^2)|^2
  - \frac{\Delta_T^2}{m_N^2} |\mathcal{I}'(\xi,\ \Delta^2)|^2 \right).
\end{equation}

Here the factor $\mathcal{C}$ reads:
\begin{eqnarray}
  {\cal C}= {(4 \pi \alpha_s)}^3
  \frac{f_N^2 f_\psi}{f_\pi} \, \frac{10}{81},
\label{eq:xsect_formula}
\end{eqnarray}
\noindent
where $f_\pi=93$~MeV is the pion weak decay constant; $f_\psi=(413
\pm 8)$~MeV is the normalization constant of the non-relativistic
light-cone wave function of heavy quarkonium; $\alpha_s$ stands for
the strong coupling. The two functions\linebreak $\mathcal{I}(\xi,\
\Delta^2)$ and $\mathcal{I}'(\xi,\ \Delta^2)$ denote the convolutions
of the hard scattering kernels with the $\pi N$ TDAs and (anti)nucleon
DAs. In our studies we employ the estimate of the cross section from
Eq.~(\ref{eq:CS_def_delta2}) within the simple cross channel nucleon
exchange model for the $\pi N$ TDAs~\cite{Pire:2011xv}. In this model
the convolution integrals $\mathcal{I}, \mathcal{I}'$ read as:
\begin{eqnarray}
  && {\cal I}(\xi, \Delta^2)= \frac{ f_\pi \, g_{\pi NN} m_N (1-\xi)
     } { (\Delta^2-m_N^2) (1+\xi )} M_0; \nonumber \\
  && {\cal I}'(\xi,
     \Delta^2)= \frac{ f_\pi \, g_{\pi NN} m_N } { (\Delta^2-m_N^2) } M_0,
\end{eqnarray}
\noindent
where $f_N=(5.0 \pm 0.5)$~GeV$^2$ is the nucleon wave function
normalization constant; $g_{_{\pi NN}} \simeq 13$ is the
phenomenological pion-nucleon coupling; $M_0$ is the standard
convolution integral of nucleon DAs occurring in the expression for
the \jpsitopbarp\ decay width within the pQCD approach of
Ref.~\cite{Chernyak:1987nv}:
\begin{eqnarray}
  \Gamma(\jpsitopbarp)= {(\pi \alpha_s)}^6 \frac{1280
  f_\psi^2 f_N^4 }{243 \pi {\bar{M}^9}} |M_0|^2.
\label{eq:Charm_dec_width}
\end{eqnarray}

In order to compute the value of the cross section given by
Eq.~(\ref{eq:CS_def_delta2}), one has to employ the phenomenological
solutions for the leading twist nucleon DAs to compute the convolution
integral $M_0$ and to specify the appropriate value of the strong
coupling $\alpha_s$ for the characteristic virtuality of the
process. Unfortunately, some controversy on this issue exists in
recent literature (see {\it e.g.} the discussion in Chapter~4 of
Ref.~\cite{Brambilla:2004wf}). There are several classes of
phenomenological solutions for the leading twist nucleon DAs:

\begin{itemize}
\item one class (usually referred as the Chernyak-Zhitnitsky-type
  solutions) contains nucleon DAs which differ considerably from the
  asymptotic form of the leading twist nucleon DAs. Such DAs require
  $\alpha_s \sim 0.25$ to describe the experimental charmonium decay
  width $\Gamma(\jpsitopbarp)$ from Eq.~(\ref{eq:Charm_dec_width});
\item another class of solutions (e.g. the Bolz-Kroll
  solution~\cite{Bolz:1996sw} and Braun-Lenz-Wittmann NLO
  model~\cite{Braun:2006hz}) contains nucleon DAs which instead are
  rather close to the asymptotic form and require $\alpha_s \sim 0.4$
  to reproduce the experimental $\Gamma(\jpsitopbarp)$.
\end{itemize}

For our rough cross section estimates, intended for the feasibility
studies, we follow the prescription from Ref.~\cite{Pire:2013jva}, and
employ the results for the $\sigrxnshort$ cross section with the value
of $\alpha_s$ fixed by the requirement that the given phenomenological
solution for the nucleon DAs reproduces the experimental
$\jpsitopbarp$ decay width. In this case the simple nucleon pole model
for $\pi N$ TDAs results in the same $\sigrxnshort$ cross section
predictions for any input phenomenological nucleon DA solution.

\begin{figure}[hbpt]
  \includegraphics[width=\columnwidth]{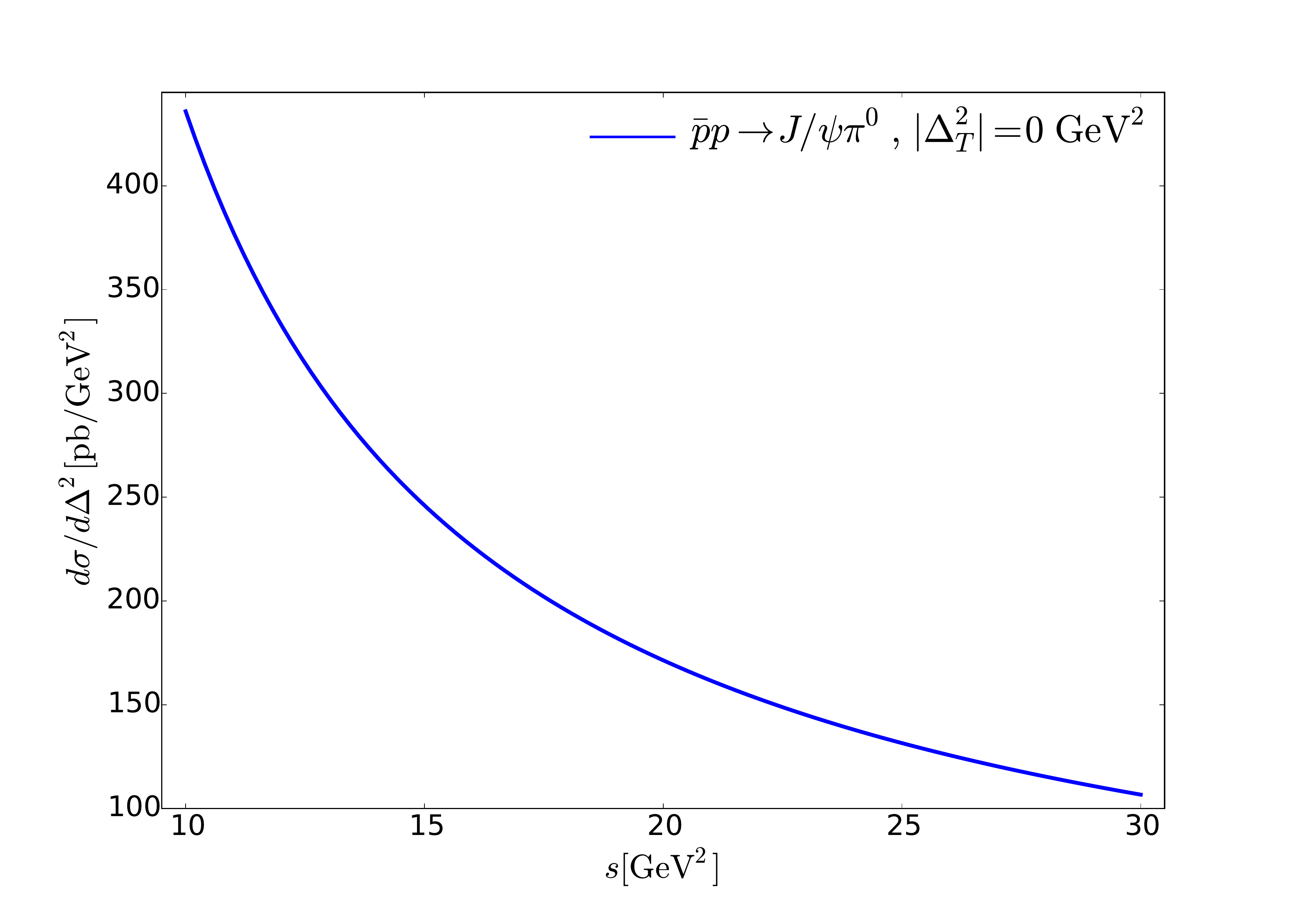}
  \caption{\label{fig:tda_wsq_dep} The $s$ dependence of the
    $\sigrxnshort$ differential cross section at $|\Delta_T^2|=0$~GeV/$c^2$
    as predicted by the calculations based on the TDA formalism given
    in Ref.~\cite{Pire:2013jva}.}
\end{figure}

\begin{figure}[hbpt]
  \includegraphics[width=\columnwidth]{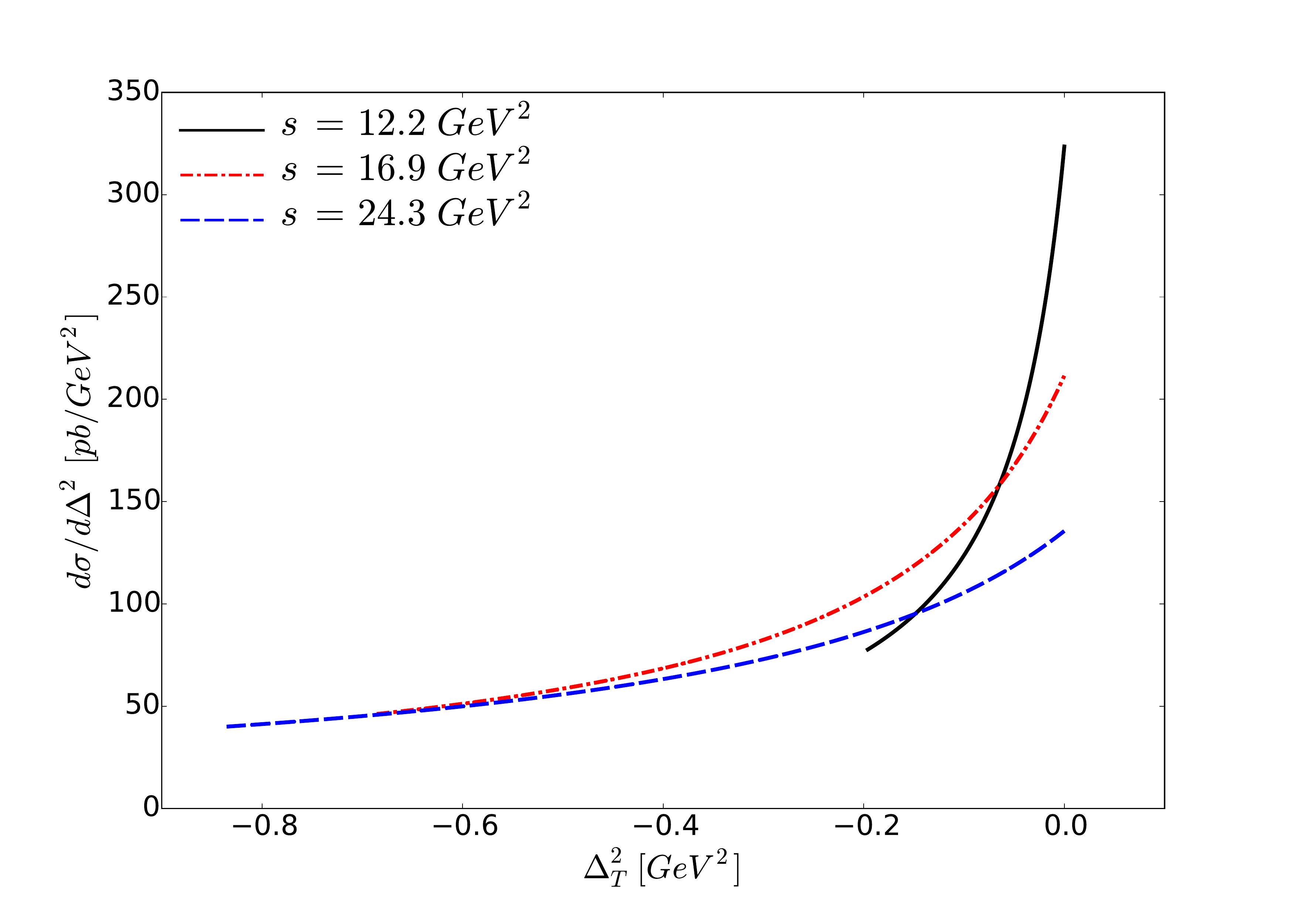}
  \caption{\label{fig:deltsq_dep} The $\Delta_T^2$ dependence of the
    differential cross section predicted by the TDA model for
    $\sigrxnshort$ at three incident $\pbar$ momenta. The curves have
    been limited to the validity range of the respective collision
    energy. The different colored lines show the dependence for the
    three collision energies considered for the study here:
    $s~=~12.3$~GeV$^2$ \plfm{solid line}, $s~=~16.9$~GeV$^2$
    \plfm{dash-dotted line}, $s~=~25.4$~GeV$^2$ \plfm{dashed line}.}
\end{figure}

Figure~\ref{fig:tda_wsq_dep} shows the prediction of the $s$
dependence of the differential cross section of $\sigrxnshort$ at
$|\Delta_T^2|=0$ which decreases only by a factor of about 4 between
the lowest to highest CMS collision energies that will be available at
\PANDA\@. Figure~\ref{fig:deltsq_dep} shows the $\Delta_T^2$
dependence of the cross section for the three collision energies which
are used for the studies addressed here, each of them limited to its
validity range. It is possible to compare this prediction to the E835
measurement of the $\sigrxn$ cross
section~\cite{Armstrong:1992ae,Joffe:2004ce,Andreotti:2005vu}, taken
at an incident $\pbar$ momentum of \bmomu{0} that corresponds to
$s~=~12.3$~GeV$^2$ or $\sqrt{s}~=~3.5$~GeV. This value of $\sqrt{s}$
is 25~MeV below the threshold for the production of the $h_c$
resonance. The reported values lie in the range from 90~pb to
230~pb. Integrating the differential cross section from the TDA model
at $s~=~12.3~$GeV$^2$ over its validity range
($-0.092<t~[$GeV$^2]<0.59$, which is smaller than the full
kinematically accessible range), we find 206.8~pb, combining
near-forward and near-backward kinematics. This value is on the upper
end of the measured range by E835. However, given that the majority of
the production rate from the TDA model prediction lies within the
validity range, this result gives a degree of confidence that the
model can be used as a basis for a feasibility study within its
validity range.

Let us stress that the validity of the factorized description of the
reaction~(\ref{eq:reac}) in terms of $\pi N$ TDAs and nucleon DAs has
only been conjectured. The corresponding collinear factorization
theorem has never been proven explicitly. Therefore, one of the
important experimental challenges is to establish evidence for the
validity of this description.  In general, there are several essential
marking signs for the onset of the collinear factorization regime for
a given hard exclusive reaction. The most obvious one is the
characteristic scaling behavior of the cross section with the relevant
virtuality $1/Q^2$. However, for the reaction~(\ref{eq:reac}) this
feature is of little use, since the virtuality is fixed by the mass of
the heavy quarkonium. Another opportunity is to look for the specific
polarization dependence.  For the case of the nucleon-antinucleon
annihilation into $J/\psi$ in association with a forward (or backward)
neutral pion it is the transverse polarization of the $J/\psi$ that is
dominant within the collinear factorized description in terms of $\pi
N$ TDAs. This dominating contribution manifests through the
characteristic $(1+\cos^2 \theta_\ell^*)$ distribution of the decay
leptons in the lepton pair CMS scattering angle $\theta_\ell^*$.  The
dominance of the corresponding polarization has to be verified by
means of a dedicated harmonic analysis.

\subsection{\label{sec:evt_gen}Event Generator}

\begin{figure*}[hbpt]
  \includegraphics[width=0.33\textwidth]{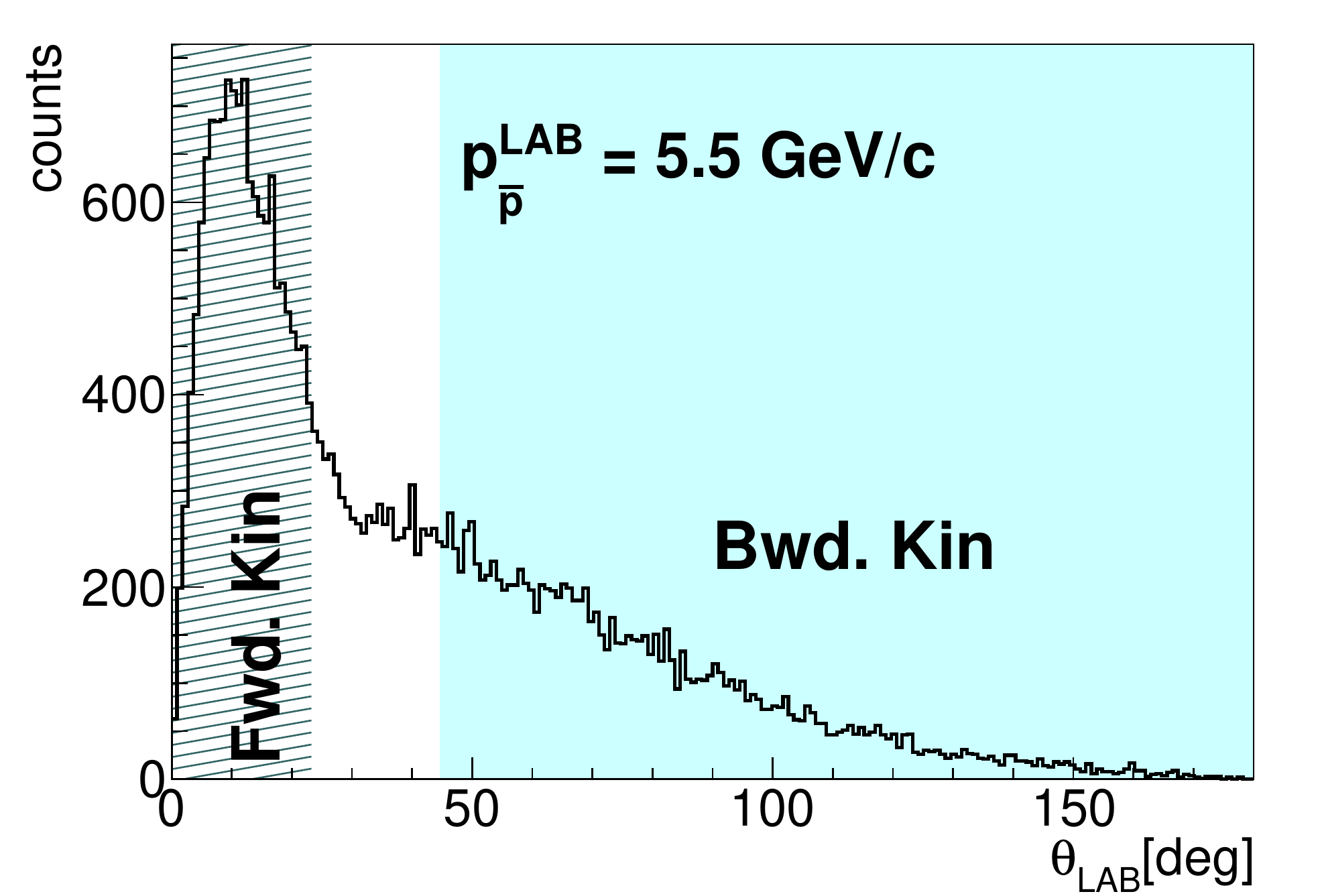}
  \includegraphics[width=0.33\textwidth]{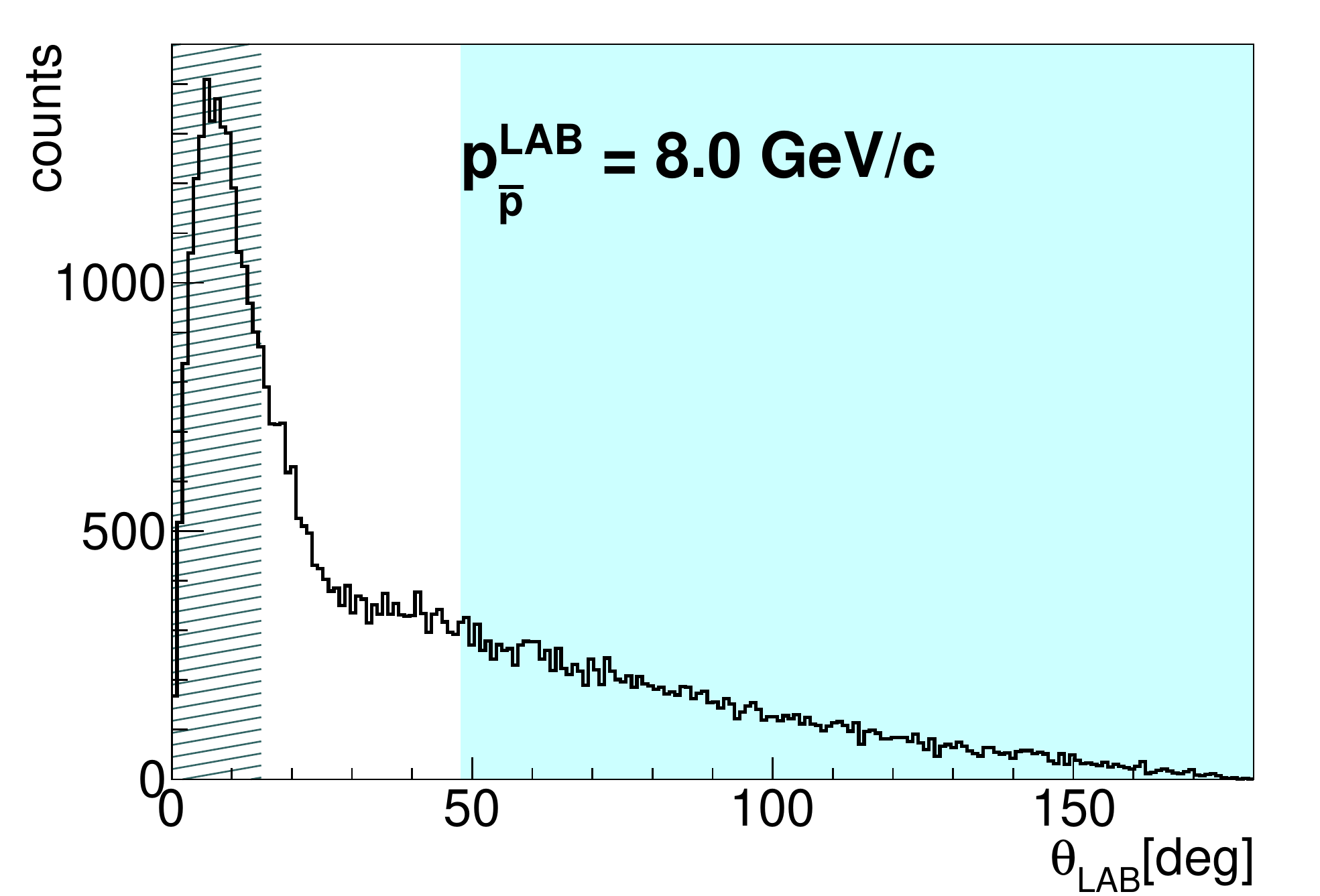}
  \includegraphics[width=0.33\textwidth]{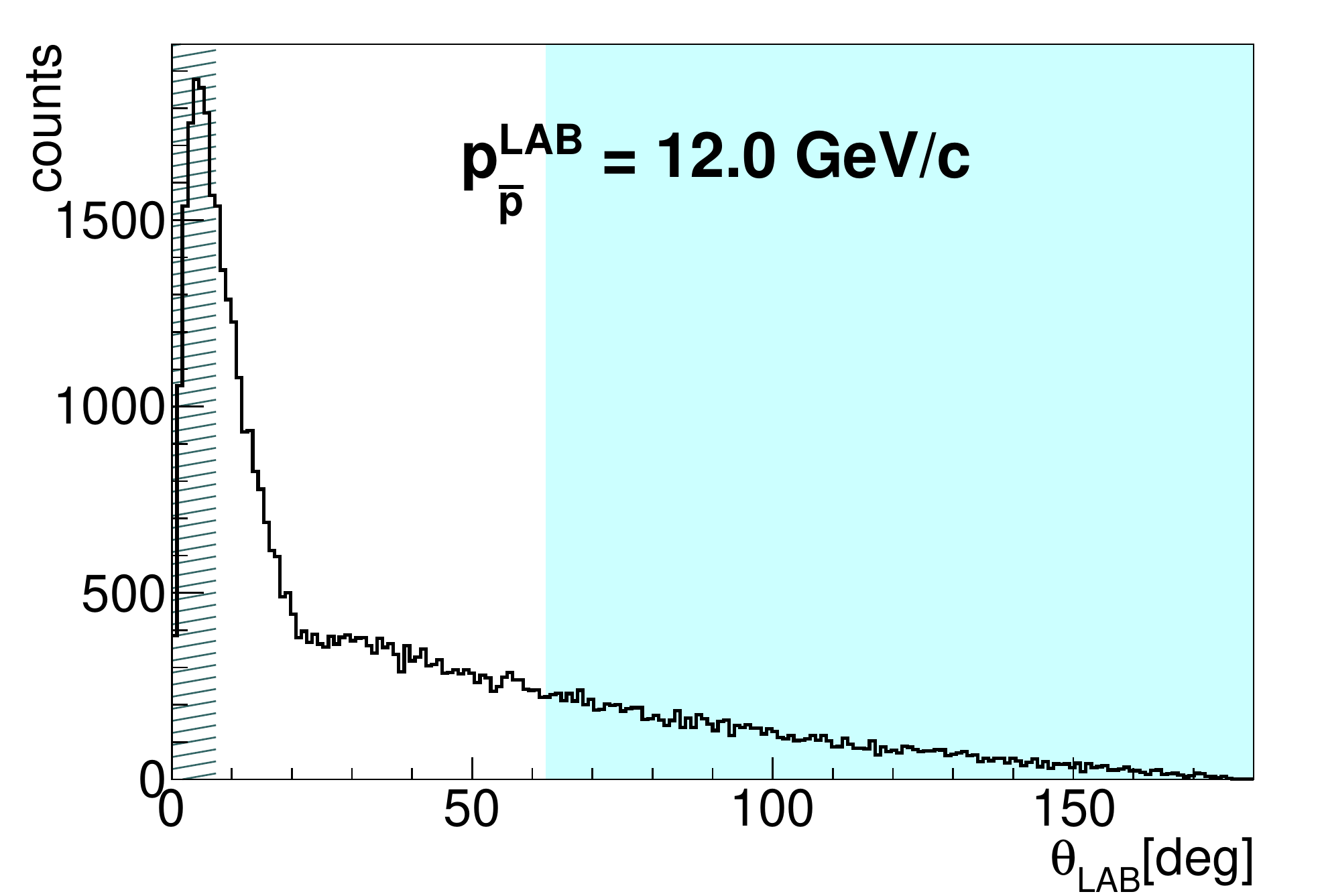}
  \caption{\label{fig:angular_dist_sig} Angular distribution of the
    $\piz$ in the lab frame from the event generator based on the TDA
    formalism in Ref.~\cite{Pire:2013jva} at three incident $\pbar$
    momenta. The validity ranges of the TDA model in terms the lab
    frame pion emission angle are shown for the near-forward
    kinematics \plfm{hatched} and near-backward kinematics \plfm{solid
      fill}. }
\end{figure*}

A Monte Carlo (MC) event generator for the signal reaction was
implemented by relying on Eq.~(\ref{eq:xsect_formula}). The angular
distribution of $\piz$s in the lab frame from this generator is shown
in Fig.~\ref{fig:angular_dist_sig}. The entire near-forward kinematic
validity range of the reaction is concentrated in a polar angle window
below $\approx$~30$^\circ$ (and even smaller window at higher
collision energy), whereas the near-backward kinematic validity range
occupies a window above $\approx$~45$^{\circ}$ (larger at higher
collision energy) extending all the way to $180^\circ$. This has
implications for the signal reconstruction efficiency as will be shown
in Section~\ref{sec:eff_corr}.

\section{\label{sec:bg_prop}Background Properties}

In the context of the \PANDA\ detector setup, the signal reaction
$\sigrxn$ has multiple background sources with varying degree of
importance. These background sources are the subject of discussion in
this section. For each source, rate estimates are given and in cases
where a full MC is warranted, the details about the event generators
that were used to simulate events are presented.

\subsection{Three Pion Production $\bgrxn$}

The $\bgrxn$ reaction is an important background source since the
cross section is orders of magnitude larger than that of the signal
and the possibility to misidentify the charged pion pair as an
electron-positron pair. This reaction has been studied in the past at
various incident $\pbar$ momenta. Despite the limited statistics that
were collected, the results from these early measurements tabulated in
Ref.~\cite{Flaminio:101631} provide a valuable benchmark for this
study.

\begin{figure}[hbpt]
  \includegraphics[width=\columnwidth]{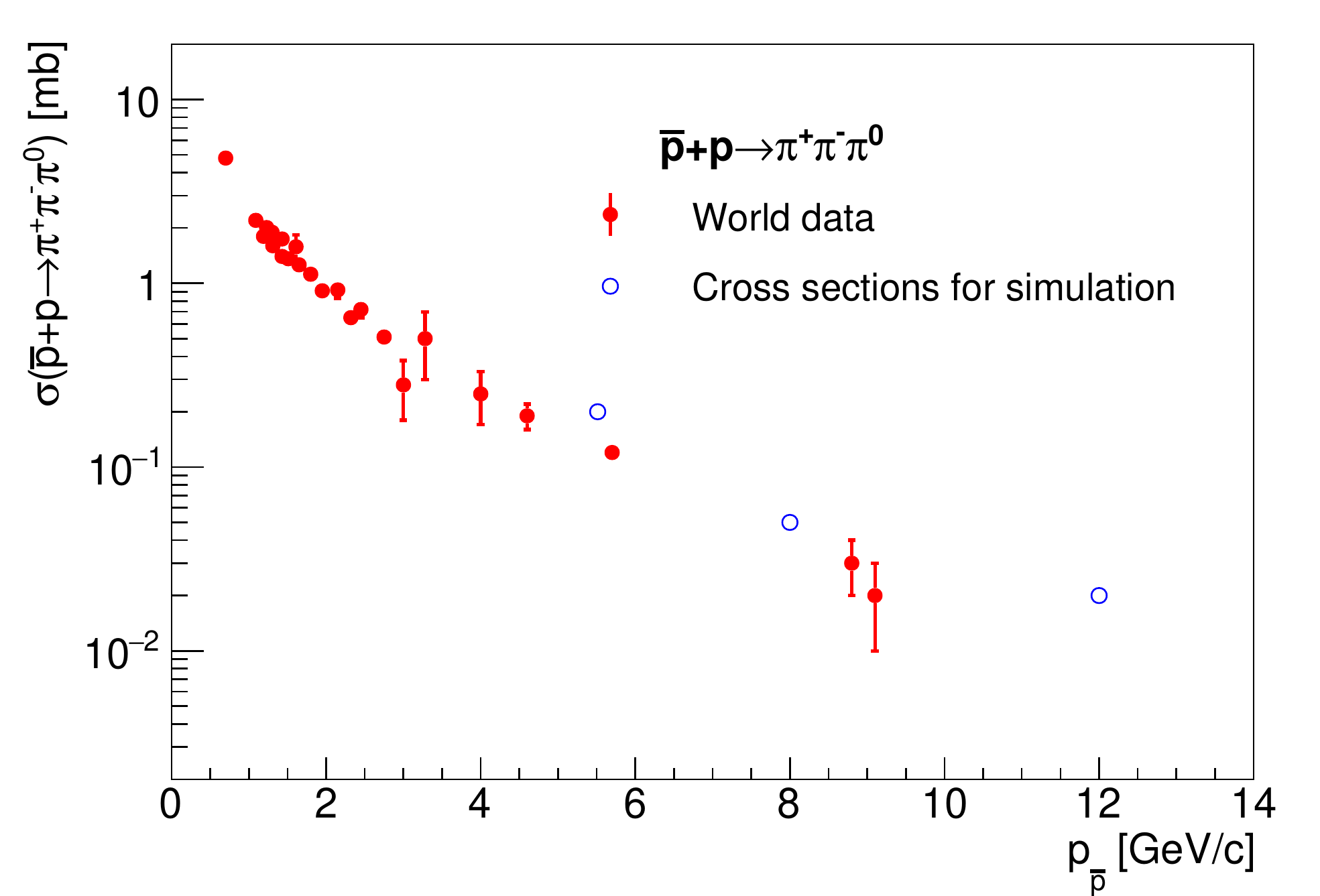}
  \caption{\label{fig:bg_xsect_world}The existing world data of the
    cross section of $\bgrxn$ \plfm{full circles} as reported in
    Ref.~\cite{Flaminio:101631}, plotted as a function of the incident
    $\pbar$ momentum. The three lab momenta chosen for this study and
    the corresponding cross sections used in the simulation of this
    background source are shown as open circles.}
\end{figure}

As shown in Fig.~\ref{fig:bg_xsect_world}, these results point to a
steep decline of the total cross section as a function of incident
$\pbar$ momentum. For our background simulations, we used total cross
sections slightly larger than the interpolation between the nearest
existing data points. Very few experiments collected enough statistics
to give high quality spectra for this reaction. As a result, the
feasibility study relies on the hadronic event generator DPM (Dual
Parton Model)~\cite{Capella1994225} to simulate the shape of the
spectra, since the model is constrained by taking into account the
sparse experimental differential distributions. In particular, it
includes the production of $\rho$ and $f_2$ resonances in agreement
with experimental observation~\cite{Bacon:1973ax}.

Table~\ref{tab:rates} shows the $\jpsipiz$ and $\pipmpiz$ cross
sections in a mass window of 2.8~--~3.3~GeV/$c^2$ for the charged pion
pair, within the validity ranges of the TDA model
(cf. Table~\ref{tab:validity}). The validity range includes both the
near-forward and near-backward kinematic approximation zones. The last
column of the table gives the approximate signal to background ratio
of produced event rates taking into account the $\jpsitoepem$
branching fraction of 5.69\%. The ratio of signal to background event
production rate is of the order $10^{-7}$ to $10^{-6}$. The fact that
the signal $\epem$ invariant mass distribution is peaked allows to
gain a factor of about 10 in signal to background ratio (S/B) before
any PID information has been used. The rejection of the rest of the
background has to come mostly from PID\@. The resulting strong
requirement on the electron-pion discrimination power will make PID
cuts a crucial component of the analysis.

\begin{table}[hbpt]
  \caption{\label{tab:rates}Signal ($\sigrxn$) cross section and
    production yields, background ($\bgrxn$) cross sections and signal
    over background ratio. The cross sections have been integrated
    over the validity range of the model (both near-forward and
    near-backward). The branching ratio for $\jpsitoepem$ and the mass
    cut on the $\pipm$ pair in the range 2.8~--~3.3~GeV/$c^2$ have
    been taken into account. The production yields for the signal are
    integrated for a luminosity of 2~\fbi, corresponding to about five
    months at full luminosity.}
  \begin{ruledtabular}
    \begin{tabular}{ccccc}
      $\mompbar$  & $\Delta\sigma^{\jpsipiz}_{valid}$ & $N^{\epempiz}_{valid}$ & $\Delta\sigma^{\pipmpiz}_{valid+m\_cut}$ & $S/B_{\pipmpiz}$ \\
      (GeV/$c$)     & (pb)                &  (2~\fbi) & (mb) &   \\
      \colrule
      \bmom{0} & 207 & 24.6k & 8.2$\times$10$^{-3}$ & 1.5$\times$10$^{-6}$\\
      \bmom{1} & 281 & 33.3k & 1.6$\times$10$^{-3}$ & 1.0$\times$10$^{-5}$\\
      \bmom{2} & 200 & 23.7k & 3.28$\times$10$^{-4}$ & 3.6$\times$10$^{-5}$\\
    \end{tabular}
  \end{ruledtabular}
\end{table}

\subsection{Multi-pion Final States ($N_{\pi}\geq 4$)}

Multi-pion ($N_{\pi}\geq 4$) final states with at least one $\pipm$
pair are also potential sources of background to the $\sigrxn$
channel, if one charged pion pair is misidentified as an $\epem$
pair. They have cross sections that are up to a factor 15 higher than
three pion production, but larger rejection factors can be achieved
due to the different kinematics. To confirm this, we performed
detailed simulation studies for the $\pipmpizpiz$ and $\pipmpipmpiz$
channels, and used the results to conclude on the rejection capability
for channels with higher number of pions. The cross sections for the
simulation of multi-pion final states is estimated by scaling the
$\pipmpiz$ cross sections discussed in the previous section, with the
scaling factor derived from DPM
simulations. Table~\ref{tab:dpm_2to1_piz_ratio} gives the ratio of
cross sections between those multi-pion final states ($\pipmpizpiz$
and $\pipmpipmpiz$), and the three-pion final state ($\pipmpiz$).

\begin{table}[hbpt]
  \caption{\label{tab:dpm_2to1_piz_ratio} The ratio of cross sections
    of four and five pion final states that could potentially be
    misidentified as signal to that of the three pion final state,
    extracted from the DPM hadronic model simulations. The ratios are
    given at three incident $\pbar$ momenta.}
  \begin{ruledtabular}
    \begin{tabular}{ccc}
      $\mompbar$ (GeV/$c$)
      & \(\frac{\sigma(\pbarp\to\pipmpizpiz)}{\sigma(\pbarp\to\pipmpiz)}\)
      & \(\frac{\sigma(\pbarp\to\pipmpipmpiz)}{\sigma(\pbarp\to\pipmpiz)}\) \\
      \colrule
      \bmom{0} &  2.9 & 8.0 \\
      \bmom{1} &  3.2 & 11.0 \\
      \bmom{2} & 3.4 & 15.0 \\
    \end{tabular}
  \end{ruledtabular}
\end{table}

\subsection{\label{sec:pizpizjpsi}$\bgtwopizshort$ with $\jpsi\to\epem$}

For the cross section of the $\bgtwopiz$ channel, which is not
predicted by the DPM model, we assumed a scaling to the $\jpsipiz$
channel according to:
\begin{equation}
  \frac{\sigma(\pbarp\to\jpsipizpiz)}{\sigma(\pbarp\to\jpsipiz)}
  \approx
  \frac{\sigma(\pbarp\to\pipmpizpiz)}{\sigma(\pbarp\to\pipmpiz)},
\end{equation}
\noindent
where the $\pipmpizpiz$ to $\pipmpiz$ ratios were calculated based on
the DPM event generator output. The results for $\bgtwopiz$ are
35.3~pb, 52.7~pb and 40.7~pb for beam momenta of \bmomu{0},
\bmomu{1} and \bmomu{2}, respectively. Although there is no existing
measurement of the $\bgtwopizshort$ cross section to confirm these
assumptions, we provide arguments below, which suggest that they are
reasonably conservative.

The E760 collaboration reported in Ref.~\cite{Armstrong:1992ae} the
non-observation of a signal for $\bgtwopiz$ at c.m.\@ energies close
to the $h_c(1P)$ mass of 3.5~GeV/$c^2$. This determines an upper limit
for the cross section of 3~pb, which is about a factor 10 below our
assumption at this energy. Calculations of non-resonant channels for
the $\pbarp\to\jpsi\pipm$ reaction at the $X(3872)$ energy have been
performed with an hadronic model~\cite{Chen:2008cg}, yielding a cross
section of about 60~pb for $\pbarp\to\jpsi\pipm\to\epem\pipm$ at
$\sqrt{s}$ around 3.872~GeV. With the assumption of a factor two
smaller cross section for $\bgtwopiz$ based on isospin coefficients,
this calculation is consistent with our assumption. These
calculations, outlined in Ref.~\cite{Chen:2008cg}, are however likely
to have been overestimated due to the absence of vertex cut-off form
factors, as in Ref.~\cite{Wiele:2013vla} for the case of
$\sigrxnshort$. Finally, the cross section for the production of the
$\bgtwopiz$ channel via feed-down from a resonance might in some cases
be expected to be lower than, or of the same magnitude, as our
assumptions. This is obviously the case for resonances with charge
conjugation $C=1$, e.g.\@ the $X(3872)$, which do not decay into
$\jpsi$ plus any number of $\piz$s. But charmonium states with $C=1$
might also contribute to the $\jpsipizpiz$ channel with cross sections
that are lower or of the same order of magnitude. For example, a
prediction of 30~pb for the reaction $\pbarp\to
Y(4260)\to\jpsipizpiz\to\epempizpiz$ at $\sqrt{s}=4.260$~GeV was
provided in Ref.~\cite{PandaPhysBook09c}, with the assumption that,
having the same quantum numbers as the $\psi(2S)$, the $Y(4260)$
decays into $\jpsipizpiz$ with the same branching ratio as what was
reported in Ref.~\cite{Negrini:2004kt}. The $Y(4260)$ resonance will
probably have a very low branching fraction for decays into the
$\jpsipiz$ and $\pipmpiz$ channels, and will therefore not contribute
significantly to the signal and to the other backgrounds.

\subsection{Di-electron Continuum: $\sigrxnepem$ }

The production of a $\gamma^*$ with an associated $\piz$ is another
process which can be used to demonstrate the universality of the
TDAs~\cite{Singh:2014pfv}. When the invariant mass of the
electron-positron pair is near the $\jpsi$ mass, this channel
represents a background source to the $\sigrxn$ channel, which can not
be rejected in the analysis procedure since the particles in final
state are identical to the signal. A similar TDA formalism to the one
used here for the prediction of the $\jpsipiz$ channel can also be
used to estimate the differential cross section for $\sigrxnepem$, as
demonstrated in Ref.~\cite{Singh:2014pfv}. The same predictions can be
used to integrate the cross section numerically over the corresponding
validity domains at each collision energy
(cf. Table~\ref{tab:validity}).

After setting the range $8.4<Q^2~[GeV^2]<10.1$ to match the window
around the $\jpsi$ mass, which will be used for the analysis, cross
sections of 13.6~fb, 21.6~fb and 24.8~fb are obtained at
$s~=~$12.3~GeV$^2$, $s~=~$16.9~GeV$^2$ and $s~=~$24.3~GeV$^2$,
respectively. Taking into account the branching ratio of
$\jpsi\to\epem$ (5.94\%), this results in contamination on the
$10^{-3}$ level and therefore has not been considered for further
simulations.

\subsection{Hadronic Decays of \jpsi}

The reaction $\sigrxnshort$, with the $\jpsi$ decaying into $\pipm$,
where the $\pipm$ pair is subsequently misidentified as a $\epem$
pair, is another potential source of background. It is highly
suppressed by the branching fraction of $\jpsi$ into $\pipm$
($\approx~10^{-4}$), and the low probability of misidentifying the
pions as electrons (\emph{cf.}  Section~\ref{sec:pid_eff}).

Similarly, $\pbarp\to\jpsipiz$ events with a hadronic $\jpsi$ decay
that can mimic the signal's final state (for example
$\jpsi\to\pipm\piz$ or $\jpsi\to\gamma\pipm$) is heavily suppressed by
the probability to identify the $\pipm$ pair. With the same production
cross sections as the signal, and with very low probability of
misidentifying the $\pipm$ pair as an $\epem$ pair, such final states
have negligible detection rates, and therefore will not be fully
simulated.

\section{\label{sec:simul}Simulations and Analysis}

For the present feasibility study, full MC simulations were performed
on events from both the signal and background event generators
described in the previous sections. In this section details of the
analysis procedure will be provided. After a brief overview of the
analysis methodology, a discussion of the PID and selection procedure,
the reconstruction of $\piz$ and $\epem$ pairs, as well as the use of
kinematic fits to gain further rejection for one class of background
reaction where PID alone is not sufficient is provided. The section
concludes with global signal to background ratios and signal purity
for each background type included in the simulation study.

\subsection{Brief Description of the Method}

The analysis starts with the generation of signal and background
events as described in Section~\ref{sec:sig_prop}, followed by the
transport of tracks in GEANT4, where the geometry of the \PANDA\
detector has been implemented. The simulation of the detector response
and digitization of the signals follows this step, at which point
tracks and neutral particles can be reconstructed.

Signal events are then passed to the full event selection chain,
including PID and analysis cuts. This ensures the most realistic
description possible of the reconstructed signal, including
statistical fluctuations. The number of signal events to simulate was
picked to correspond exactly to the expected signal counts shown in
Table~\ref{tab:rates} within the validity domain. By directly plotting
spectra with tracks that pass all identification cuts, the plots will
reflect the expected statistical accuracy.

For the background events, a different approach was followed. To check
the feasibility of the measurement, the residual background
contamination needs to be understood with a good precision in each bin
used for the extraction of the physics observable (\emph{e.g.\@}
differential cross section distribution in $t$). This would not be
possible if the cuts are directly applied due to the small number of
background events that would pass all cuts. Therefore, a weight
proportional to the product of single charged track misidentification
probabilities is applied to each event instead of direct application
of cuts as in the case of signal event simulation. The charged pion
misidentification probability is parameterized as a function of
momentum of the track. The photon selection is done by direct
application of the cuts, just as in the case of the signal event
analysis.

\begin{figure*}[hbpt]
  \includegraphics[width=0.33\textwidth]{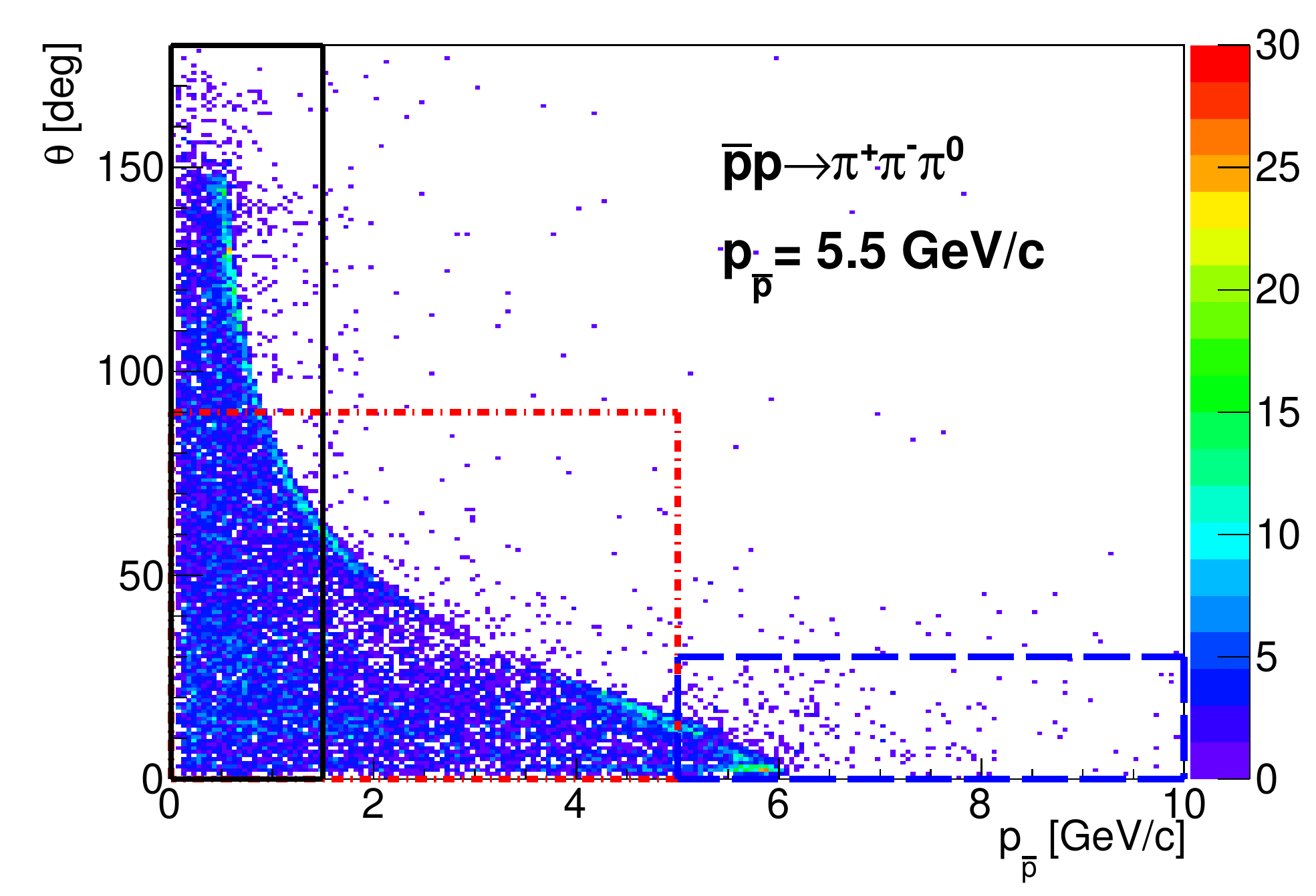}
  \includegraphics[width=0.33\textwidth]{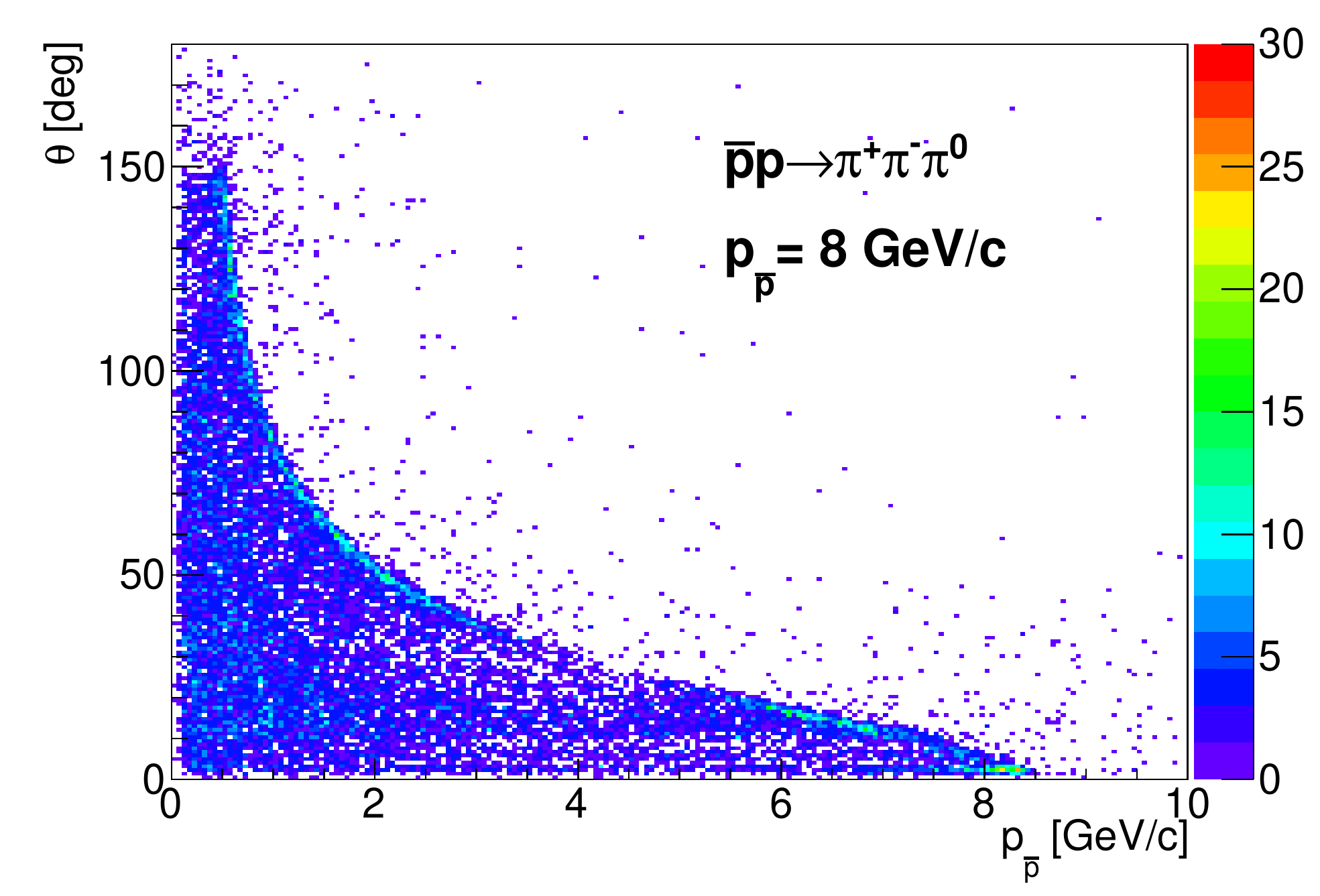}
  \includegraphics[width=0.33\textwidth]{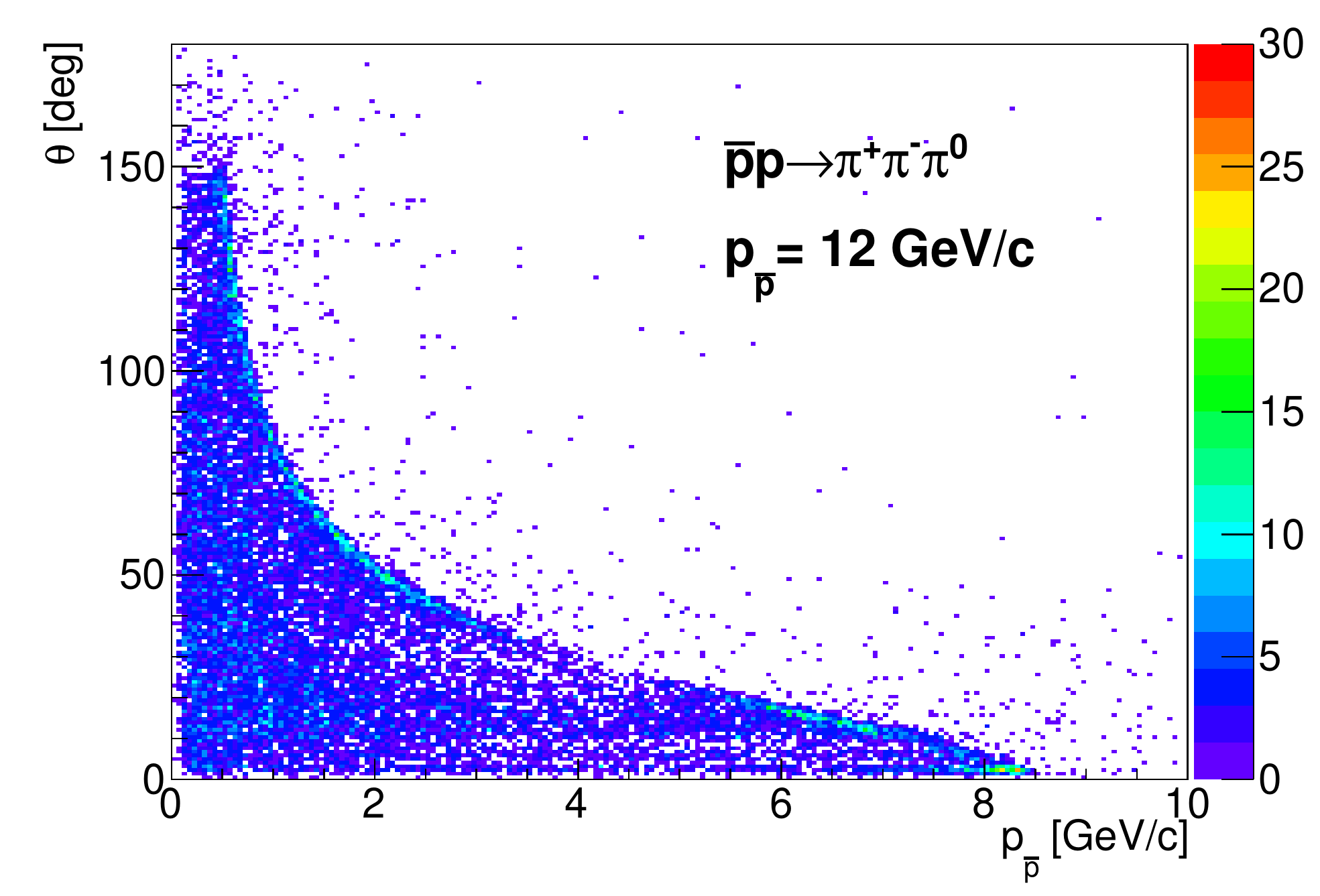}\\
  \includegraphics[width=0.33\textwidth]{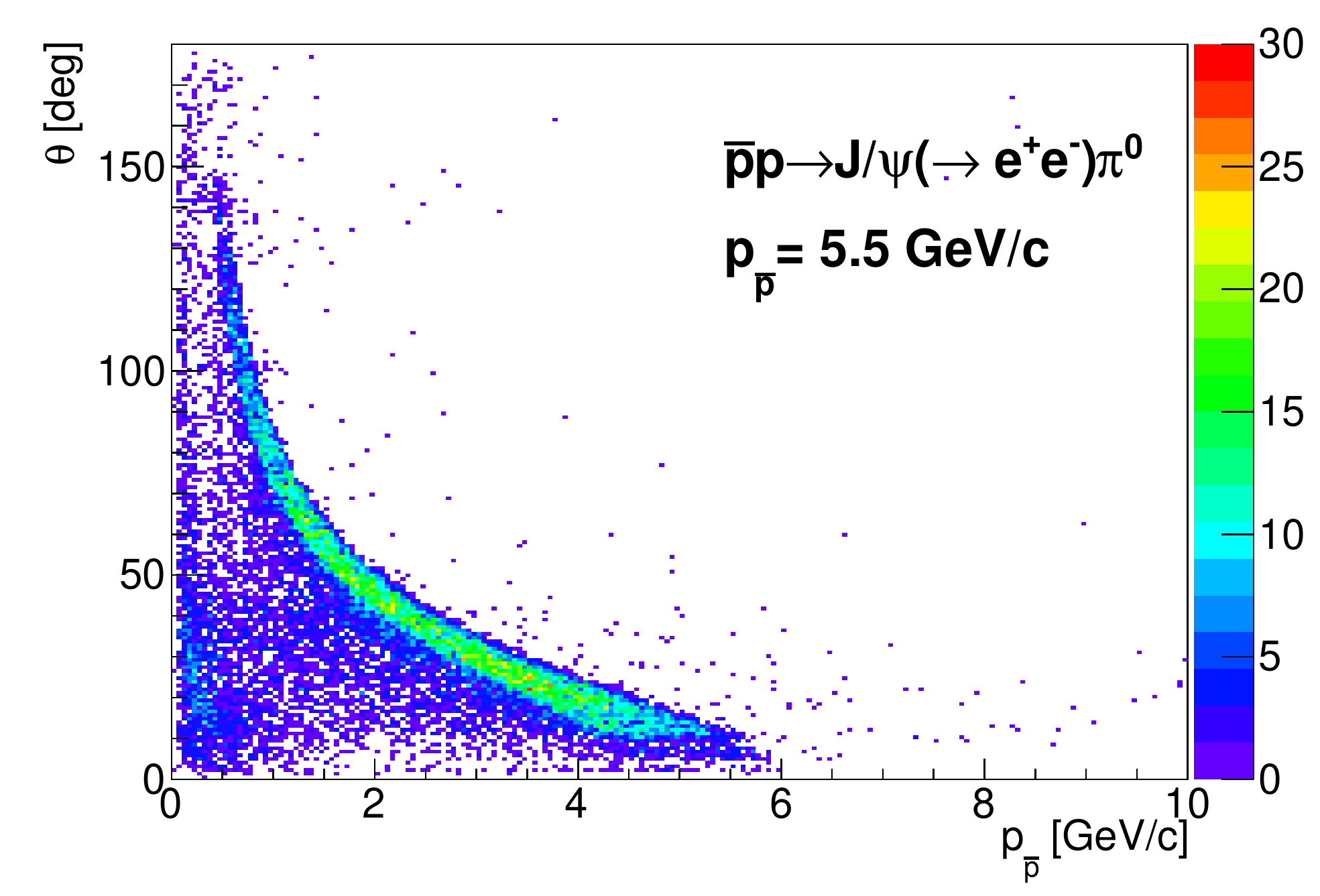}
  \includegraphics[width=0.33\textwidth]{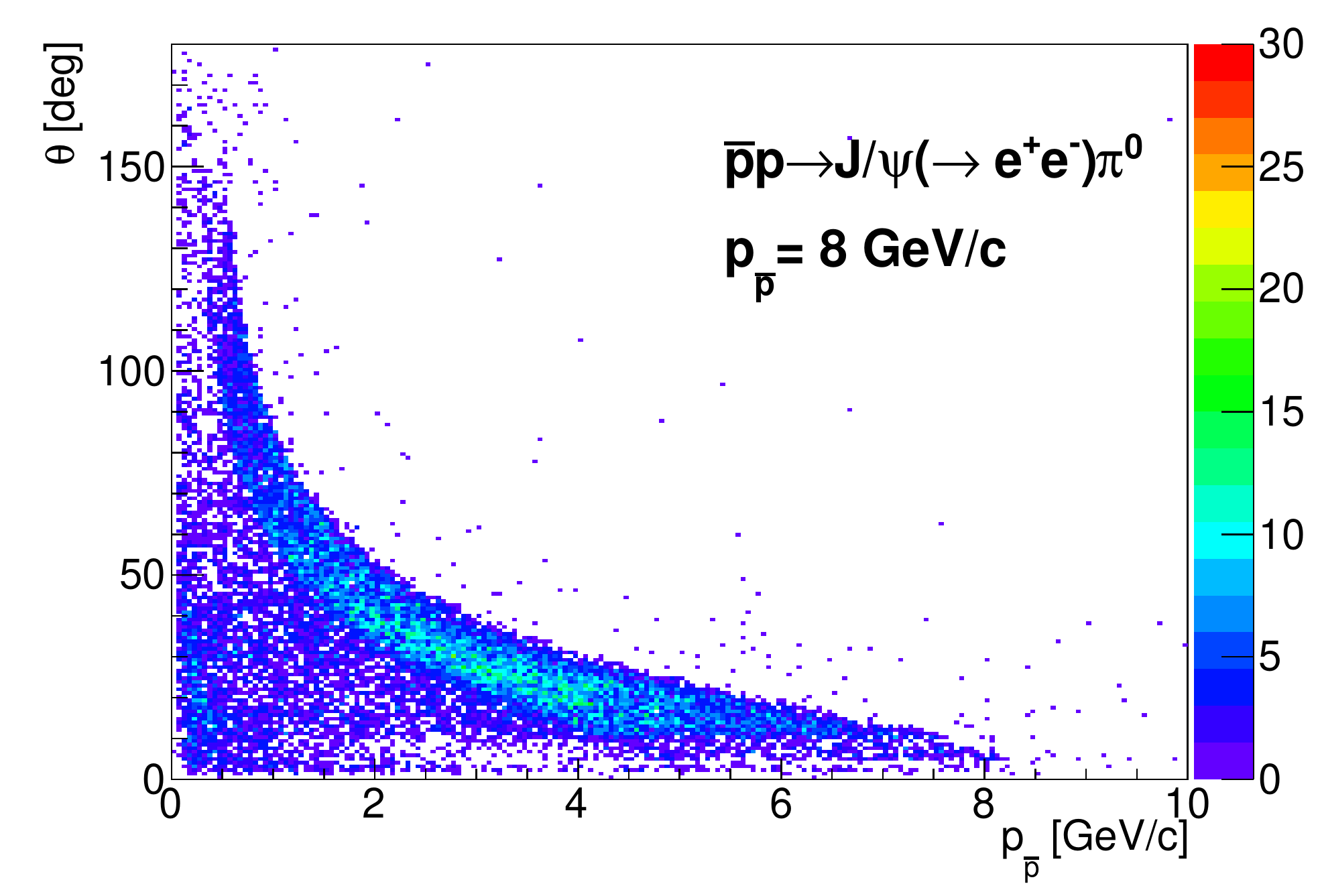}
  \includegraphics[width=0.33\textwidth]{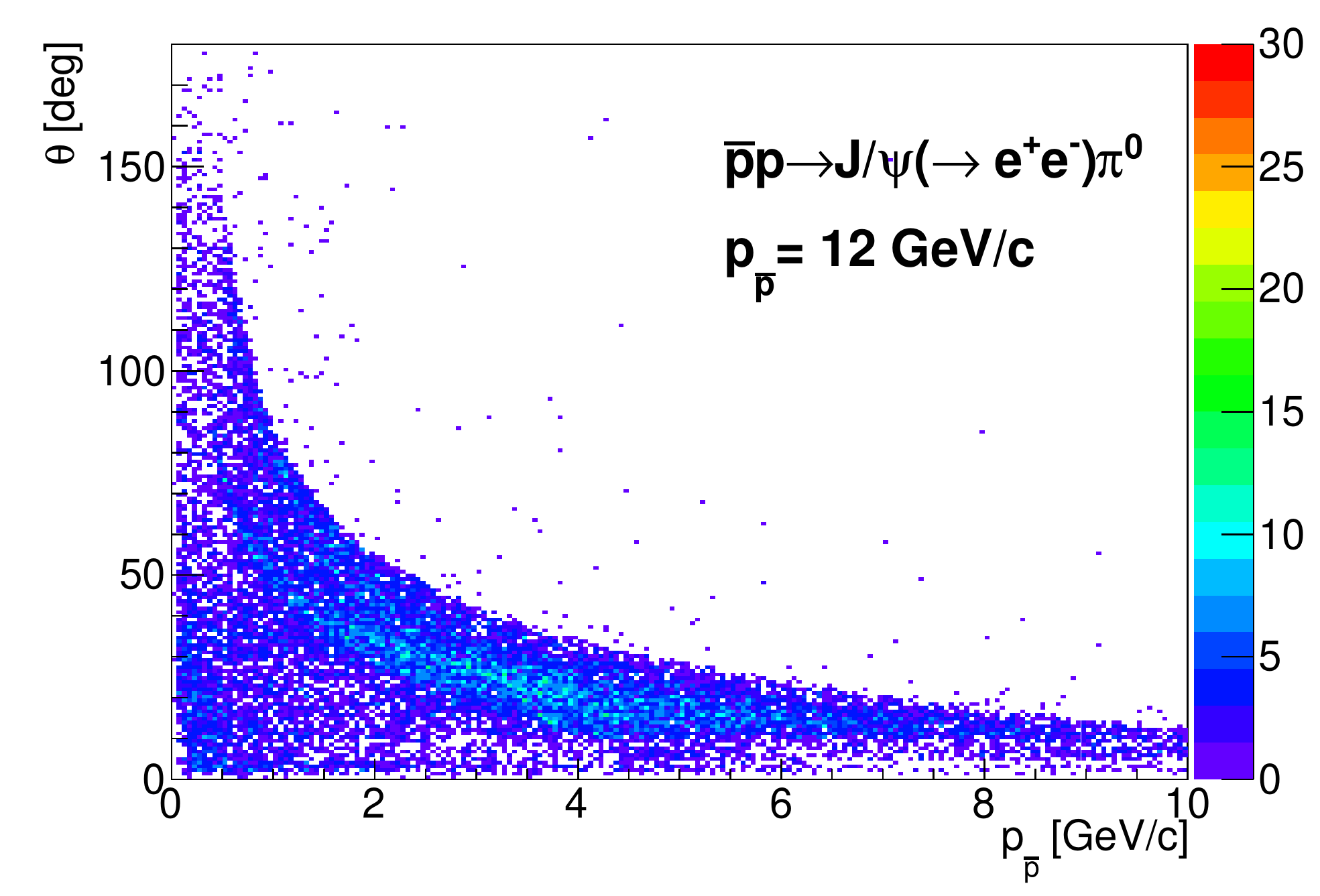}\\
  \caption{\label{fig:kin_dists} The polar angle vs momentum
    distribution of reconstructed single charged tracks in full MC
    simulation of the background \plfm{top row} and signal
    \plfm{bottom row} event generator output. The simulations were
    performed at the three $\pbar$ incident momenta chosen for this
    study. The boxes in the top left panel show the regions in which
    high statistics flat single track simulations were generated for
    low \plfm{full line}, intermediate \plfm{dash-dotted line} and
    high \plfm{dashed line} momentum tracks for use in the efficiency
    and rejection studies. }
\end{figure*}

Figure~\ref{fig:kin_dists} shows the polar angle versus momentum
distributions of reconstructed final state tracks in the background
(top row) and signal (bottom row) simulations at the three different
$\pbar$ incident momenta used in this study.

\subsection{\label{sec:pid_eff}PID Efficiency}

For the investigation of PID efficiency, single particle simulations
of electrons, positrons, positive and negative pions were performed in
the three kinematic zones depicted by red, black and blue boxes in
Fig.~\ref{fig:kin_dists}. This is to focus the simulation to values of
$p$ and $\theta$ that matter most for this study. In total,
$5\times10^6$ single $e^+$ and $e^-$ events each were used for the
estimation of the electron efficiency. To obtain a good precision on
the very low pion misidentification probability, a larger statistics
of $25\times10^6$ each for single $\pip$ and $\pim$ events were
used. The following section discusses in more detail how these
efficiencies were determined. For the sake of simplicity in the
remainder of the discussion, the word electron is used to refer to
both electrons and positrons, except when the distinction is
necessary.

As mentioned above, one of the critical aspects of this analysis is
the reduction of the hadronic background using PID cuts, since the
$\pipmpiz$ final state is kinematically very similar to the
$\jpsipiz\to\epempiz$ final state when the $\pipm$ invariant mass is
close to the $\jpsi$ invariant mass. To this end, the PID should be
used in the most effective way possible to maximize pion rejection
until the cost to electron efficiency is prohibitive. This section
describes the selection cuts of the analysis.

\subsubsection{Combined PID Probability}
The starting point for the application of global PID is the
detector-by-detector PID probability information. To simplify the
process of combining information from various detectors, the relevant
PID variables from a given detector subsystem $i \in \{subsys\} =
\{\detlist\}$ is used to determine an estimate of the probability
$\prob^{ID}_{i}(j)$ that a given track is one of the five charged
particles species that can be identified by \PANDA, where $j \in
\{\splist\}$. The combined probability that a given track is of type
$j$ is then given by:
\begin{equation}
  \prob^{ID}_{comb}(j) = \frac{\prob^{ID}(j)}{1+\prob^{ID}(j)},
  \label{eq:prob_comb}
\end{equation}
where:
\begin{equation}
  \prob^{ID}(j) = \prod_{i\in \{subsys\}} \frac{\prob^{ID}_{i}(j)}{1-\prob^{ID}_{i}(j)}.
\end{equation}

\begin{figure}[hbpt]
  \includegraphics[width=\columnwidth]{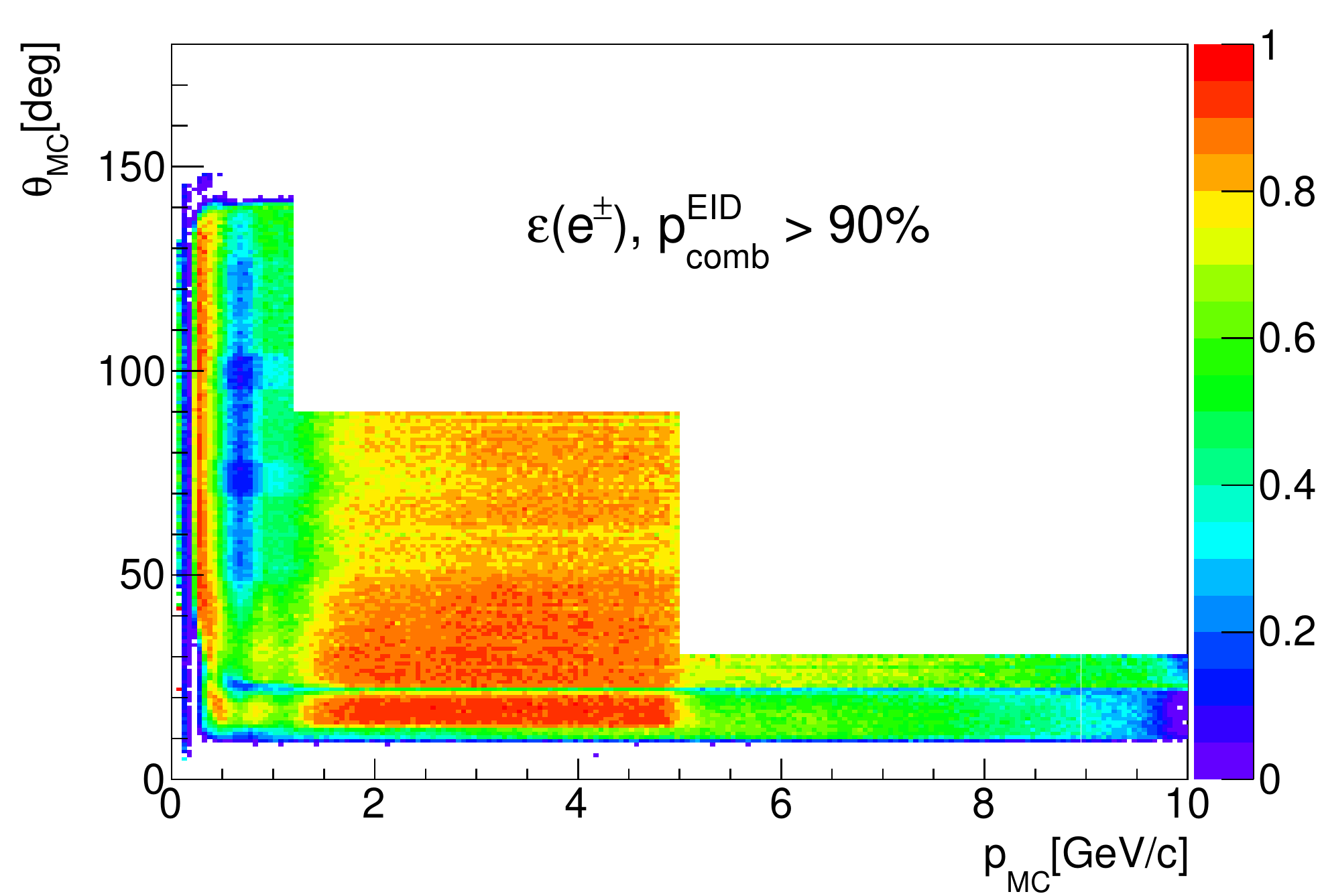}
  \caption{\label{fig:eff_eid} The EID efficiency with a probability
    threshold of 90\% for electrons and positrons as a function of MC
    polar angle $\theta_{MC}$ and MC momentum $p_{MC}$.}
\end{figure}

\begin{figure}[hbpt]
  \includegraphics[width=\columnwidth]{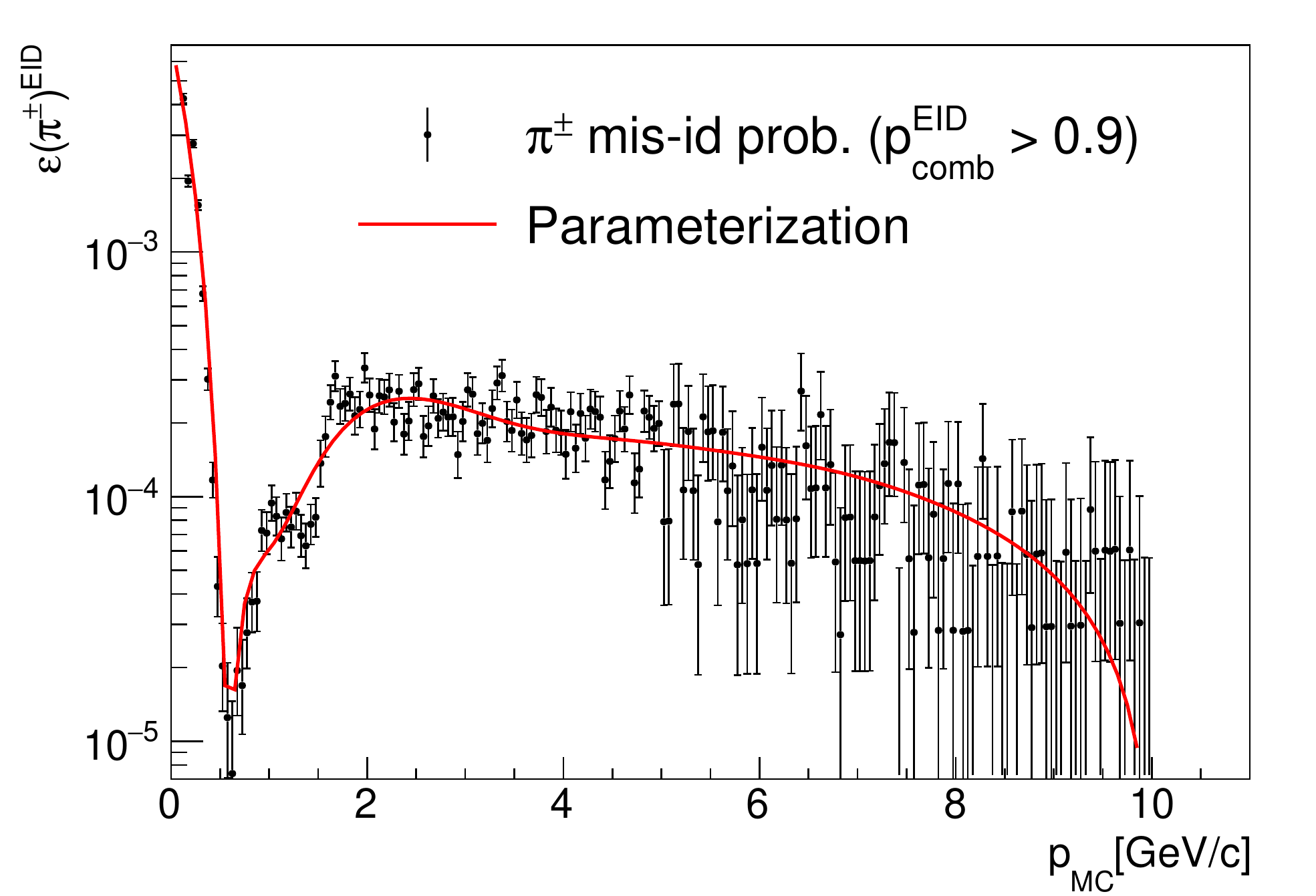}
  \caption{\label{fig:eff_eid_pi} Misidentification probability for
    charged pions as a function of true MC momentum for a combined EID
    probability threshold of 90\%. The efficiency parameterization is
    shown by the full line. }
\end{figure}

Equation~(\ref{eq:prob_comb}) ensures the proper normalization of the
probabilities assigned for each track: $\sum_{j}{\prob^{ID}_{comb}(j)}
= 1$. With the combined probability in hand, a cut is applied at an
appropriate value for the identification of any given species. In the
present case, a sufficiently high threshold on $\probeidcomb$ is
imposed to select electrons and reject all other species. Here we show
in Fig.~\ref{fig:eff_eid} the global PID efficiency
$\varepsilon_{eid}^{e^{\pm}}(\theta_{MC},\ p_{MC})$ as a function of
the true MC polar angle $\theta_{MC}$ and momentum $p_{MC}$ with a cut
of $\probeidcomb>0.9$, based on a simulation of $10^7$ electrons and
positrons. Figure~\ref{fig:eff_eid_pi} shows the pion
misidentification probability $\varepsilon_{eid}^{\pi^{\pm}}(p_{MC})$
as a function of $p_{MC}$ for the same cut based on a simulation of
$5\times10^7$ charged pions.

\subsubsection{Optimization of the Cut on $\probeidcomb$}

\begin{figure}[hbpt]
  \includegraphics[width=\columnwidth]{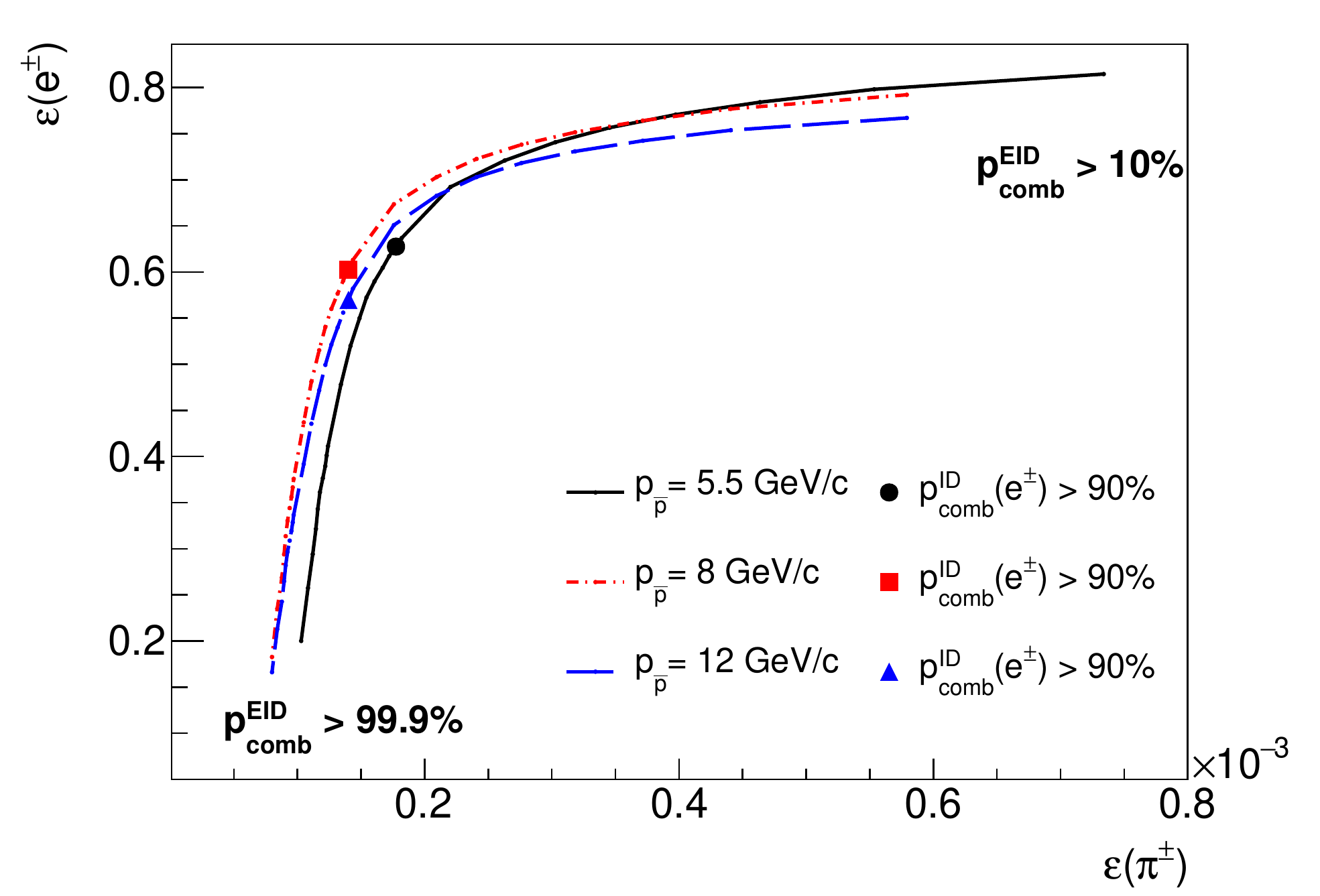}
  \caption{\label{fig:roc}ROC curves for $\probeidcomb$ cuts showing
    the efficiency to identify an electron vs the probability to
    misidentify charged pions at the three different $\pbar$ incident
    momenta, \bmomu{0} \plfm{solid line}, \bmomu{1} \plfm{dash-dotted
      line} and \bmomu{2} \plfm{dashed line}. The full points with the
    corresponding color show the efficiencies and misidentification
    probabilities for $\probeidcomb~>~90\%$ at the respective beam
    momenta.}
\end{figure}

To determine an optimal cut on $\probeidcomb$ for EID, the
relationship between average efficiency of EID $\varepsilon(e^{\pm})$
versus average misidentification probability for charged pions
$\varepsilon(\pi^{\pm})$ was studied as a function of the cut on
$\probeidcomb$, and plotted in Fig.~\ref{fig:roc} as a ROC (Receiver
Operating Characteristics) curve, which shows the performance of the
classifier as the discrimination threshold is
varied. $\varepsilon(e^{\pm})$ and $\varepsilon(\pi^{\pm})$ are
determined by taking the bin-by-bin weighted average of the electron
efficiency shown in Fig.~\ref{fig:eff_eid} and pion misidentification
probability shown in Fig.~\ref{fig:eff_eid_pi}. The weight for each
bin is set to the content of the corresponding bin in the kinematic
distributions shown in Fig.~\ref{fig:kin_dists}. One can observe that
there is a significant gain in charged pion rejection with relatively
small loss in EID efficiency up to a cut of $\probeidcomb>90\%$,
beyond which the rejection gain no longer justifies the associated
loss in efficiency. The cut of $\probeidcomb>90\%$ is therefore chosen
for the remainder of this analysis. It should be noted that the
difference of the ROC curves between different incident $\pbar$
momenta comes from the differences in the momentum and angular
distributions of single tracks in the signal and background events
that were used as a weight to average the efficiencies bin by bin.

\subsubsection{\label{sec:pid_appli}Application of the $\probeidcomb$
  Cut}

The application of the cut on $\probeidcomb$ is straightforward for
the signal simulations. The cut is applied on a track-by-track basis
to every reconstructed electron, and the spectra are constructed from
those tracks that pass the cut. This approach is however not realistic
for the background simulations, since the rejection is very high
($\approx~10^{-4}$ per charged pion). An unrealistically large number
of background events need to be simulated to produce sufficient
statistics in each bin after all cuts are applied. For this reason a
different approach is adopted. The single pion misidentification
probability shown in Fig.~\ref{fig:eff_eid_pi} is parameterized by a
function $f^{\varepsilon}_{\pi^\pm}(p)$ to smooth out the statistical
fluctuation, and subsequently used as a weight for each background
event based on the product of the values of
$f^{\varepsilon}_{\pi^\pm}(p)$ at the respective true MC momenta of
the two identified pions, $p_{\pip}$ and $p_{\pip}$:
\begin{equation}
  w(p_{\pip},p_{\pim}) = \frac{N^{BG}}{N^{MC}_{evt}}f^{\varepsilon}_{\pi^\pm}(p_{\pip})f^{\varepsilon}_{\pi^\pm}(p_{\pim}),
  \label{eq:bg_wt}
\end{equation}
\noindent
where $N^{MC}_{evt}$ is the number of full $\bgrxn$ reaction events
that were simulated, and $N^{BG}$ is the number of background events
that are expected from 2~\fbi\ integrated luminosity based on the
cross sections given in column 4 of Table~\ref{tab:rates}. The
constant factor $N^{BG}/N^{MC}_{evt}$ ensures the proper normalization
of the background spectra.

\subsection{\label{sec:piz_rec}$\piz$ Reconstruction}

Neutral pions are reconstructed through their two-photon decay
channel, in the invariant mass spectrum formed by combining all
photons within an event into $\gamma\gamma$ pairs. A cluster from the
EMC with a minimum reconstructed energy of 3~MeV is considered to
originate from a photon if there is no charged track candidate whose
extrapolation to the EMC falls within a 20~cm radius from the EMC
cluster. The invariant mass spectra show a contribution from
combinatorial $\gamma\gamma$ pairs, which can be reduced by relying on
the kinematic correlation of $\piz$ decay photons that the
combinatorial $\gamma\gamma$ pairs do not display.

\begin{figure*}[hbpt]
  \includegraphics[width=0.9\textwidth]{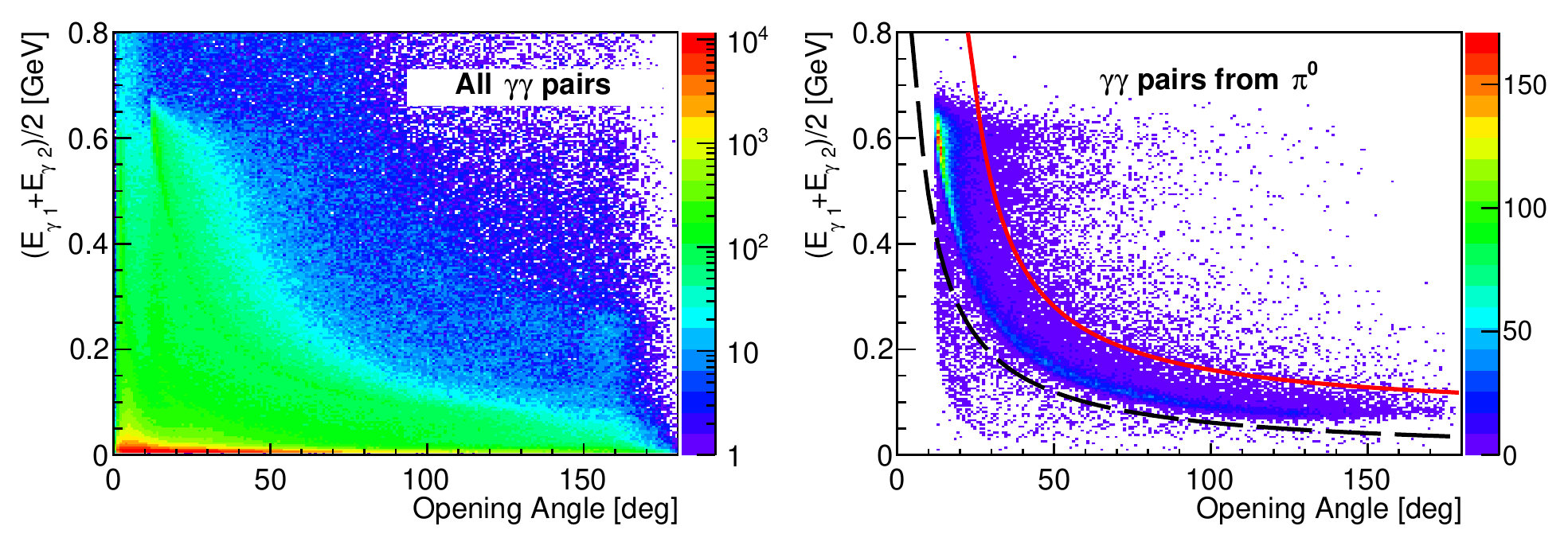}
  \caption{\label{fig:piz_decay_kin} The average reconstructed energy
    of a photon pair versus its opening angle for all $\gamma\gamma$
    pairs within an event \plfm{left panel} compared to $\gamma\gamma$
    pairs stemming from $\piz$ decay \plfm{right panel}, in a
    simulation of $\sigrxn$ events at $\mompbar=$ \bmomu{0}. The full
    and dashed lines in the right panel show the upper and lower
    bounds of the cut described in Eq.~(\ref{eq:pi0_sel}), with
    $a^L_0=0$, $a^L_1=0.11$, $a^L_2=-0.05$, $a^U_0=0.07$, $a^U_1=0.14$,
    and $a^U_2=0.21$. }
\end{figure*}

Figure~\ref{fig:piz_decay_kin} shows the correlation of the
reconstructed average photon energy to the opening angle in the lab
frame between two photons, from all $\gamma\gamma$ pairs including
those from combinatorial pairs on the left, and for $\gamma\gamma$
pairs that decay from a $\piz$ on the right, in $\sigrxn$ events
simulated within the validity domains of the TDA model.

\begin{figure*}[hbpt]
  \includegraphics[width=0.9\textwidth]{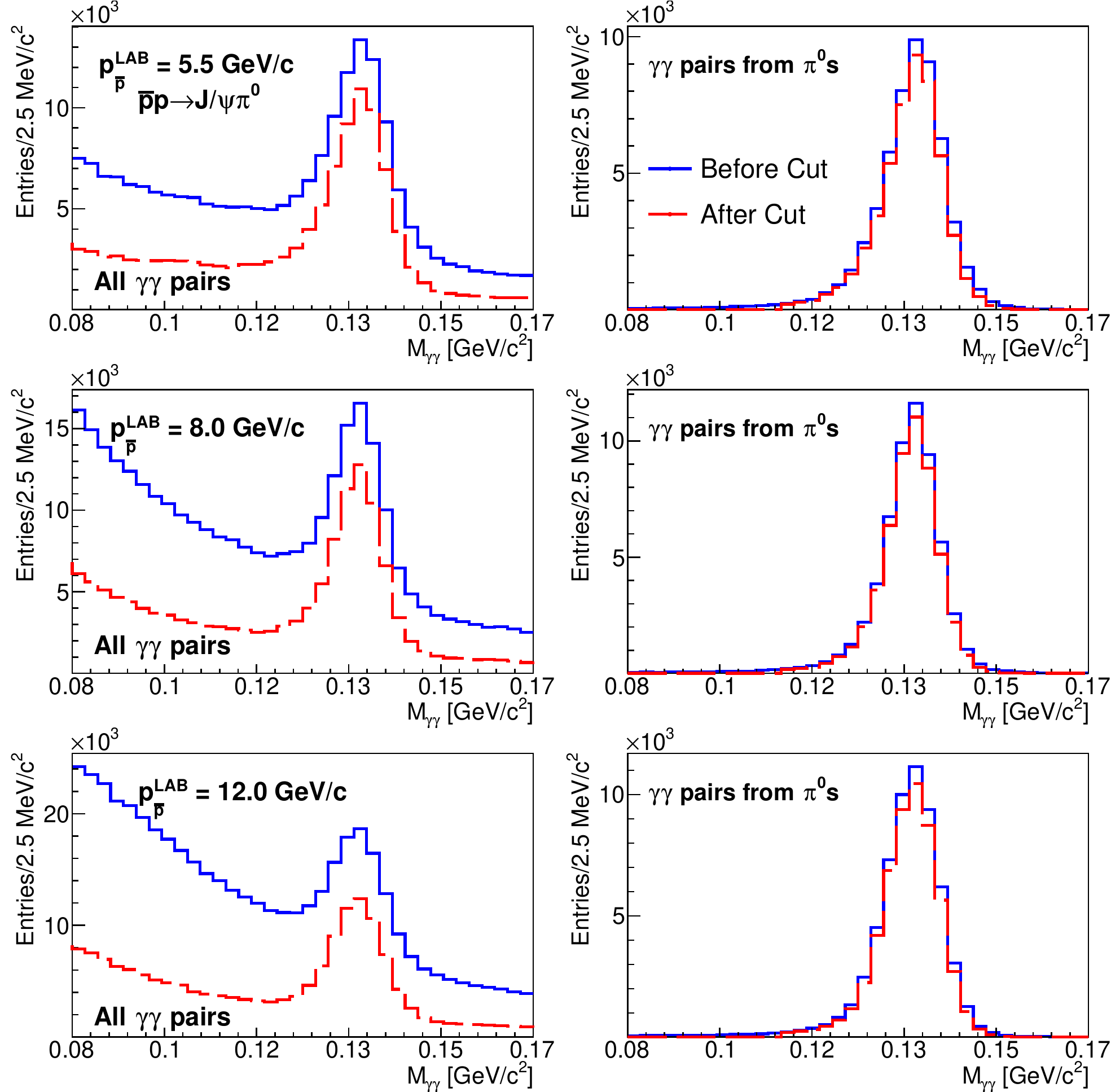}
  \caption{\label{fig:gg_invm} Two photon invariant mass spectra for
    the signal reaction $\sigrxn$. The left column shows the invariant
    mass spectra for all $\gamma\gamma$ pairs in the event, before the
    cut given by Eq.~(\ref{eq:pi0_sel}) \plfm{solid line} and after
    the cut \plfm{dashed line}. The right column shows the
    corresponding invariant mass distributions for reconstructed
    photon pairs from $\piz$ decay. Each row corresponds to a
    different $\pbar$ incident momentum: \bmomu{0} \plfm{top row},
    \bmomu{1} \plfm{middle row} and \bmomu{2} \plfm{bottom row}.}
\end{figure*}

The following cuts are applied to the data:
\begin{equation}
  \label{eq:pi0_sel}
  f_L(OA) \ < \ \frac{E_{\gamma 1}+E_{\gamma 2}}{2} \ < \
  \begin{cases}
    \infty \ \ \  (OA \leq a^U_2) \\
    f_U(OA) \ \ \ (OA > a^U_2)
  \end{cases}
\end{equation}
\noindent
with:
\begin{equation}
  \nonumber
  f_L(x) = a^L_0 + \frac{a^L_1}{x - a^L_2}, \ \ and \ \
  f_U(x)  = a^U_0 + \frac{a^U_1}{x - a^U_2},
\end{equation}
\noindent
where OA is the opening angle between the two photons, $E_{\gamma 1}$
and $E_{\gamma 2}$ are the energies of the two photons, and $a^L_i$
and $a^U_i$, with $i=0,1,2$ are coefficients of the parametrization
determined independently for each collision energy. Thus, it is
possible to reduce the combinatorial background to a few percent while
keeping an efficiency larger than 90\% for pairs where both photons
stem from $\piz$ decays. The effect of this cut is shown in
Fig.~\ref{fig:gg_invm} in a simulation of $\sigrxn$ events, for all
photon pairs on the left and for photons originating from $\piz$
decays on the right. In addition to this, an invariant mass cut of 110
$<~m_{\gamma\gamma}~<$ 160~MeV/c$^2$ is applied on the two photon
system.

\subsection{\label{sec:jpsi_rec}$\jpsi$ Reconstruction}

\begin{figure}[hbpt]
  \includegraphics[width=\columnwidth]{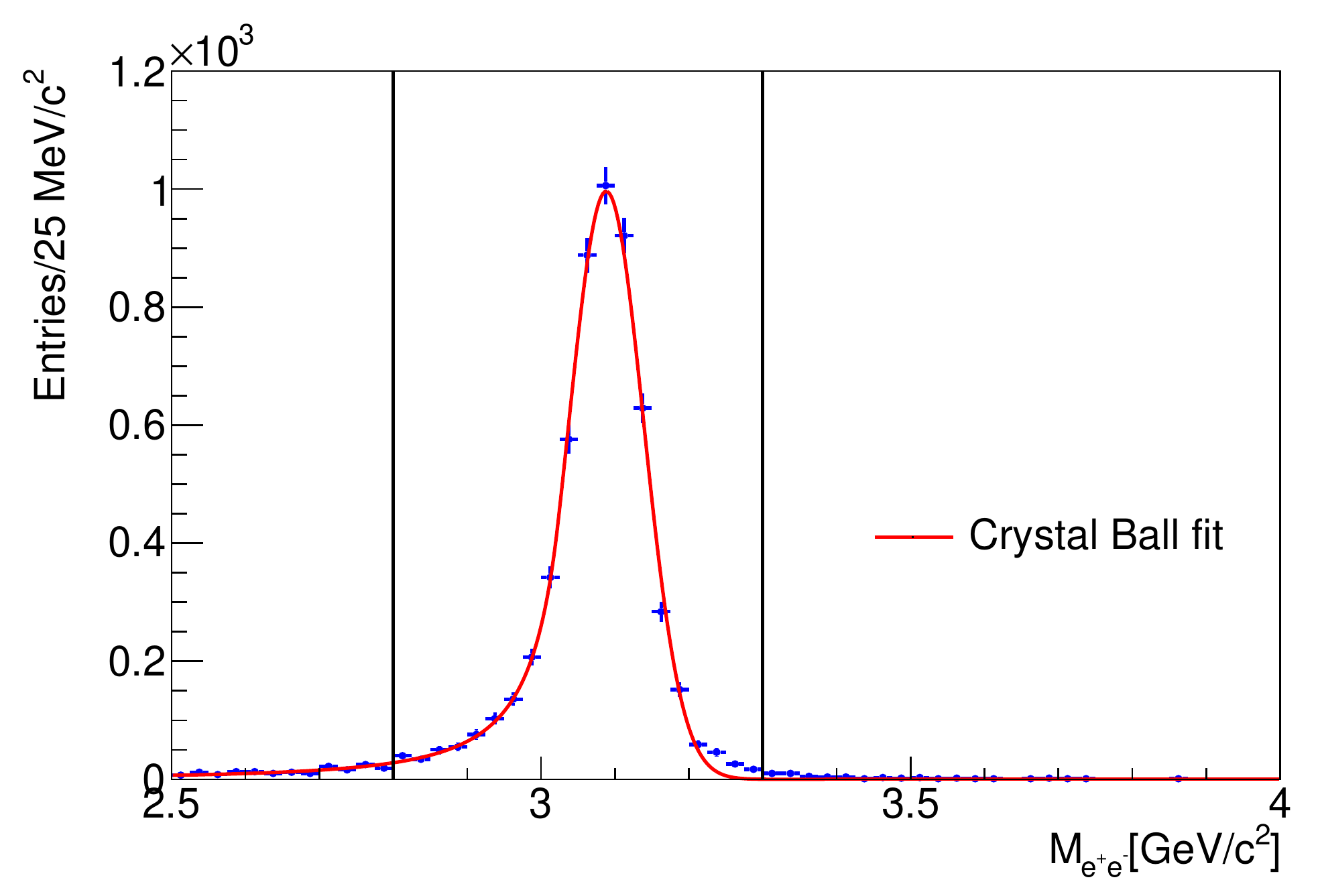}
  \caption{\label{fig:jpsi_peak} $\epem$ invariant mass spectrum (data
    points) fit with a Crystal Ball function (solid line) for the
    reaction $\sigrxn$ at a beam momentum of
    $\mompbar=$~\bmomu{0}. The solid vertical lines denote the
    2.8~<~$M_{\epem}$~[GeV/$c^2$]~$<$~3.3 mass window used in this
    analysis for the selection of $\jpsi$ candidates.}
\end{figure}

The reconstruction of $\jpsi$ candidates is accomplished by pairing
all positive charged candidates passing EID cuts with all negative
charged candidates that also pass the EID
cuts. Figure~\ref{fig:jpsi_peak} shows the invariant mass spectrum of
$\epem$ pairs after application of the EID cuts, while requiring the
presence of at least one reconstructed $\piz$ in the event. The mass
distribution can be described satisfactorily by a Crystal Ball
function~\cite{Oreglia:1980cs}. A fit to the mass distribution has a
peak at $(3.088 \pm 0.001)$~GeV/$c^2$ and a width of $(51.3 \pm
1.0)$~MeV/$c^2$. The solid vertical lines show the mass window
\mjpsiwin\ which is used as a selection to reconstruct $\epem$ pairs
from $\jpsi$.

\subsection{\label{sec:sig_rec}$\jpsi\piz$ Signal Reconstruction}

\begin{figure}
  \vskip 1.5mm
  \includegraphics[width=0.96\columnwidth]{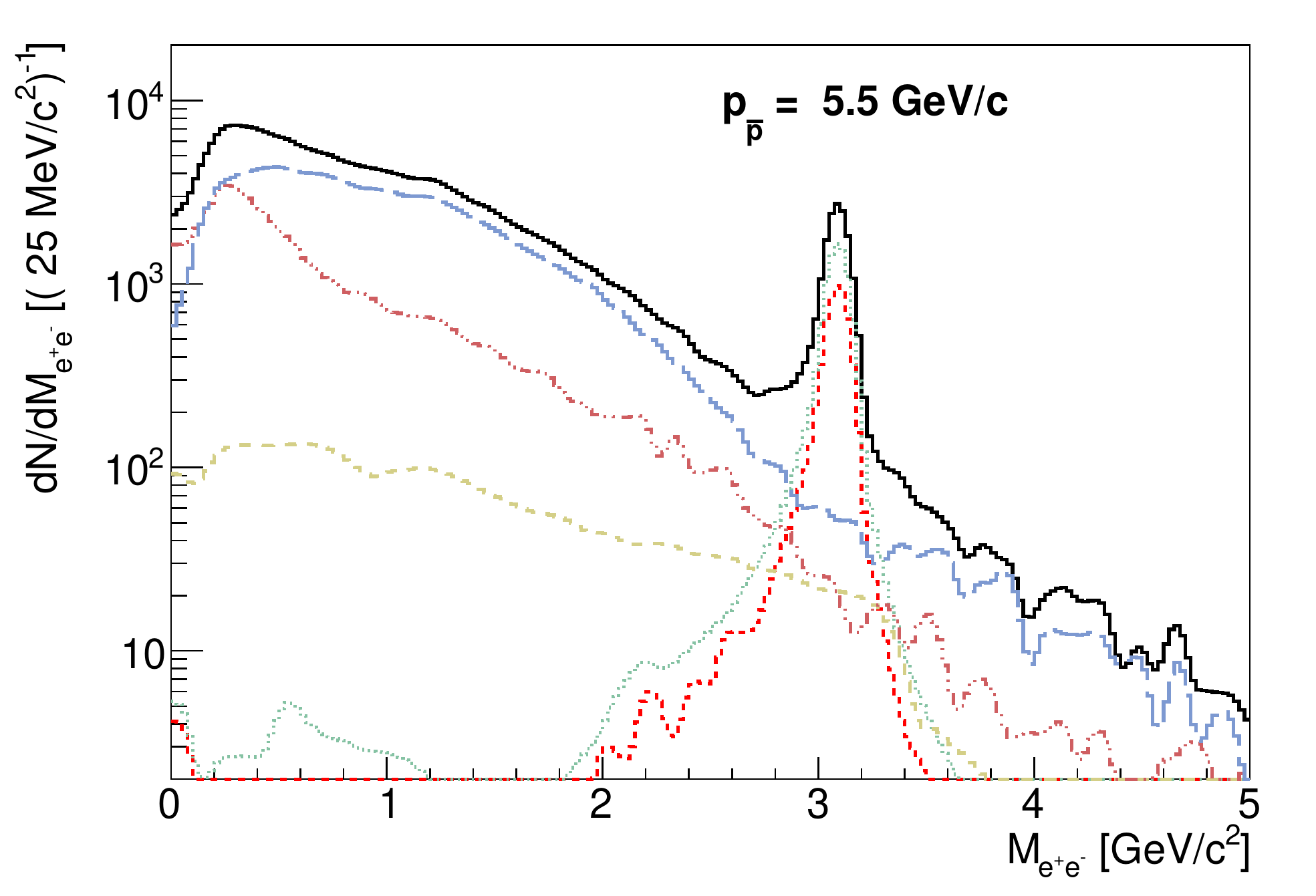}\\
  \includegraphics[width=0.96\columnwidth]{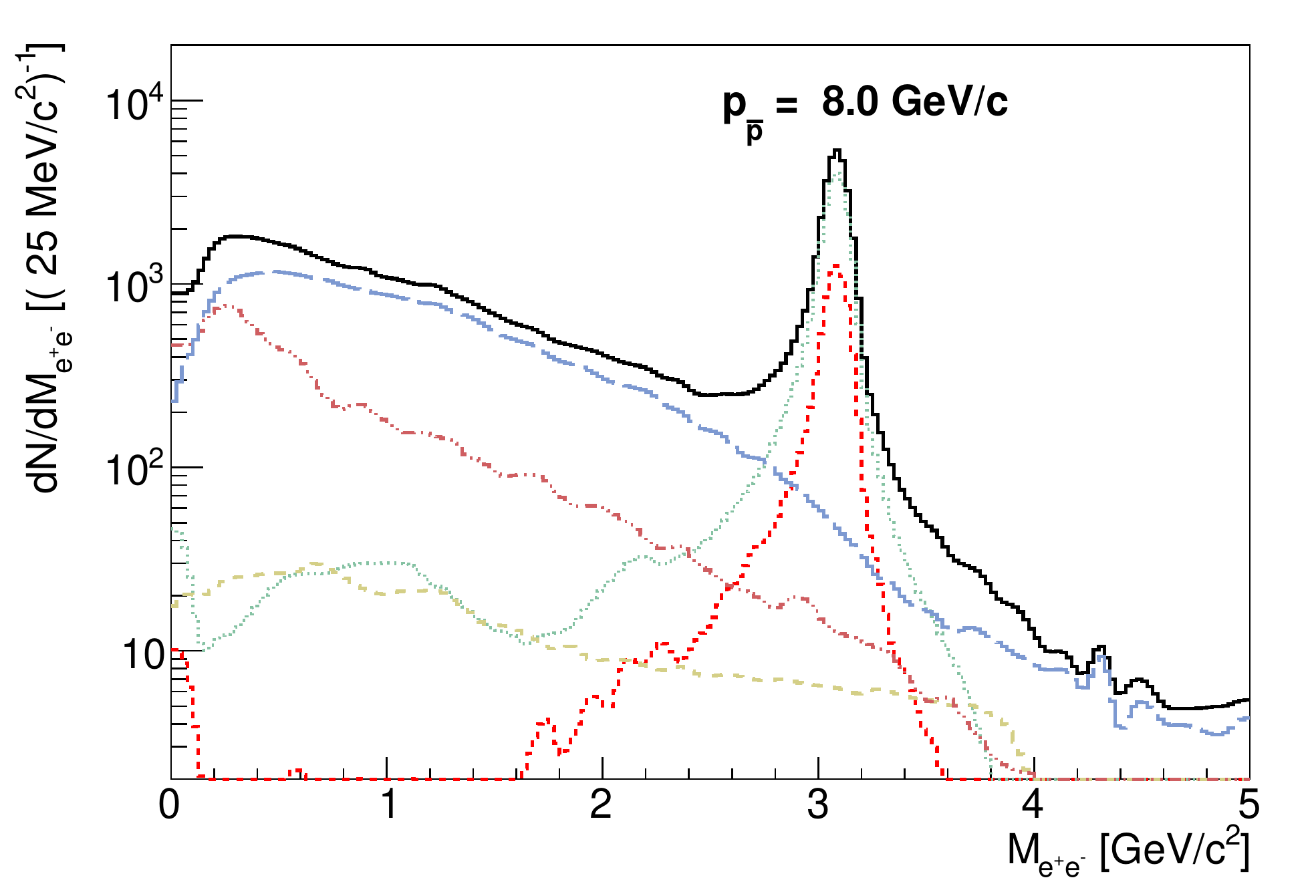}\\
  \includegraphics[width=0.96\columnwidth]{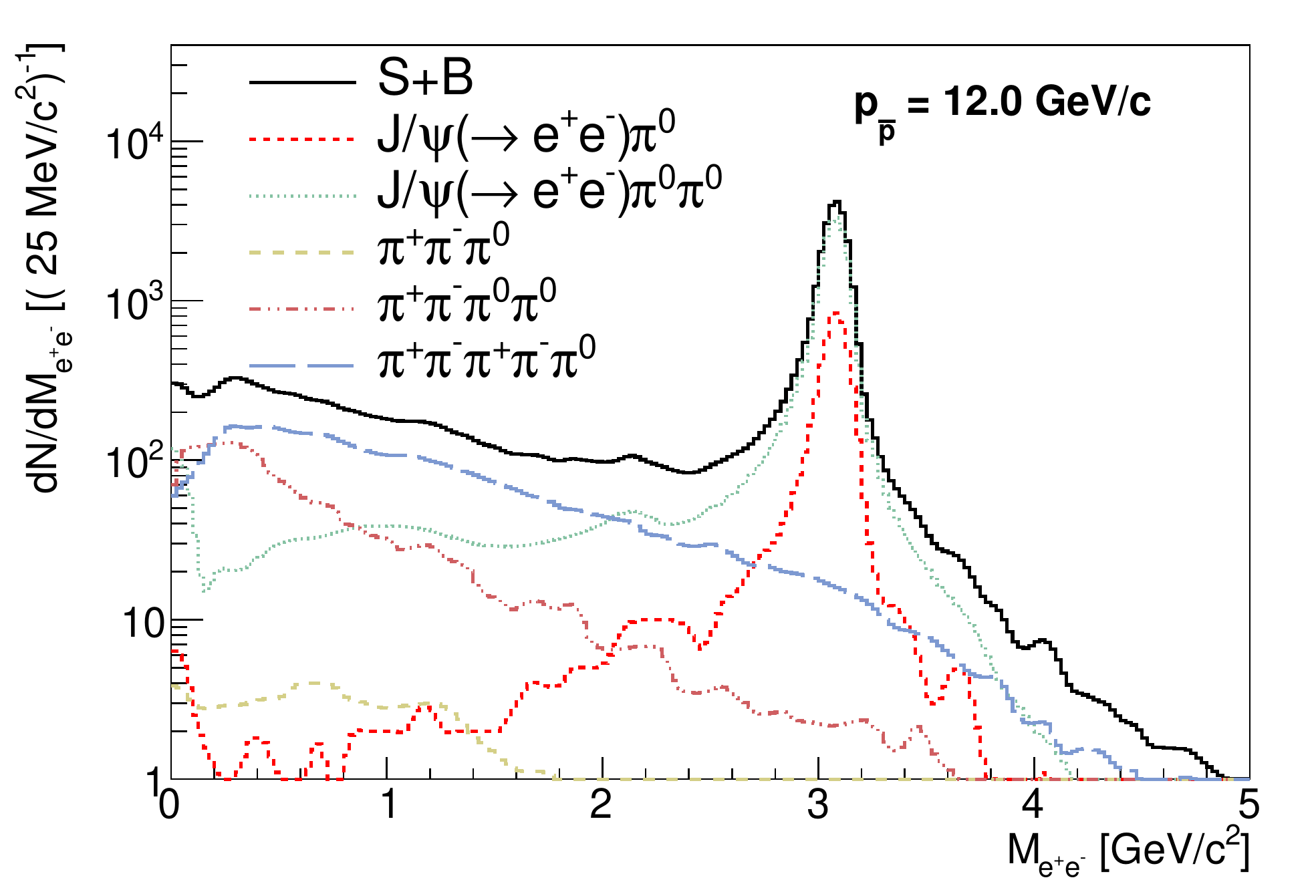}
  \caption{\label{fig:bgsrc_bef_kinfit} Dilepton invariant mass
    distributions of the simulated signal and background sources after
    selection of one $\piz$ and one $\epem$ pair. The three beam
    momenta used for this study are shown: \bmomu{0} \plfm{top panel},
    \bmomu{1} \plfm{middle panel} and \bmomu{2} \plfm{bottom
      panel}. Individual contributions from the various sources of
    background discussed in the text are also plotted with the line
    style depicted in the legend. $\pipmpiz$, $\pipmpizpiz$ and
    $\pipmpipmpiz$ are generated using DPM, whereas $\jpsipizpiz$ is
    generated using the phase-space (PHSP) model. The combined
    contribution (S+B) of the signal (S) and all background (B)
    reactions is shown by the solid line. Events of the signal channel
    $\jpsipiz$ are generated using the TDA model-based generator.}
\end{figure}

Finally, the full event is reconstructed by pairwise combining all
reconstructed $\piz$s with all $\jpsi$ candidate $\epem$ pairs in the
same event. Due to the presence of combinatorial background in both
the $\piz$ reconstruction (random $\gamma\gamma$ pairs) and the
$\jpsi$ reconstruction (random $\epem$ pairs), there could be more
than one candidate $\jpsi\piz$ pair per event. In such events, the
angle between the $\piz$ and $\jpsi$ in the CMS is calculated for each
pair and the combination closest to 180$^\circ$ (the most back-to-back
pair) is selected. Figure~\ref{fig:bgsrc_bef_kinfit} depicts the
invariant mass distributions of the signal reaction as well as all the
simulated background sources after the selection of the $\piz$ and
$\epem$ pair. The sum of contributions (S+B) from signal (S) and all
background sources (B) is shown in the same figure as the black
histogram.

Table~\ref{tab:stob_bef_kin} shows the signal to background ratios of
the different background sources simulated at this stage of the
analysis, and the first column of Table~\ref{tab:big_table} displays
the efficiency of the signal and the rejection powers of different
sources of background. The channels with a charged pion pair are
suppressed with rejection powers of the order 10$^7$. The main effect
(roughly $10^6$) comes from PID, while the remaining factor of $10$
comes from the cut on the charged pair invariant mass. As a result,
background events can be rejected to levels where they can be
subtracted when needed in the $\epem$ invariant mass spectra by a
sideband analysis, in which invariant mass regions to the right and
left side of the $\jpsi$ peak are used to estimate the background
contribution under the peak. This is discussed in more detail in
Section~\ref{sec:sig_cnt}. As expected, the $\jpsipizpiz$ channel is
selected with an efficiency similar to the $\jpsipiz$ channel. It is
therefore now the dominant background source, roughly a factor four
larger than the signal. This ratio is a direct result of the
conservatively high cross section assumption discussed in
Section~\ref{sec:pizpizjpsi}. A dedicated analysis of the
$\jpsipizpiz$ channel will be possible with \PANDA, allowing for
measurement of the cross section at the same c.m\@ energy as the
$\jpsipiz$ signal. In the mean time, we stick to the cross section
inputs as described in Section~\ref{sec:pizpizjpsi}, and propose
further analysis cuts described in the following section to reduce
this background to the percent level, keeping in mind that that they
will later be adjusted to match realistic values of the $\jpsipizpiz$
cross section.

\begin{table}[hbpt]
  \caption{\label{tab:stob_bef_kin} Signal to background ratio for
    different background sources after selection of one $\piz$ and one
    $\epem$ pair. The signal (S) and background (B) are counted within
    a window of 2.8 to 3.3~GeV/$c^2$ in the invariant mass of the
    charged pair, and inside the validity range of the TDA model.}
  \begin{ruledtabular}
    \begin{tabular}{cccc}
      S/B &  \bmomu{0} & \bmomu{1} & \bmomu{2} \\
      \colrule
      $\jpsipizpiz$  & 0.273 & 0.251  & 0.225  \\
      $\pipmpiz$      & 12.6  & 56.4   & 366.0  \\
      $\pipmpizpiz$  & 10.3  & 24.9   & 101.0  \\
      $\pipmpipmpiz$ & 4.8   & 6.76   & 15.6   \\
      combined         & 0.247 & 0.239  & 0.221  \\
    \end{tabular}
  \end{ruledtabular}
\end{table}

\subsection{Kinematic Fit}

\begin{figure}[hbpt]
  \includegraphics[width=0.97\columnwidth]{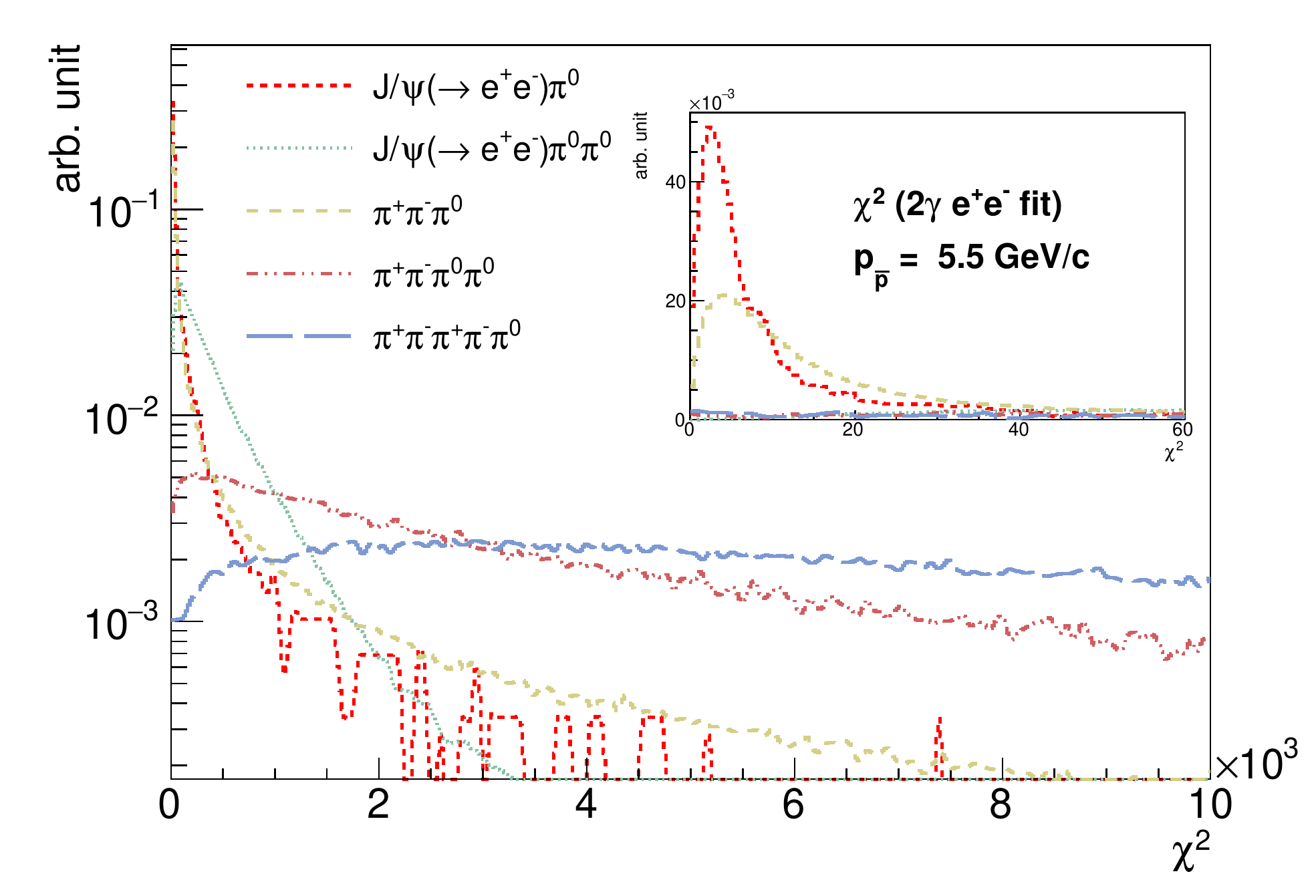}
  \includegraphics[width=0.97\columnwidth]{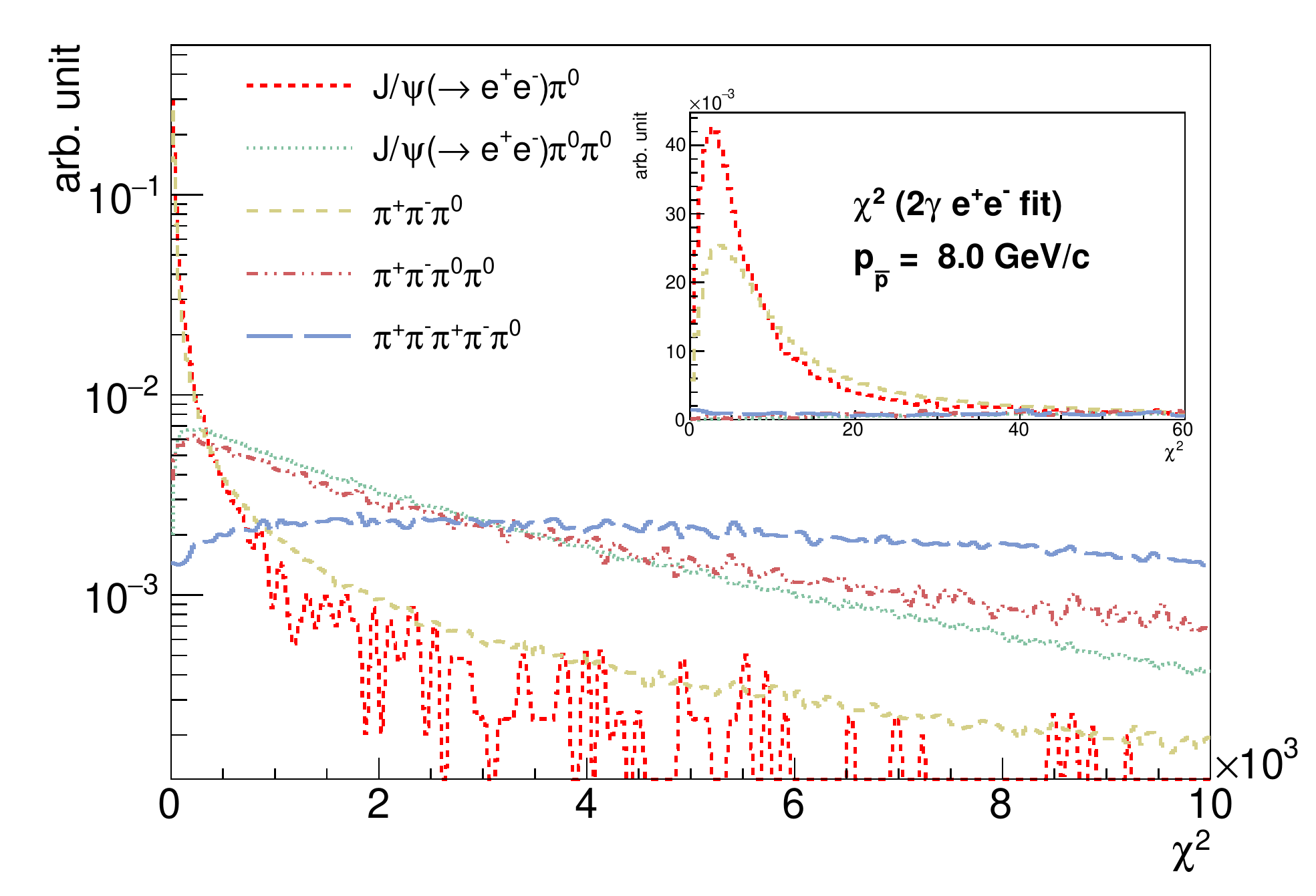}
  \includegraphics[width=0.97\columnwidth]{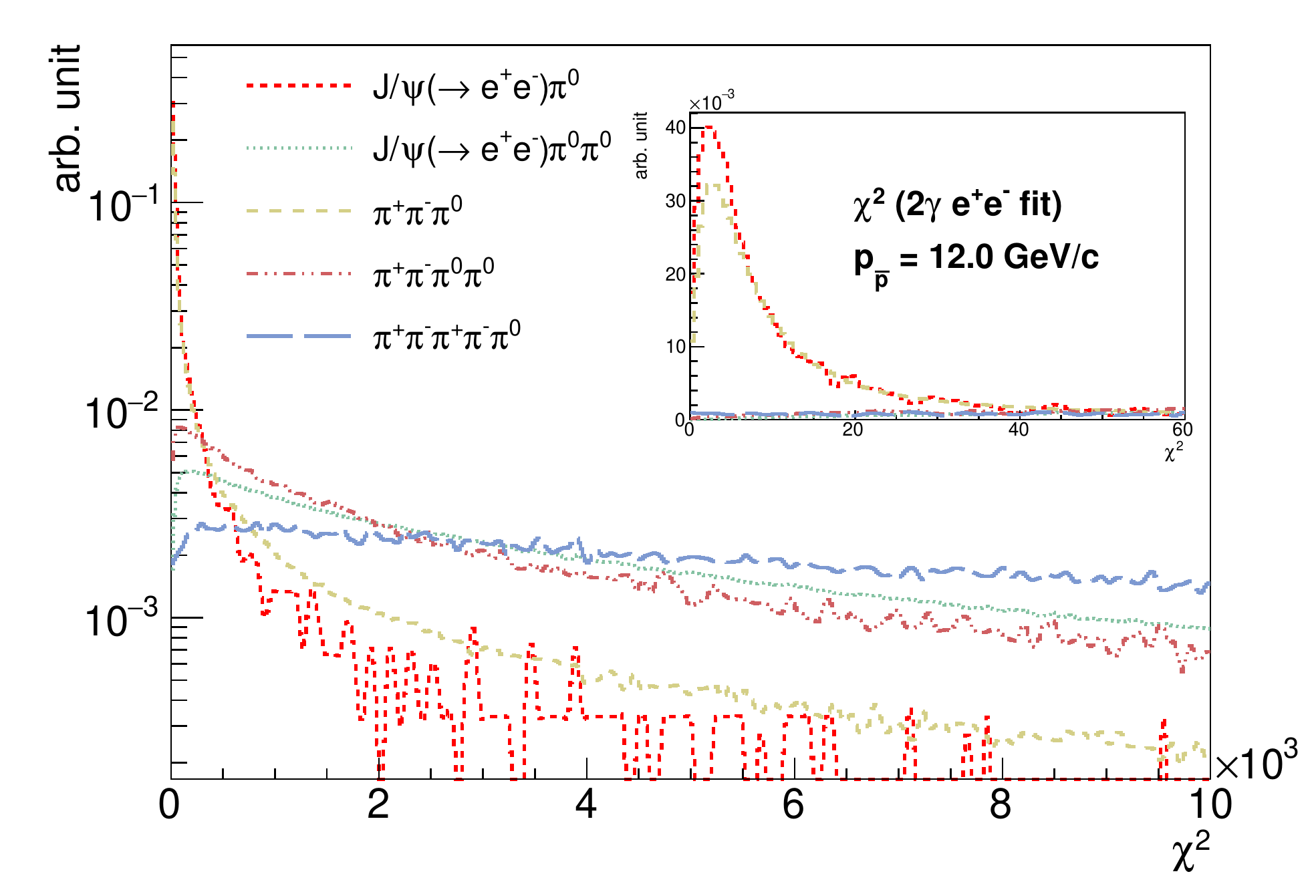}
  \caption{\label{fig:sig_hyp_chi2} The $\chi^2$ distribution of the
    kinematic fit (main plot). The inset shows a small region of
    $\chi^2$ with a linear scale. The three plots represent the
    different beam momenta studies, \bmomu{0} (top), \bmomu{1}
    (center) and \bmomu{2} (bottom). The different distributions
    represent the signal and four sources of background simulated for
    the feasibility study. The distributions are normalized to have
    the same integral for easier comparison.}
\end{figure}

The $\jpsipizpiz$ background will be reduced by exploiting the
kinematic differences to the $\jpsipiz$
channel. Figure~\ref{fig:sig_hyp_chi2} shows the $\chi^2$
distributions of a kinematic fit with four-momentum conservation
enforced as a constraint by assuming an exclusive $\tgamma\epem$ event
(denoted as $\chi^2_{\sigh}$). The plots are shown for the three beam
momenta of this study. The insets show the same plot on a linear scale
with a restricted range on the $\chi^2$ axis, demonstrating that
$\chi^2_{\sigh}$ is peaked at values compatible with the four degrees
of freedom of the kinematic fit to the $\sigrxnshort$ hypothesis. In
contrast, the $\jpsipizpiz$ events show a significantly flatter
$\chi^2_{\sigh}$ distribution extending to very large values,
providing a powerful tool for further rejection. As shown in the
second column of Table~\ref{tab:big_table}, by applying a maximum
cut-off on $\chi^2_{\sigh}$ of 20, 50 and 100 at $\mompbar=$~\bmom{0},
\bmom{1} and \bmomu{2}, respectively, it is possible to reduce the
$\jpsipizpiz$ contamination to less than 8\%, while keeping the
corresponding loss in signal efficiency to $\approx$~15~--~30\%,
depending on $\mompbar$.

\begin{figure*}[hbpt]
  \includegraphics[width=0.9\textwidth]{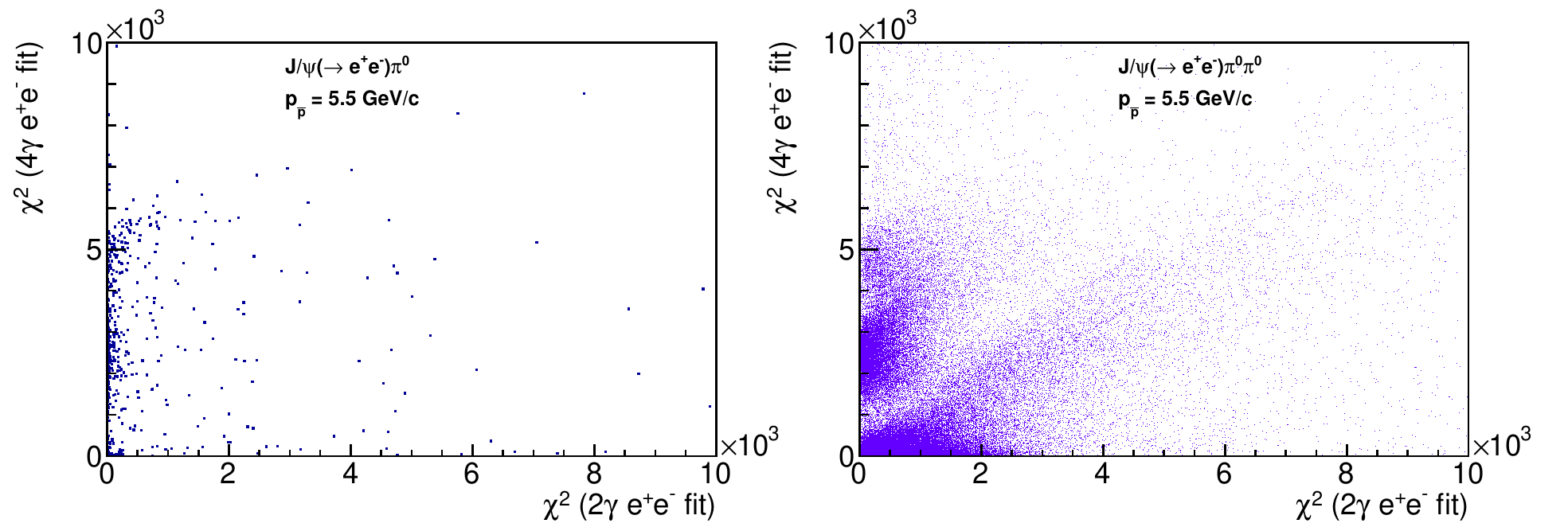}\\
  \caption{\label{fig:bg_hyp_chi2} The correlation between
    $\chi^2_{\sigh}$ and $\chi^2_{\bgh}$ for $\jpsipiz$ \plfm{left}
    and $\jpsipizpiz$ \plfm{right} events at beam momentum of
    $\mompbar=$~\bmomu{0}.}
\end{figure*}

Despite the significant improvement of the S/B ratio, additional cuts
are needed to bring the contamination by $\jpsipizpiz$ to under a few
percent at all energies. A comparison of the $\chi^2$ value of the
kinematic fit for the $2\gamma\epem$ hypothesis ($\chi^2_{\sigh}$)
with the $\chi^2$ value for the $4\gamma\epem$ hypothesis
($\chi^2_{\bgh}$) can provide additional discrimination power. The
correlation between $\chi^2_{\sigh}$ and $\chi^2_{\bgh}$ is shown in
Fig.~\ref{fig:bg_hyp_chi2}. By requiring
$\chi^2_{\bgh}~>~\chi^2_{\sigh}$ for those reconstructed $\jpsipiz$
events with an additional $\tgamma$ pair in the event lowers the
$\jpsipizpiz$ contamination down to less than 5\%, depending on
$\mompbar$, with no significant loss in signal efficiency.

\begin{figure}[hbpt]
  \includegraphics[width=\columnwidth]{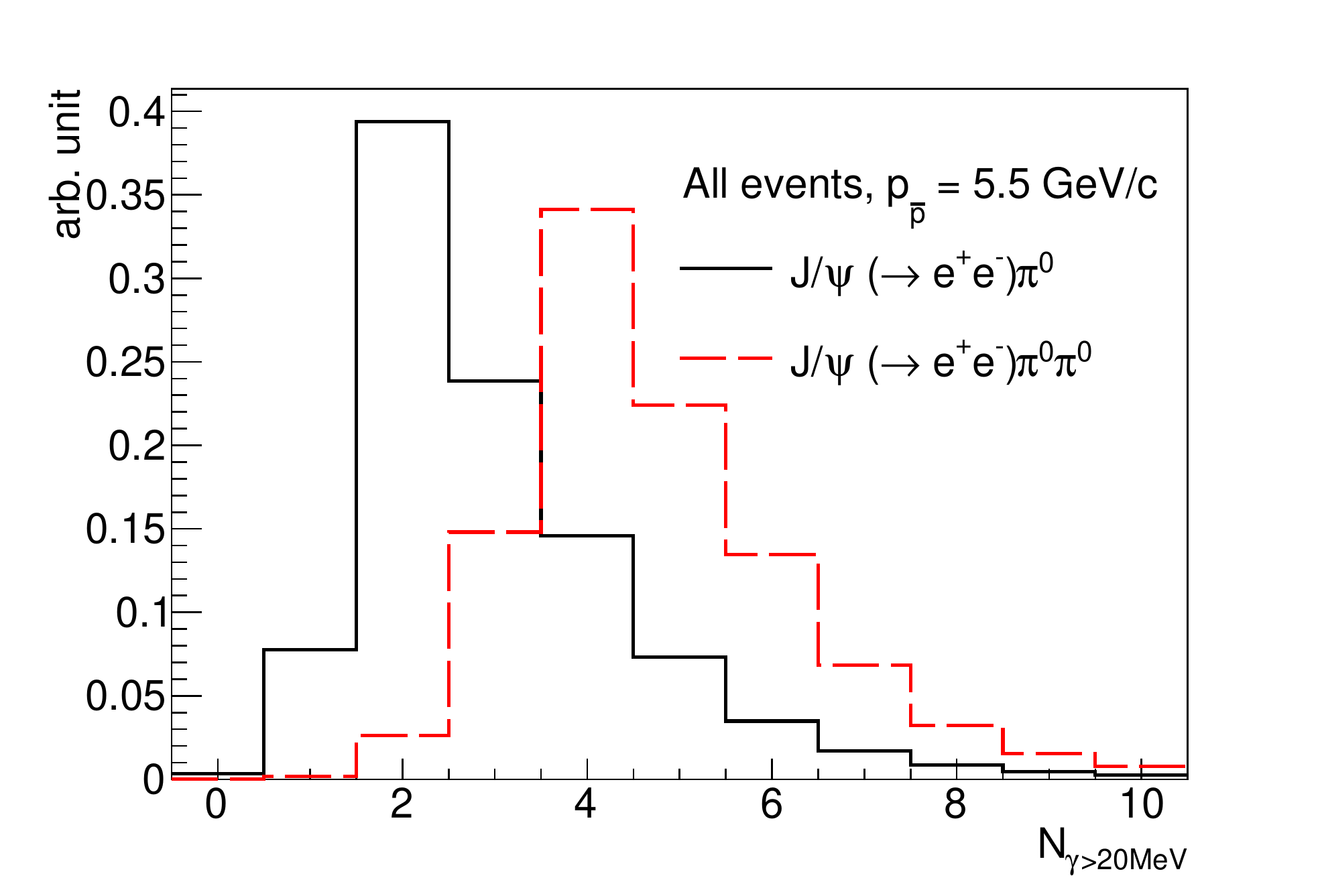}
  \caption{\label{fig:ngamma20mev} Distribution of the number of
    photons above 20~MeV per event for the $\jpsipiz$ signal
    \plfm{solid line} and $\jpsipizpiz$ background \plfm{dashed line}
    channels at beam momentum of $\mompbar=$~\bmomu{0}. }
\end{figure}

Finally, by requiring that the number of photons in the event with
energy above 20 MeV ($N_{\gamma(>20MeV)}$, shown in
Fig.~\ref{fig:ngamma20mev} for $\jpsipiz$ and $\jpsipizpiz$ events) to
be less than or equal to three, it is possible to achieve a
$\jpsipizpiz$ contamination of less than 2\% with an additional loss
of efficiency of only about 10~--~15\%.

Table~\ref{tab:big_table} summarizes key quantities, signal
efficiency, contamination from $\jpsipizpiz$ and $\pipmpiz$ and purity
from combinatorial background as well as S/B ratio including all
background sources after each step in the analysis procedure described
above, starting from the $\epempiz$ selection.

\begin{table}[hbpt]
  \caption{\label{tab:big_table}The efficiency for signal events and
    rejection factor ($\mathcal{R}$) of the two significant background
    contributions, together with background contamination
    ($\mathcal{C}$), signal purity from combinatorial background
    ($\mathcal{P}_{comb}$) and total signal to background ratio. The
    quantities are tabulated with successive application of kinematic
    cuts starting from the selection of the $\piz$ and $\epem$ pair
    (column $\epempiz$ sel.). The signal and background count rates
    are determined within a $M_{\epem}$ window of 2.8 to 3.3 GeV/$c^2$
    in the invariant mass of the charged pair. The beam momentum for
    each table is given in the top left cell.}
  \begin{ruledtabular}
    \begin{tabular}{ccccc}
      5.5~GeV/$c$ & $\epempiz$ sel. & $\chi^2_{\sigh}$ & $\chi^2_{\bgh}$ & $N_{\gamma(>20MeV)}$ \\
      \colrule
      $\rowa$ & 17.8\%   & 12.5\%   & 12.5\%   & 11.3\%   \\
      $\rowb$ & 4.5      & 360      & 590  & 1.6$\times10^{3}$  \\
      $\rowc$ & 9.0$\times10^{8}$  & 1.8$\times10^{9}$  & 1.8$\times10^{9}$  & 2.9$\times10^{9}$  \\
      $\rowd$ & 441.9\%  & 7.5\%    & 4.5\%    & 1.8\% \\
      $\rowe$ & 6.6\%    & 4.9\%    & 4.9\%    & 3.4\% \\
      $\rowf$ & 90.4\%   & 98.8\%   & 98.8\%   & 99.0\%   \\
      $\rowg$ & 0.2      & 7.7      & 10.1      & 19.6     \\
      \colrule
      \colrule
      8.0~GeV/$c$ & $\epempiz$ sel. & $\chi^2_{\sigh}$ & $\chi^2_{\bgh}$ & $N_{\gamma(>20MeV)}$ \\
      \colrule
      $\rowa$ & 15.5\%   & 12.2\%   & 12.2\%   & 10.5\%   \\
      $\rowb$ & 5.2      & 650      & 1.4$\times10^{3}$  & 3.8$\times10^{3}$  \\
      $\rowc$ & 7.6$\times10^{8}$  & 1.3$\times10^{9}$  & 1.3$\times10^{9}$  & 2.2$\times10^{9}$  \\
      $\rowd$ & 456.4\%  & 3.4\%    & 1.7\%    & 0.8\% \\
      $\rowe$ & 1.5\%    & 1.2\%    & 1.2\%    & 0.8\% \\
      $\rowf$ & 89.3\%   & 98.7\%   & 98.7\%   & 99.0\%   \\
      $\rowg$ & 0.2      & 19.6     & 33.7     & 67.5     \\
      \colrule
      \colrule
      12.0~GeV/$c$ & $\epempiz$ sel. & $\chi^2_{\sigh}$ & $\chi^2_{\bgh}$ & $N_{\gamma(>20MeV)}$ \\
      \colrule
      $\rowa$ &  9.4\%   &  7.9\%   &  7.9\%   &  6.6\%   \\
      $\rowb$ & 7.7      & 6.8      & 1.8$\times10^{3}$  & 5.3$\times10^{3}$  \\
      $\rowc$ & 1.4$\times10^{9}$  & 2.1$\times10^{9}$  & 2.1$\times10^{9}$  & 4.0$\times10^{9}$  \\
      $\rowd$ & 413.2\%  & 4.0\%    & 1.2\%    & 0.6\% \\
      $\rowe$ & 0.2\%    & 0.1\%    & 0.1\%    & 0.1\% \\
      $\rowf$ & 89.1\%   & 99.7\%   & 98.8\%   & 99.0\%   \\
      $\rowg$ & 0.2      & 21.9     & 57.9     & 159.3    \\
    \end{tabular}
  \end{ruledtabular}
\end{table}

\subsection{\label{sec:stob}Signal to Background Ratio}

The expected background contamination from all sources considered in
this study is plotted in Fig.~\ref{fig:reco_sig_rate_tu} as the shaded
histogram for each validity range at the three beam momenta. In the
worst scenario (at the lowest energy studied here at
$\mompbar=$~\bmomu{0}), the S/B ratio will be at least of the order of
15 at all values of $t$. At higher energies, the background
contamination from $\bgrxn$ is about or less than a percent, even with
the conservative estimates of the background cross section used. As
already mentioned, \PANDA\ will provide dedicated measurements of
these background channels, hence allowing for a subtraction of the
corresponding contributions. In addition, background channels with no
$\jpsi$ in the final state can be suppressed using a sideband analysis
of the invariant mass distribution of the charged pair, as will be
discussed in the next section. This gives us confidence that the
measurement should be readily feasible from the point of view of
background rejection.

\subsection{\label{sec:sig_cnt}Sideband Background Subtraction}

The number of reconstructed signal events is extracted by subtracting
the background contribution from the total number of entries within
the range \mjpsiwin. The background contribution is obtained by
integrating the polynomial component of the Crystal Ball (see
Section~\ref{sec:jpsi_rec}) plus third order polynomial function
fitted to the sum of signal and background $\epem$ invariant mass
histograms. The fitted range was set to \fitwin. The parameter for the
position of the maximum and the width of the Crystal Ball component
were fixed to values extracted from the fit to the integrated $\jpsi$
yield shown in Fig.~\ref{fig:jpsi_peak}, whereas the remaining
parameters of the function were allowed to vary freely during
fitting. The loss of signal events that fall outside of the counting
window due to the Bremsstrahlung tail of the $\jpsi$ peak is accounted
for as an analysis cut efficiency loss in the correction procedure
that will be outlined in Section~\ref{sec:eff_corr}.

\begin{figure*}[hbpt]
  \includegraphics[width=0.33\textwidth]{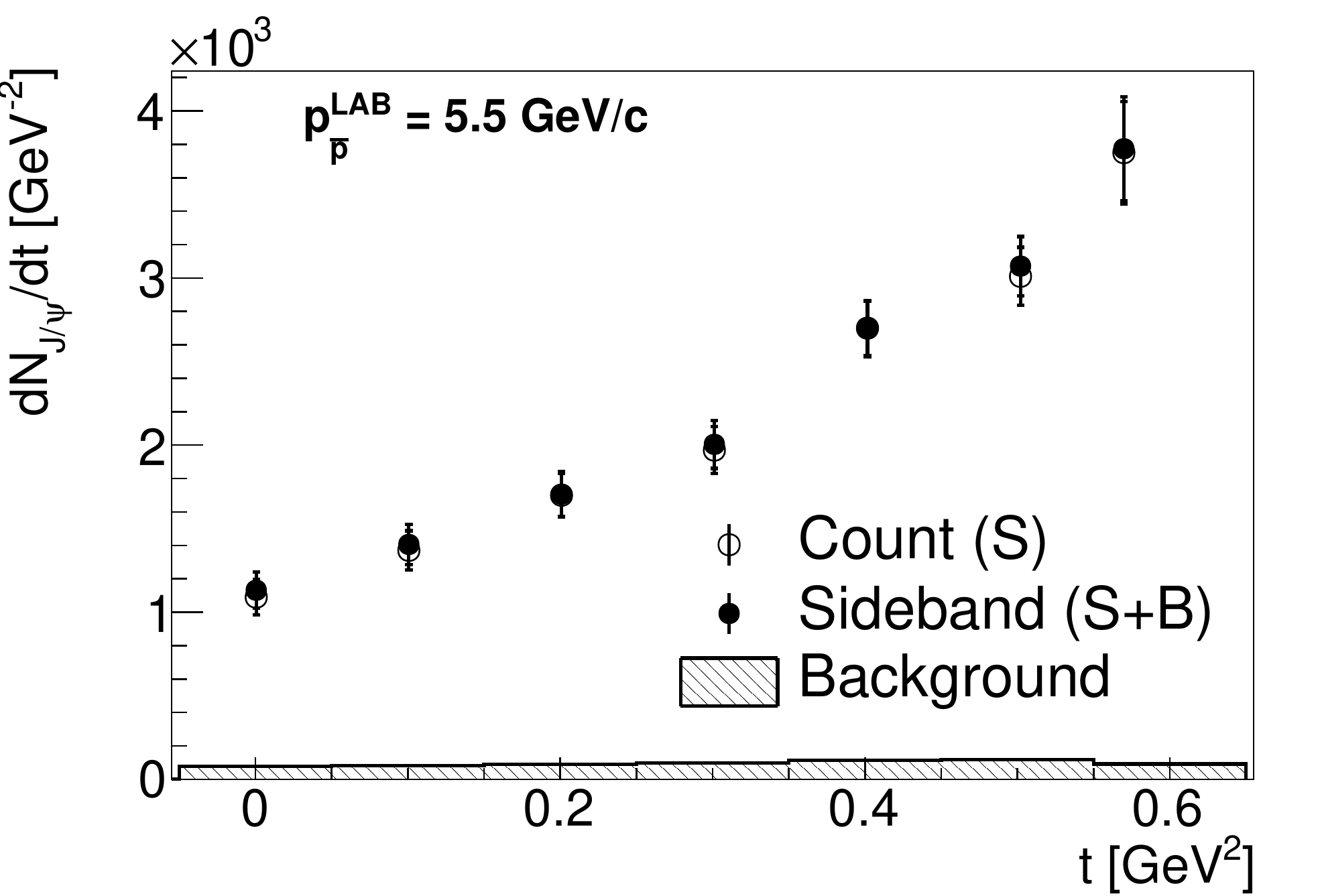}
  \includegraphics[width=0.33\textwidth]{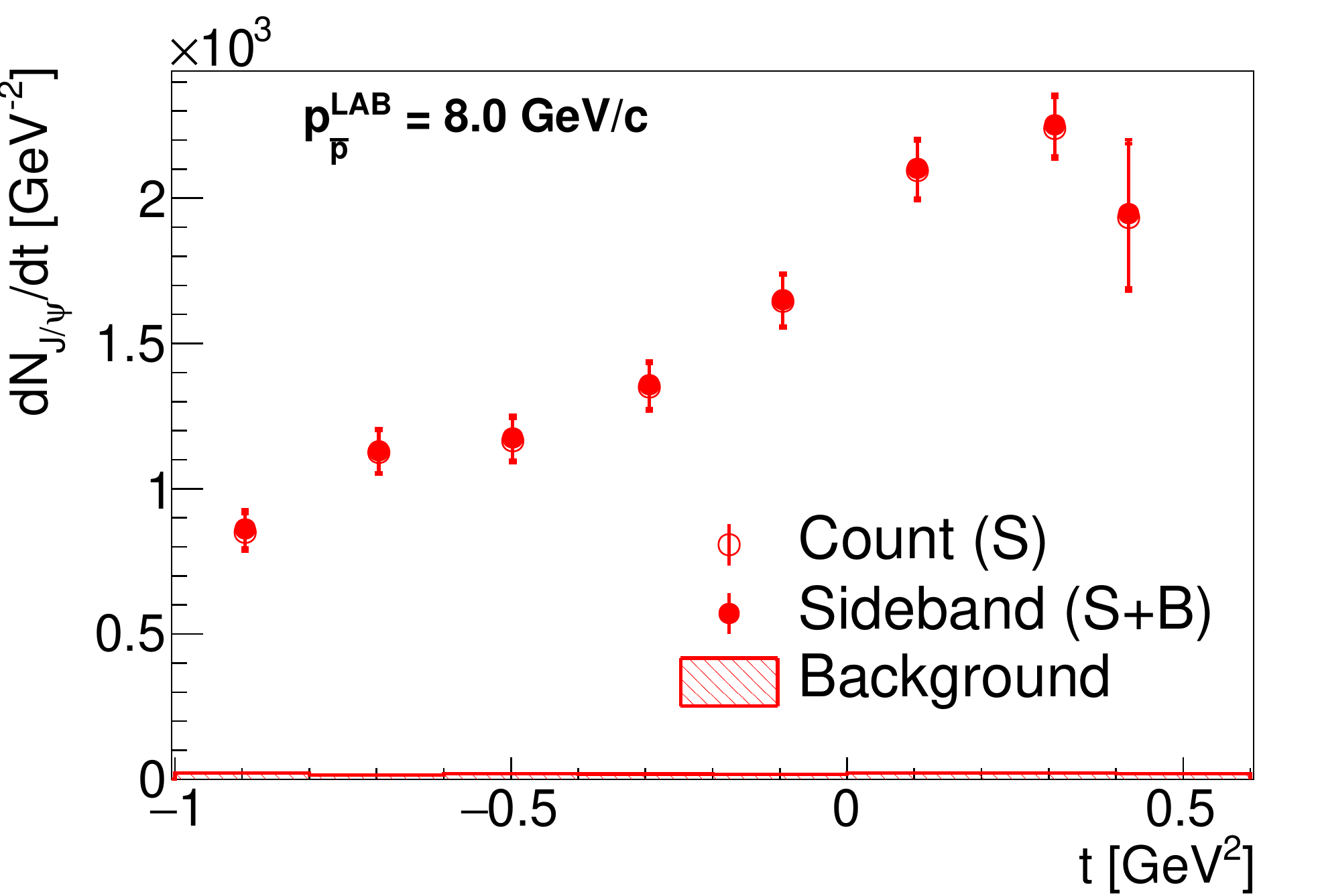}
  \includegraphics[width=0.33\textwidth]{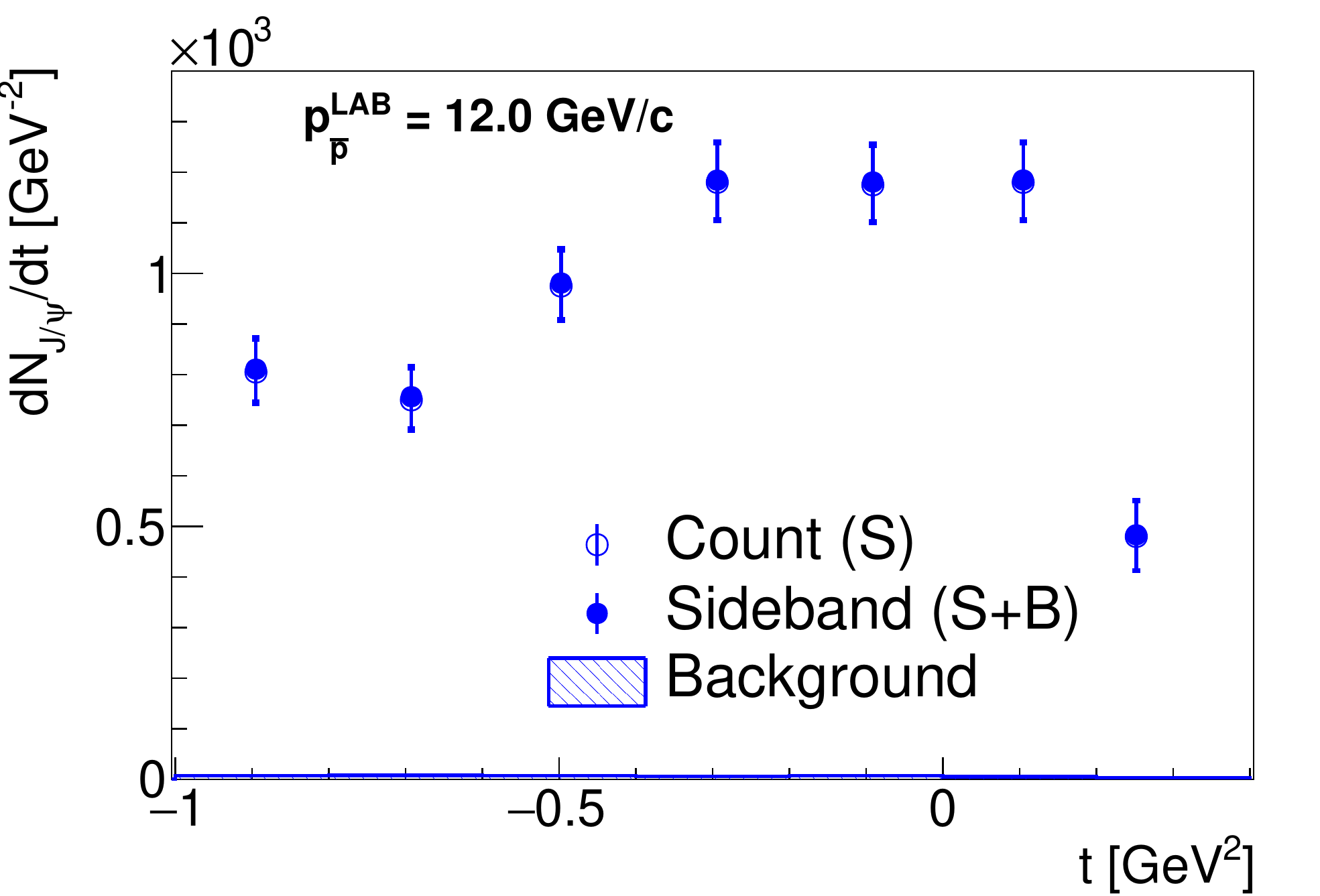}\\
  \includegraphics[width=0.33\textwidth]{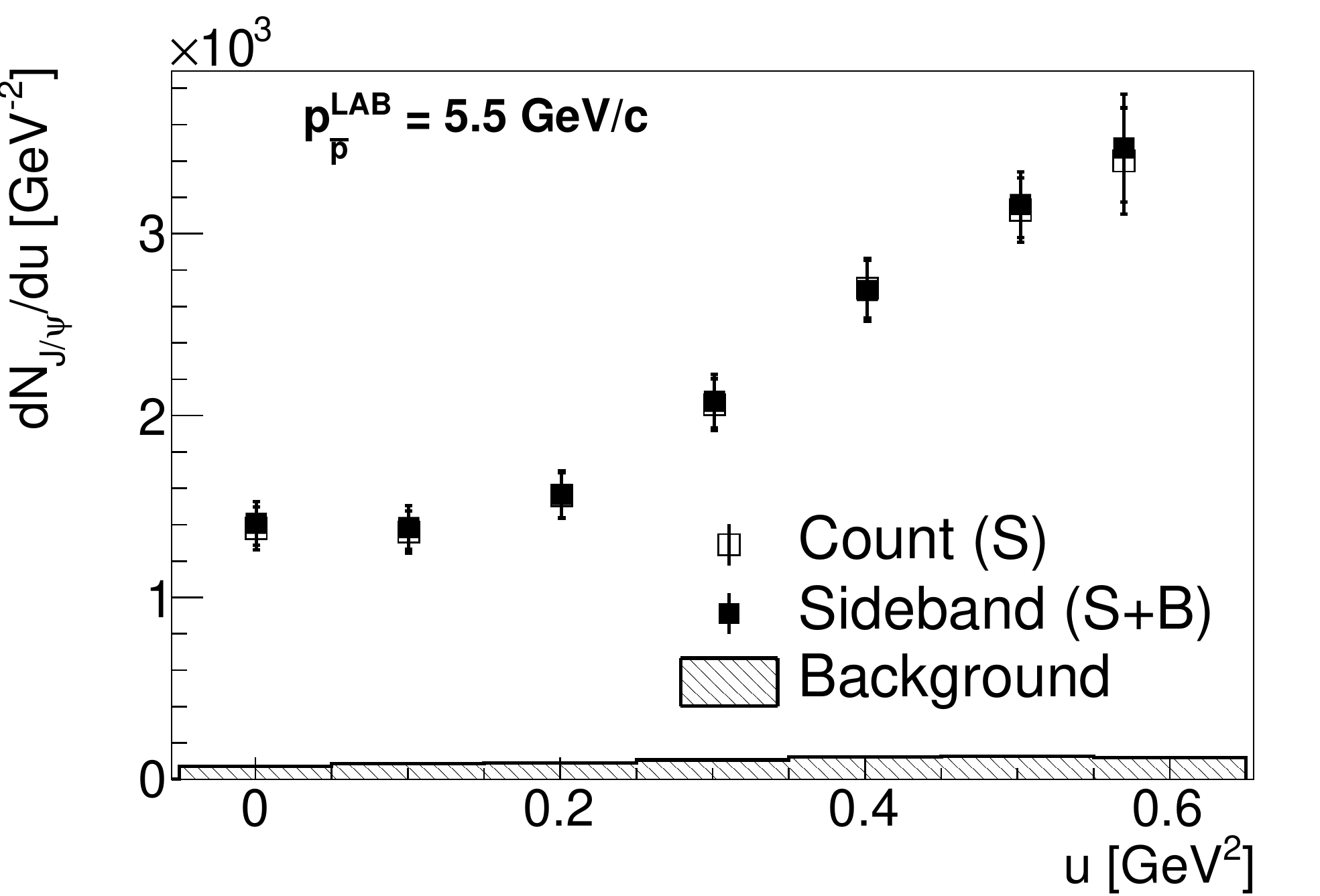}
  \includegraphics[width=0.33\textwidth]{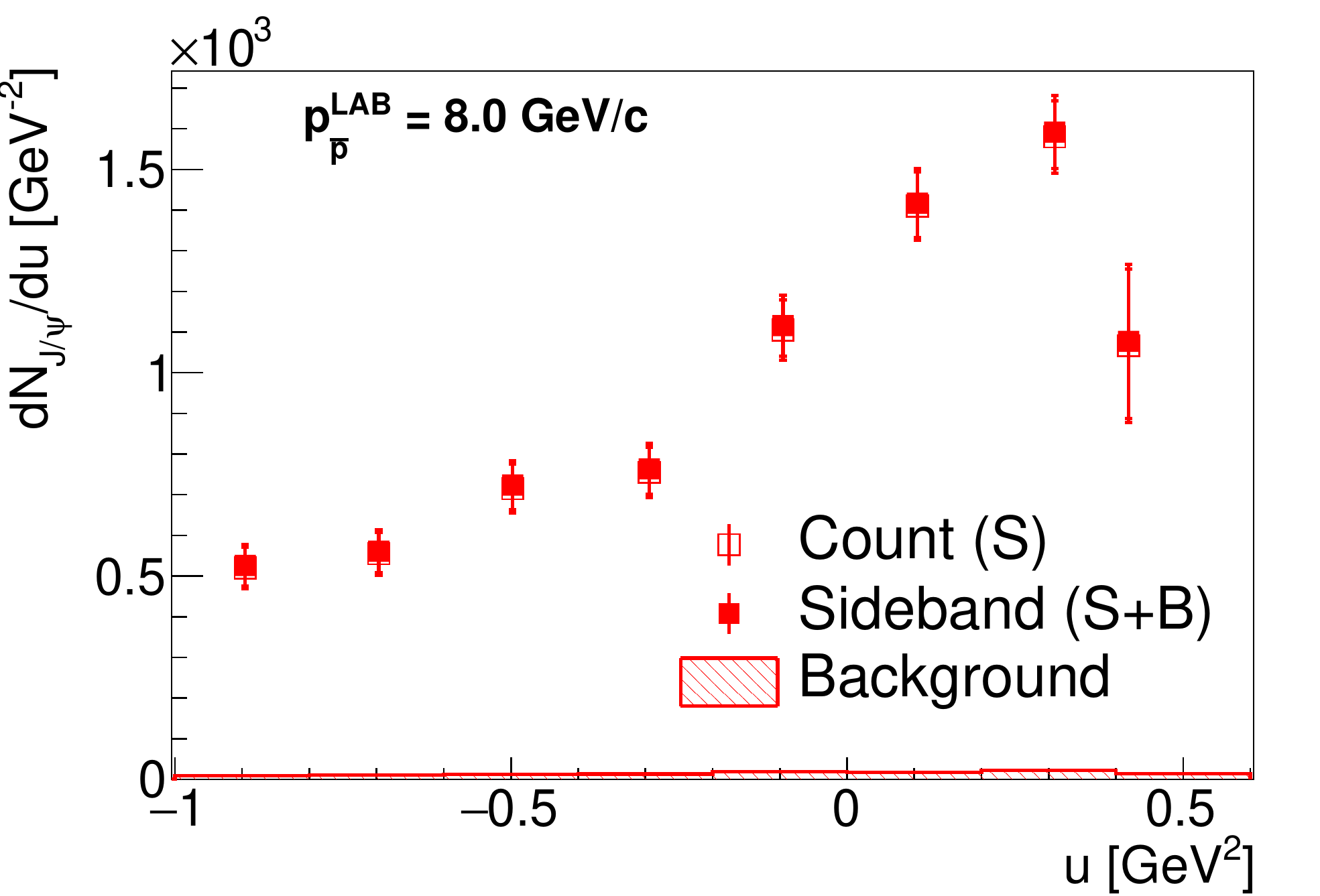}
  \includegraphics[width=0.33\textwidth]{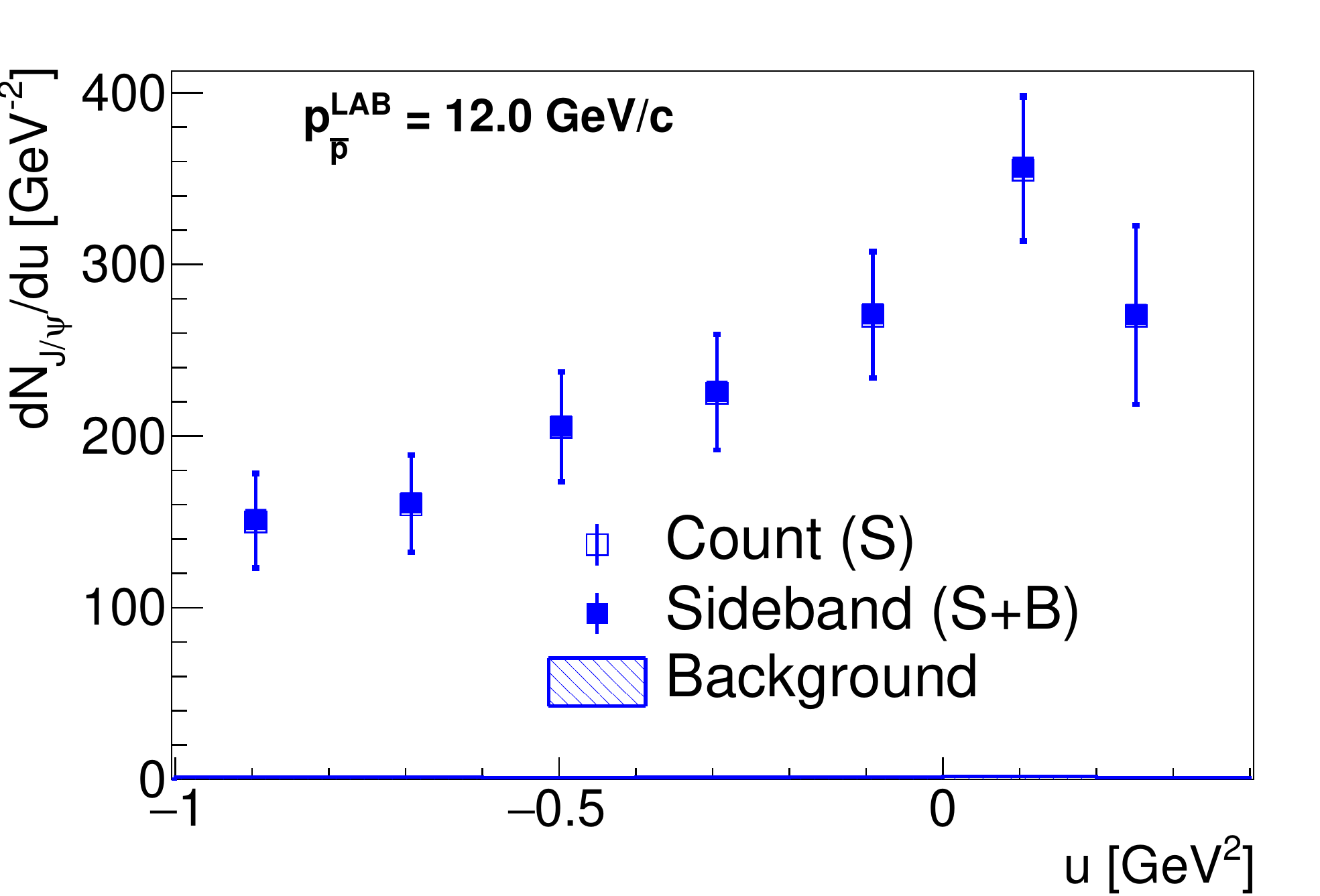}
  \caption{\label{fig:reco_sig_rate_tu} The count rate of fully
    reconstructed $\jpsi\piz$ events as a function of $t$ in the
    near-forward \plfm{top row} and as a function of $u$ in the
    near-backward \plfm{bottom row} kinematic approximation validity
    ranges for the three incident $\pbar$ momenta: \bmomu{0}
    \plfm{left column}, \bmomu{1} \plfm{middle column} and \bmomu{2}
    \plfm{right column}. The full points are obtained using the
    sideband method from foreground (S+B) histograms as explained in
    Section~\ref{sec:sig_cnt}.  The open symbols denote the count
    rates for signal only simulations (S) within the range
    \mjpsiwin. The amount of background within the same window is
    shown by the shaded histogram. The plot corresponds to an
    integrated luminosity of 2~\fbi, after application of all cuts
    (EID, $\chi^2$ from kinematic fitting, number of photons per
    event).}
\end{figure*}

The result of the subtraction procedure is summarized in
Fig.~\ref{fig:reco_sig_rate_tu} as a function of squared momentum
transfer for the near-forward and near-backward approximations. For
comparison, the count rates of $\epem$ pairs within the same $\jpsi$
mass window from the signal simulation (with no background added) are
shown as open markers. The sideband subtraction procedure
overestimates the signal count rate by roughly the amount of
contamination that comes from $\jpsipizpiz$ events, since it can only
take into account background sources such as misidentified $\pipmpiz$
events that vary smoothly.

\subsection{\label{sec:eff_corr}Efficiency Correction}

This section describes a procedure for efficiency correction of the
reconstructed signal count rate to obtain differential cross sections,
and compare the result to the TDA model that was used to generate the
signal events. The result will also serve as a guide to the
statistical uncertainties expected for this measurement. For
illustration, the $t$ variable in the near-forward kinematic
approximation is used, but the same arguments hold for the
near-backward kinematic approximation if $t$ is replaced by $u$. The
fully corrected data points at a given value of $t$ are calculated
using the equation:
\begin{equation}
  \frac{d\sigma}{dt}(t) = \frac{1}{\mathcal{L}_{int} \cdot BR(\jpsitoepem)} \frac{1}{\varepsilon(t)} \frac{N_{\jpsi\piz}(t)}{\Delta t},
  \label{eq:eff_corr}
\end{equation}
\noindent
where $\mathcal{L}_{int}$ is the integrated luminosity (2~\fbi), and
$BR(\jpsitoepem)$ indicates the branching ratio of the $\jpsitoepem$
decay channel (5.94\%). The quantity $N_{\jpsi\piz}$ is the number of
reconstructed $\jpsi\piz$ events in a particular bin in $t$, and
$\Delta t$ is the width of the bin. The location of the point on the
$t$ axis is determined by the mean $t$ value of all the entries in the
given $t$-bin at the generator level.

For the efficiency calculation, a separate simulation data set of the
signal channel was used. The efficiency correction $\varepsilon(t)$ in
a given bin $[t_{\min},\ t_{\max}]$ is defined as the ratio of the
number of reconstructed events to the number of the generated events
within that window:
\begin{equation}
  \varepsilon(t) = \frac{N^{REC}_{\jpsi\piz}(t_{\min} < t_{REC} < t_{\max})}{N^{GEN}_{\jpsi\piz}(t_{\min} < t_{MC} < t_{\max})}.
  \label{eq:eff_defn}
\end{equation}

\begin{figure}[hbpt]
  \includegraphics[width=\columnwidth]{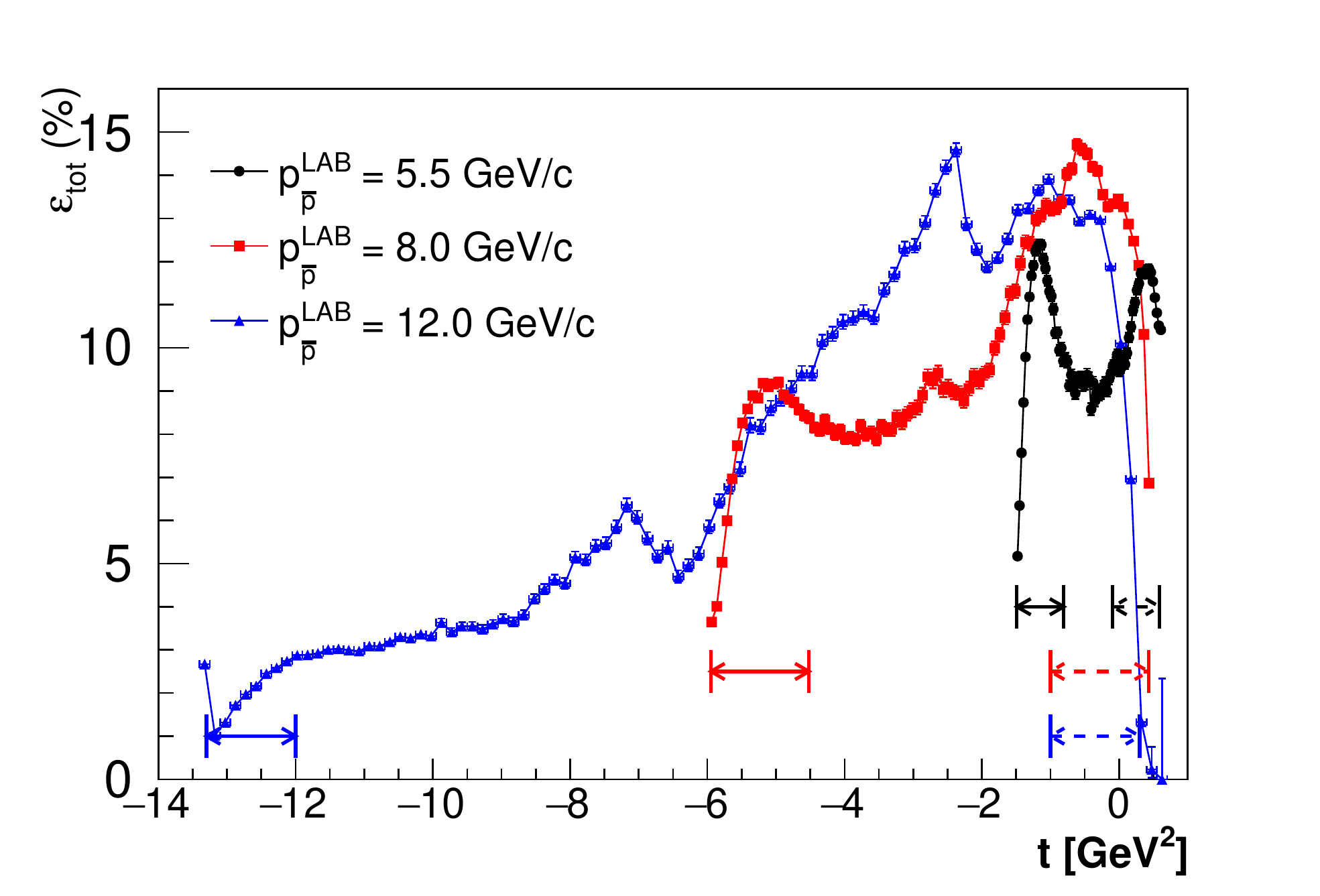}
  \caption{\label{fig:eff_sig} The overall signal reconstruction
    efficiency as a function of $t$ for the three incident $\pbar$
    momenta considered in this study: \bmomu{0} \plfm{circles},
    \bmomu{1} \plfm{squares} and \bmomu{2} \plfm{triangles}. Note
    that the distribution as a function of $u$ is a reflection of
    these distributions with respect to their individual center
    point. The arrows span the near-forward \plfm{dashed line} and
    near-backward \plfm{solid line} kinematic validity ranges of the
    TDA model. }
\end{figure}

We note that in the numerator, the number of signal events is counted
in a bin determined by the reconstructed value of $t$, denoted by
$t_{REC}$, whereas in the denominator the generated (true MC) value of
$t$, denoted by $t_{MC}$, is used. The signal reconstruction
efficiencies calculated in this manner for the three incident $\pbar$
momenta are shown in Fig.~\ref{fig:eff_sig} for the full kinematic
range accessible at each energy regardless of the validity range of
the TDA model. The efficiency as a function of $u$ is just the mirror
image of this distribution, where the point of reflection sits midway
on the full domain covered at the particular energy. Therefore, the
efficiency for the backward $\piz$ emission can be deduced from this
plot by looking at the most negative values of $t$. For clarity, the
validity ranges of the TDA model on the $t$ axis for both the
near-forward and near-backward kinematic approximation regimes are
shown as arrows at the bottom of the plot with a color that
corresponds to the efficiency histogram. As far as count rates are
concerned, this efficiency plot illustrates that our study, which is
based on the TDA model, can be extrapolated to any other model. In
particular, the reaction $\sigrxnshort$ was recently studied in a
hadronic model including intermediate baryonic
resonances~\cite{Wiele:2013vla}, where detailed predictions have been
made for the center of mass energy dependence of the cross section and
the $\piz$ angular distribution, which can be compared to the future
\PANDA\ data. While the efficiency in the near-forward and
near-backward regimes for the lowest beam momentum (\bmomu{0}) are
more or less comparable, this is not true for the higher beam momenta,
where the efficiency in the backward regime tails off to lower
values. This is due to the increasing probability for one of the
leptons from the $\jpsi$ decay to fall in the forward spectrometer
region outside the acceptance of the detectors included in this
analysis.

\section{\label{sec:results}Sensitivity for Testing TDA Model}

\subsection{\label{sec:precision}Differential Cross Sections}

\begin{figure*}[hbpt]
  \includegraphics[width=0.33\textwidth]{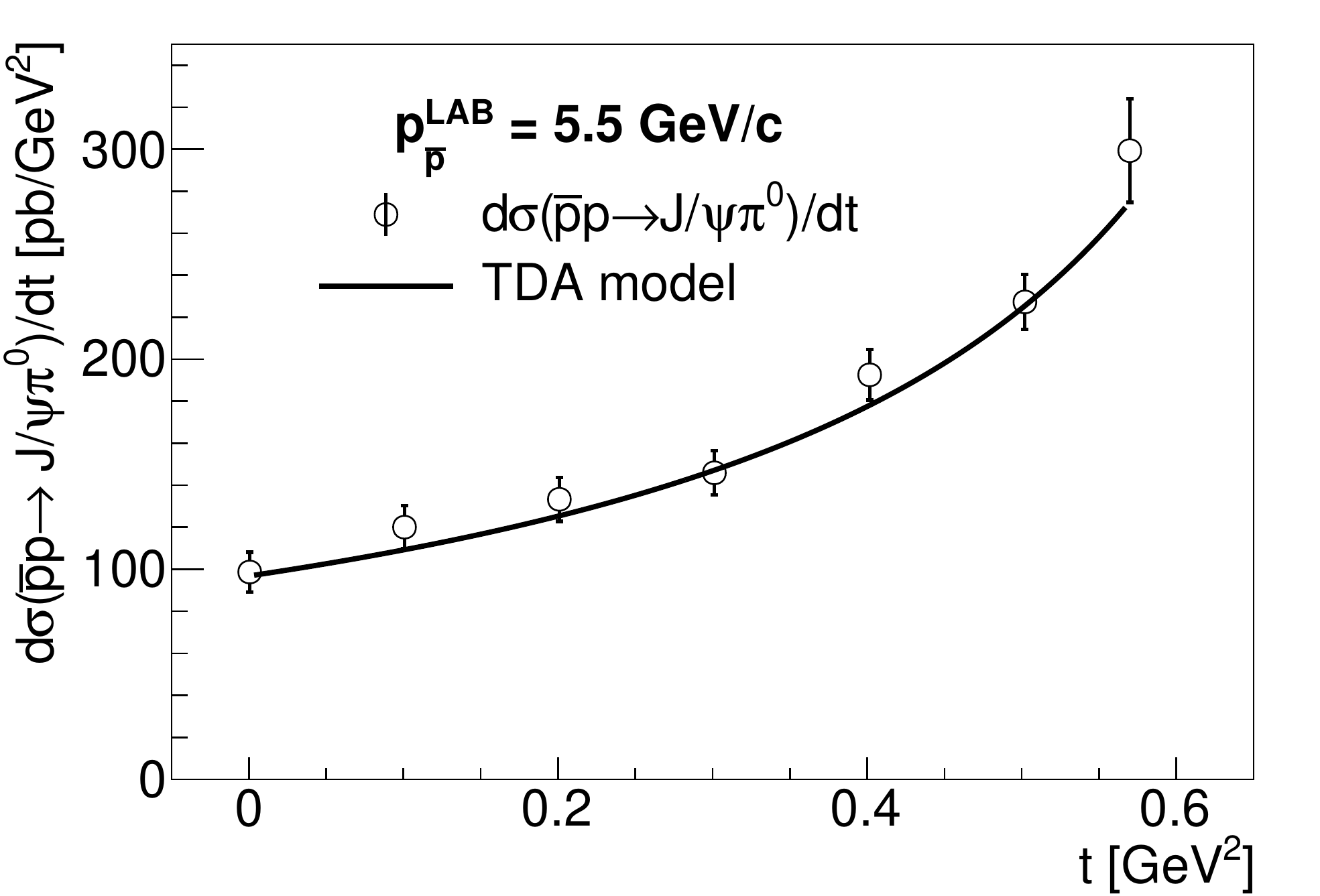}
  \includegraphics[width=0.33\textwidth]{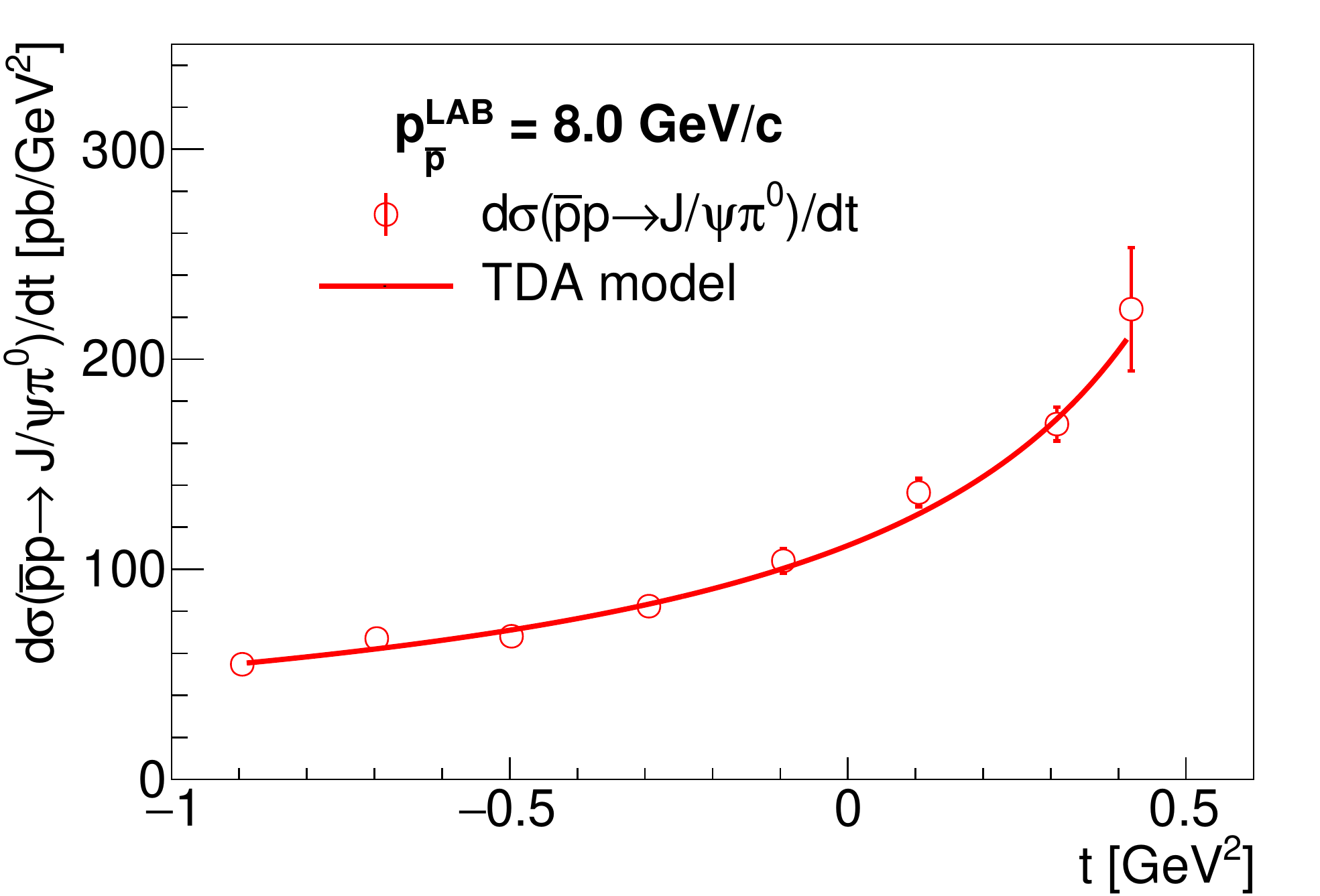}
  \includegraphics[width=0.33\textwidth]{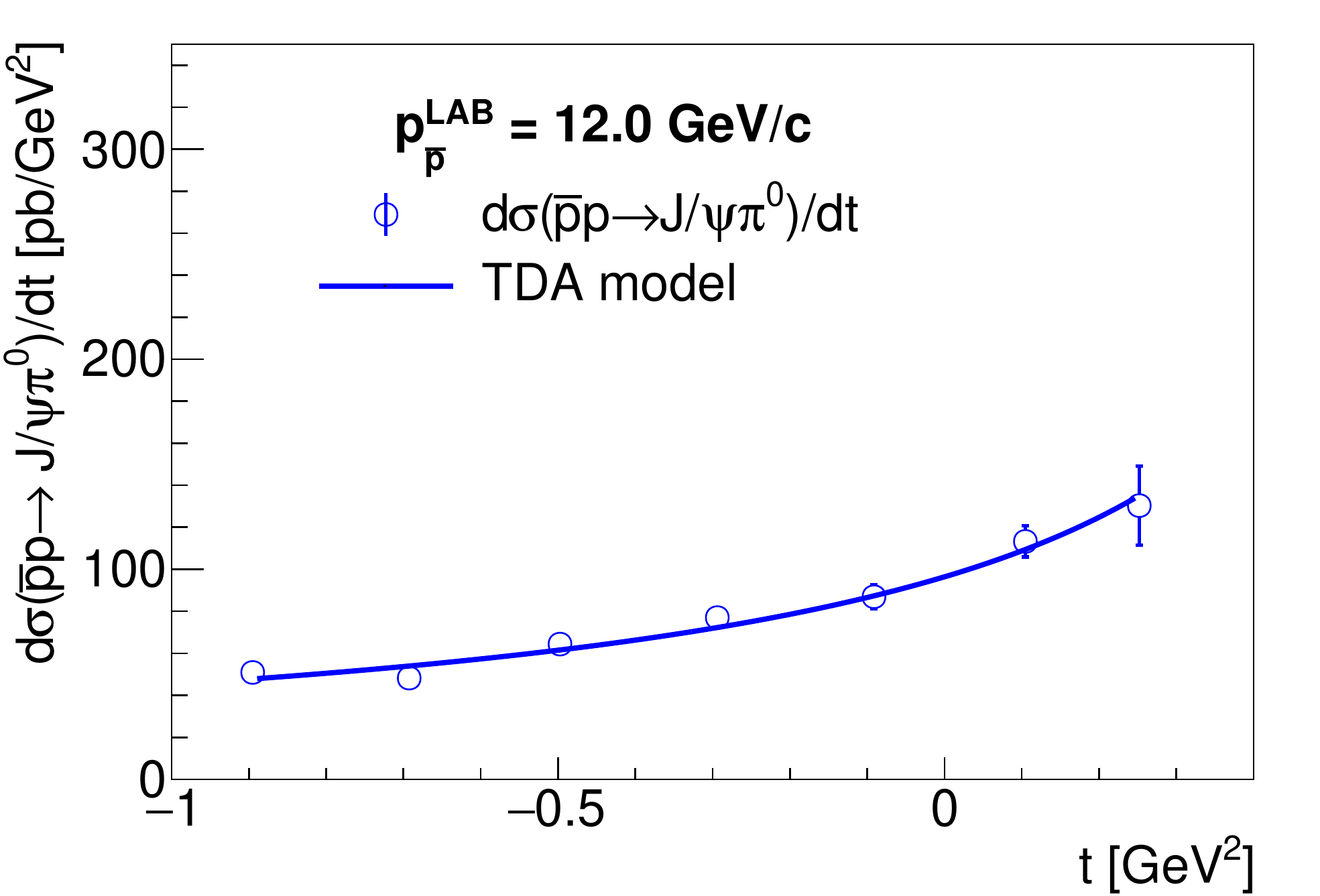}\\
  \includegraphics[width=0.33\textwidth]{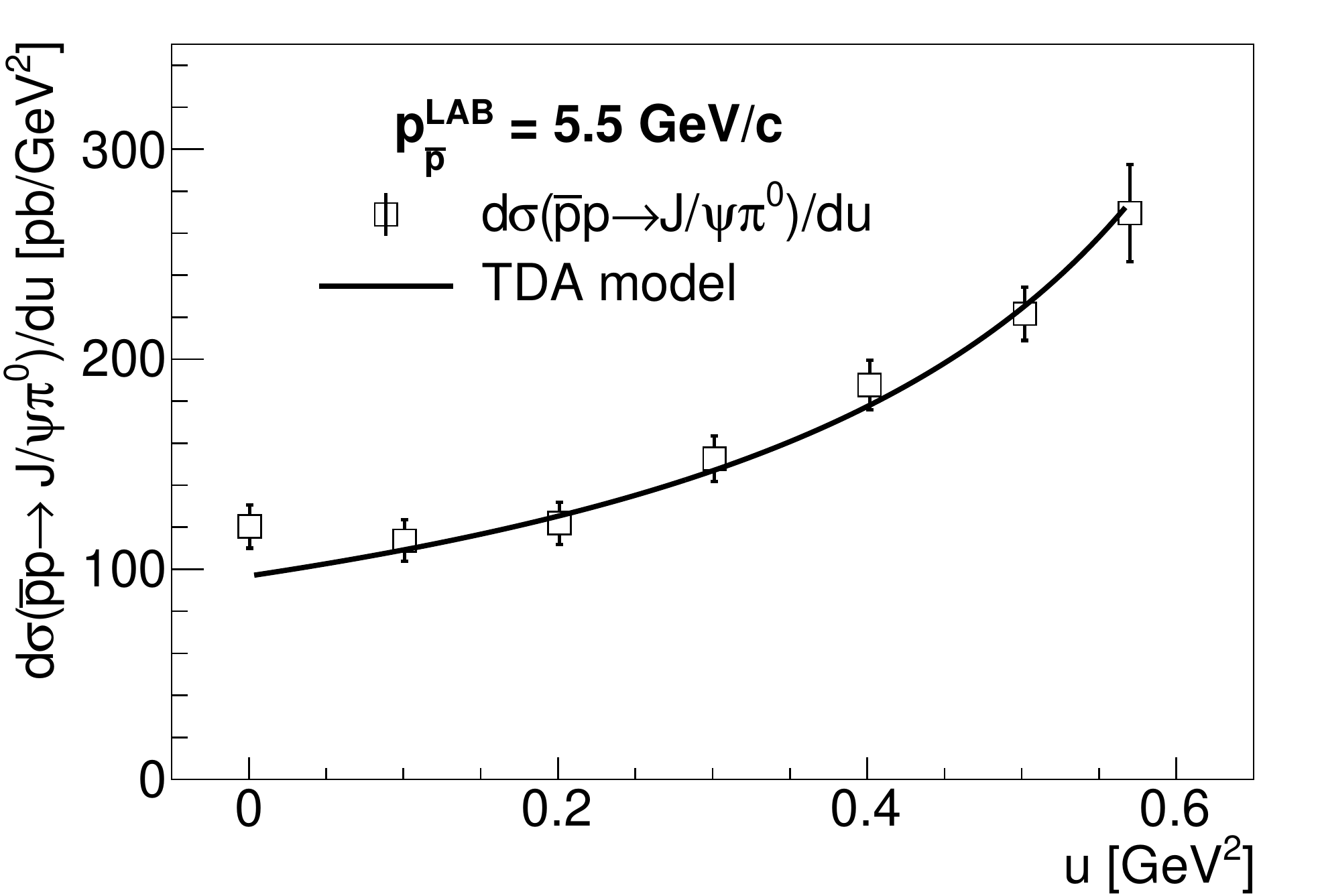}
  \includegraphics[width=0.33\textwidth]{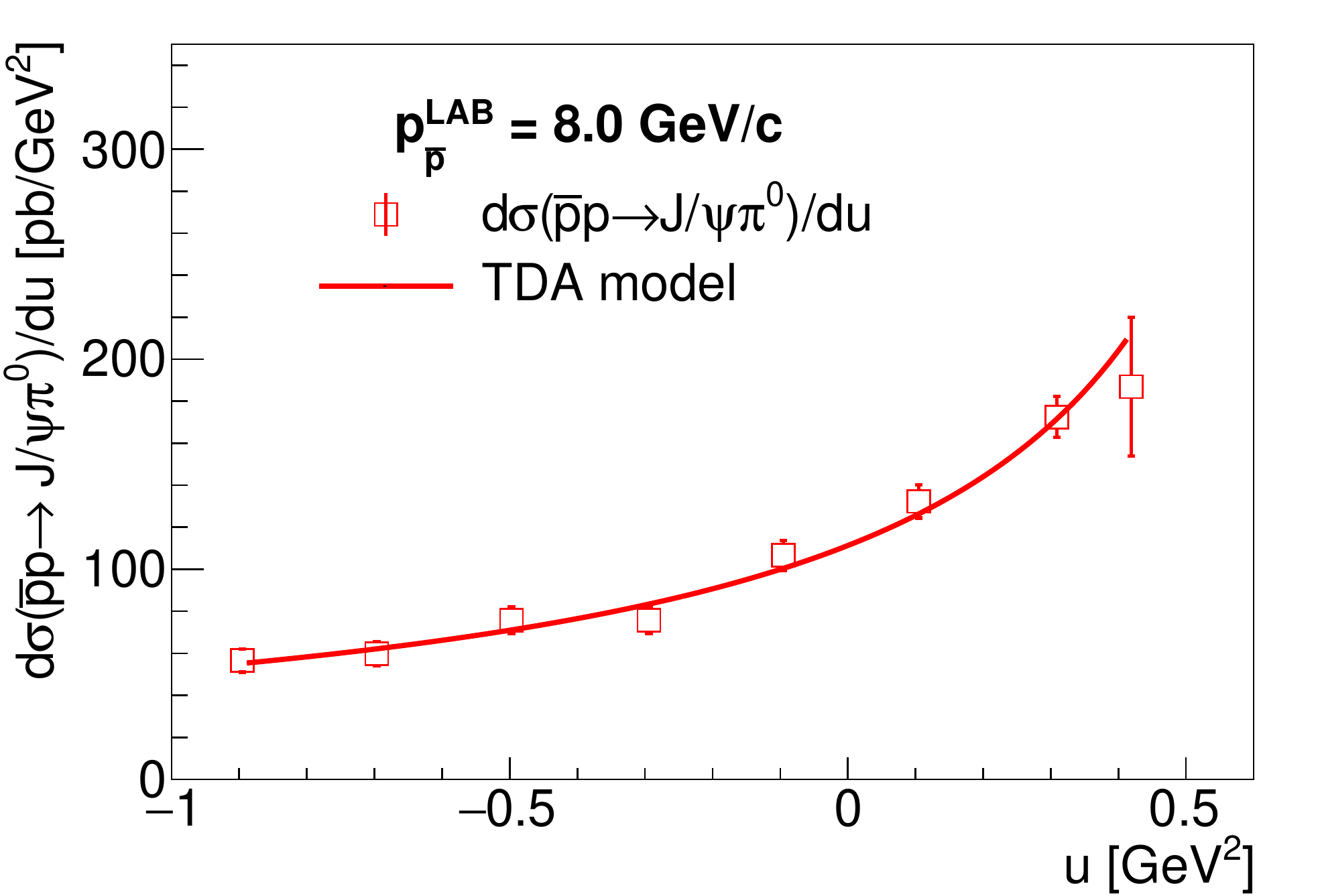}
  \includegraphics[width=0.33\textwidth]{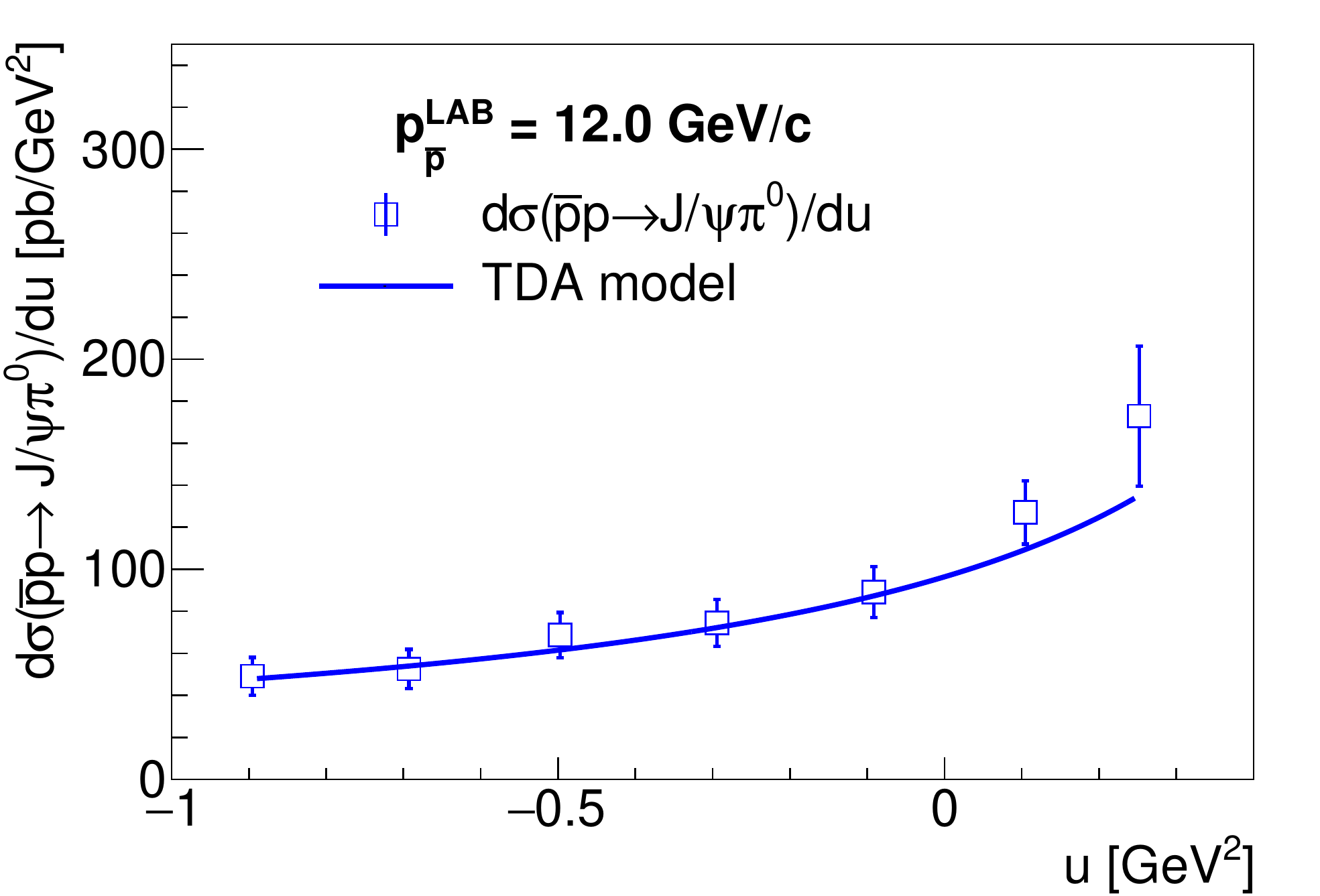}
  \caption{\label{fig:model_comp}Comparison between the cross sections
    extracted from the fully efficiency corrected yields expected from
    2~\fbi\ integrated luminosity \plfm{data points} and the TDA model
    prediction \plfm{full curves} at the three incident $\pbar$
    momenta: \bmomu{0} \plfm{left column}, \bmomu{1} \plfm{middle
      column} and \bmomu{2} \plfm{right column}. Top row:
    Near-forward kinematic approximation validity range as a function
    of $t$. Bottom row: Near-backward kinematic approximation validity
    range as a function of $u$.}
\end{figure*}

Figure~\ref{fig:model_comp} shows a comparison between the expected
precision of the measured differential cross section for an integrated
luminosity of 2~\fbi\ to the prediction of Ref.~\cite{Pire:2013jva}
that was used as the basis for the signal event generator. The
measurements have a satisfactory precision of about 8~--~10\% relative
uncertainty. This level of precision will allow a quantitative test of
the prediction of TDA models.

\subsection{\label{sec:ang_dist}$\jpsi$ Decay Angular Distributions}

The angular distribution of the $e^{+}$ and $e^{-}$ in the $\jpsi$
reference frame constitutes a key observable that could allow to test
the validity of the factorization approach. The leading twist
description of the TDA model predicts a specific form for the
differential cross section with respect to the polar emission angle of
the $e^{+}$ or $e^{-}$ in the $\jpsi$ reference frame relative to the
direction of motion of the $\jpsi$, $\epth$:
\begin{equation}
  \frac{d\sigma}{d\epth}\sim\opcthsq{}.
\end{equation}

This section gives results of an attempt to reconstruct the
differential angular distribution with an assumed 2~\fbi\ of
integrated luminosity.

\begin{figure*}[hbpt]
  \includegraphics[width=0.33\textwidth]{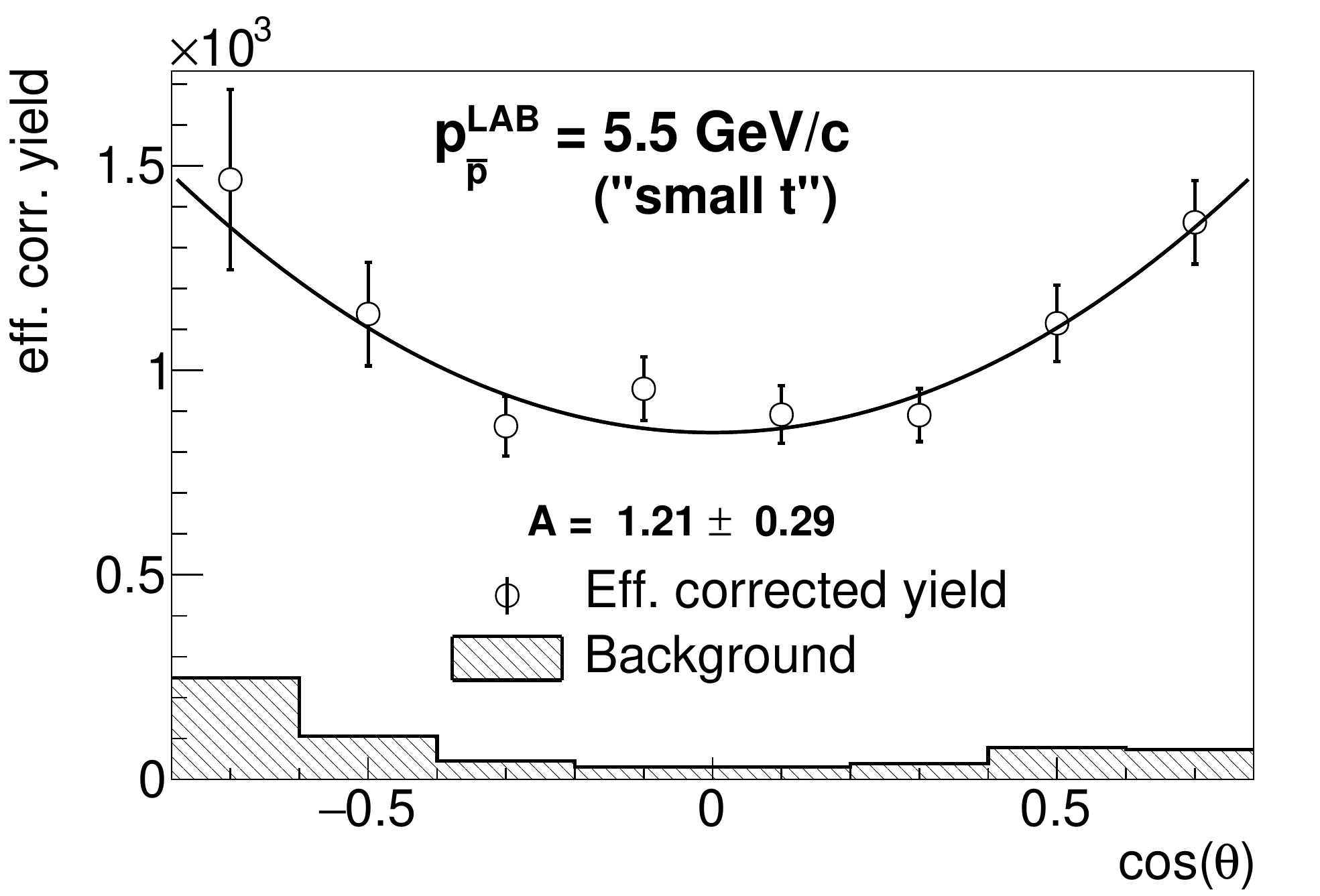}
  \includegraphics[width=0.33\textwidth]{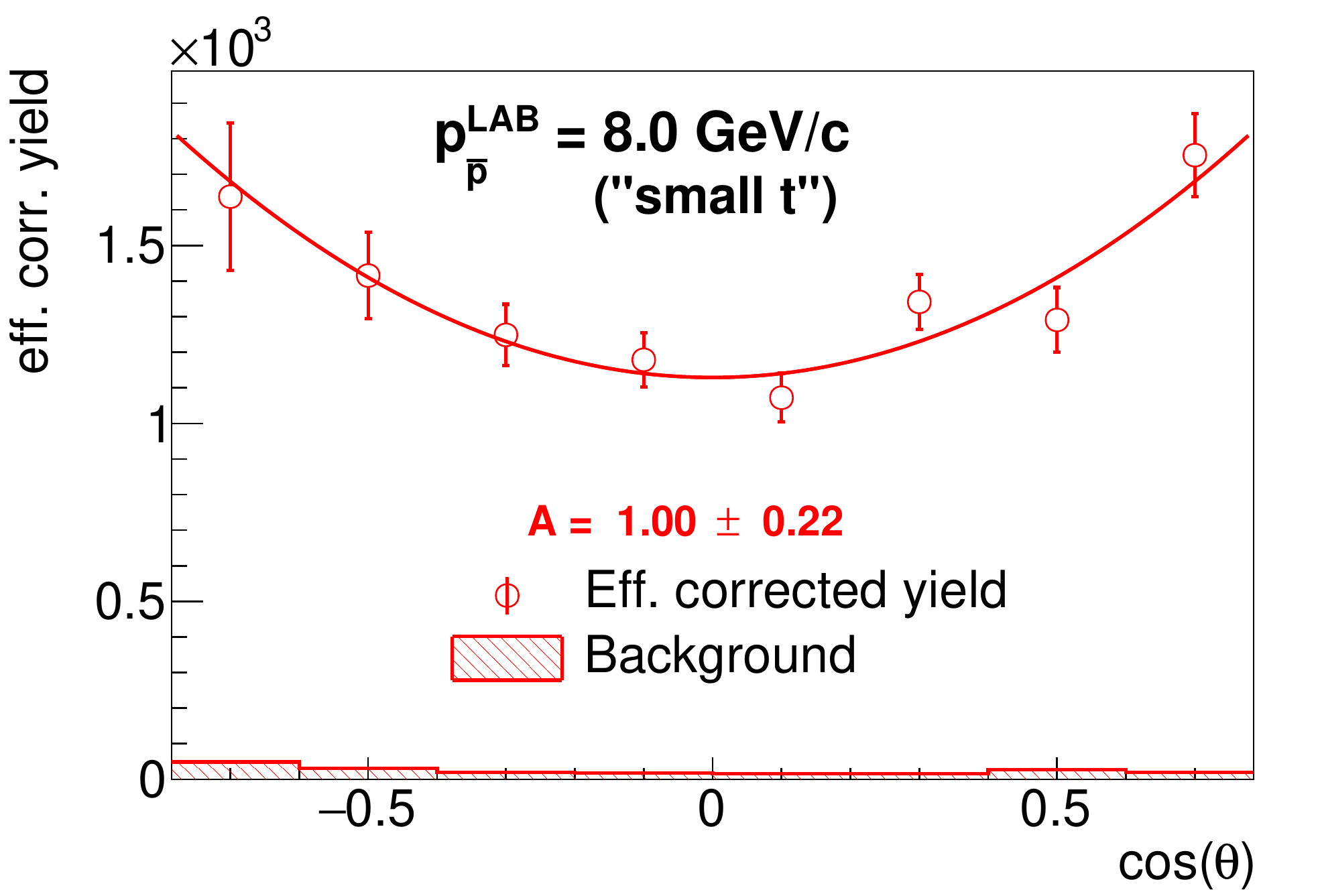}
  \includegraphics[width=0.33\textwidth]{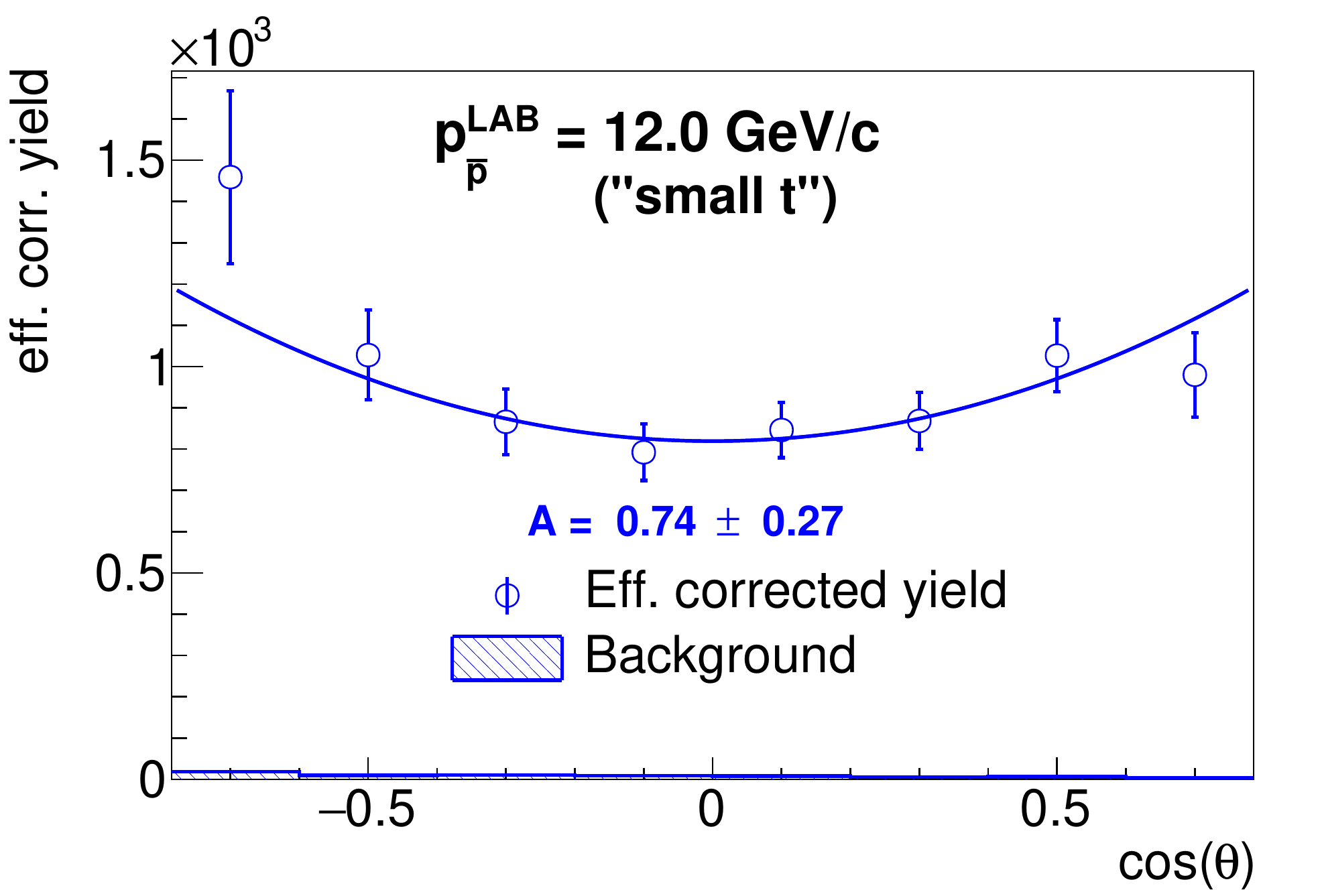}\\
  \includegraphics[width=0.33\textwidth]{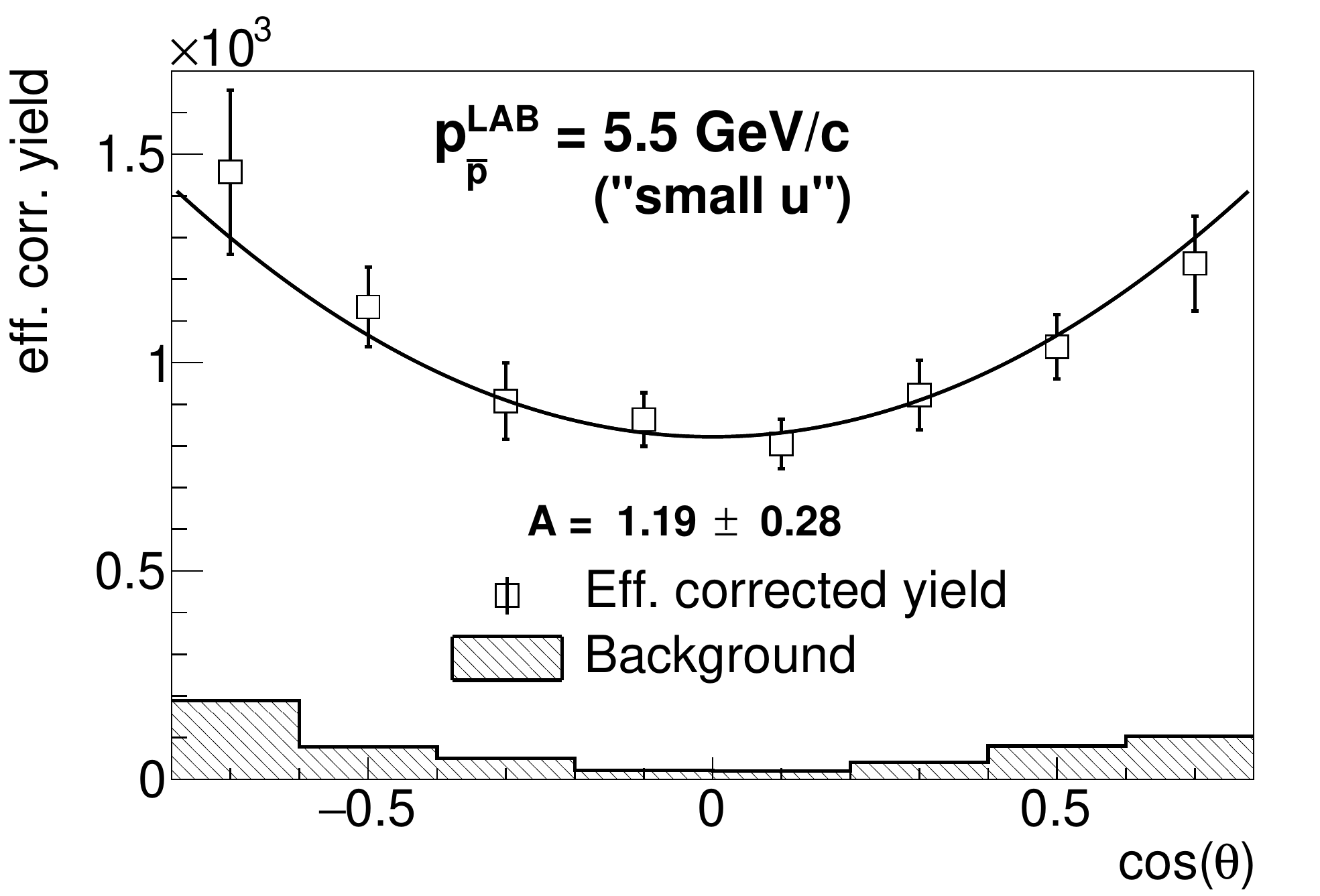}
  \includegraphics[width=0.33\textwidth]{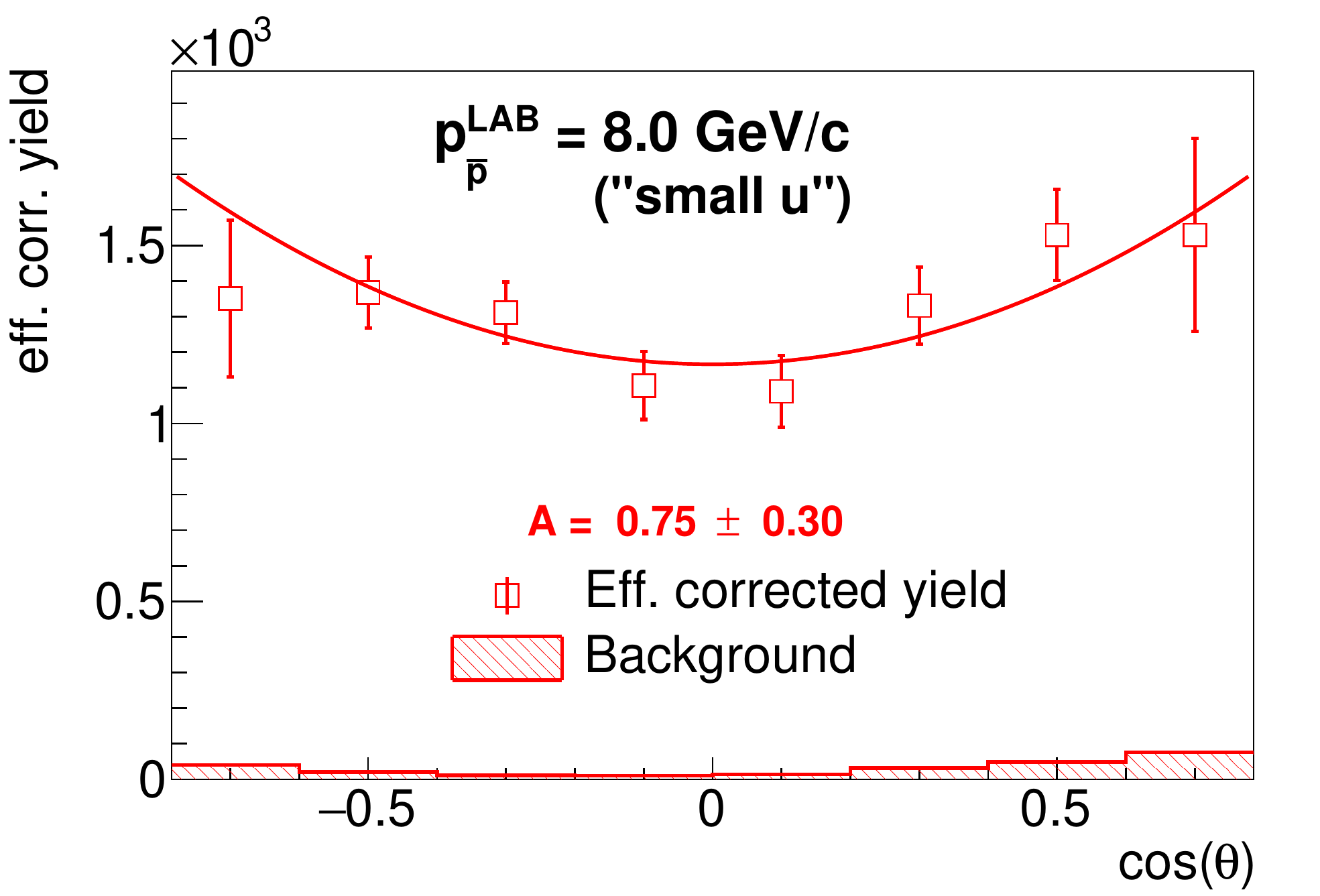}
  \includegraphics[width=0.33\textwidth]{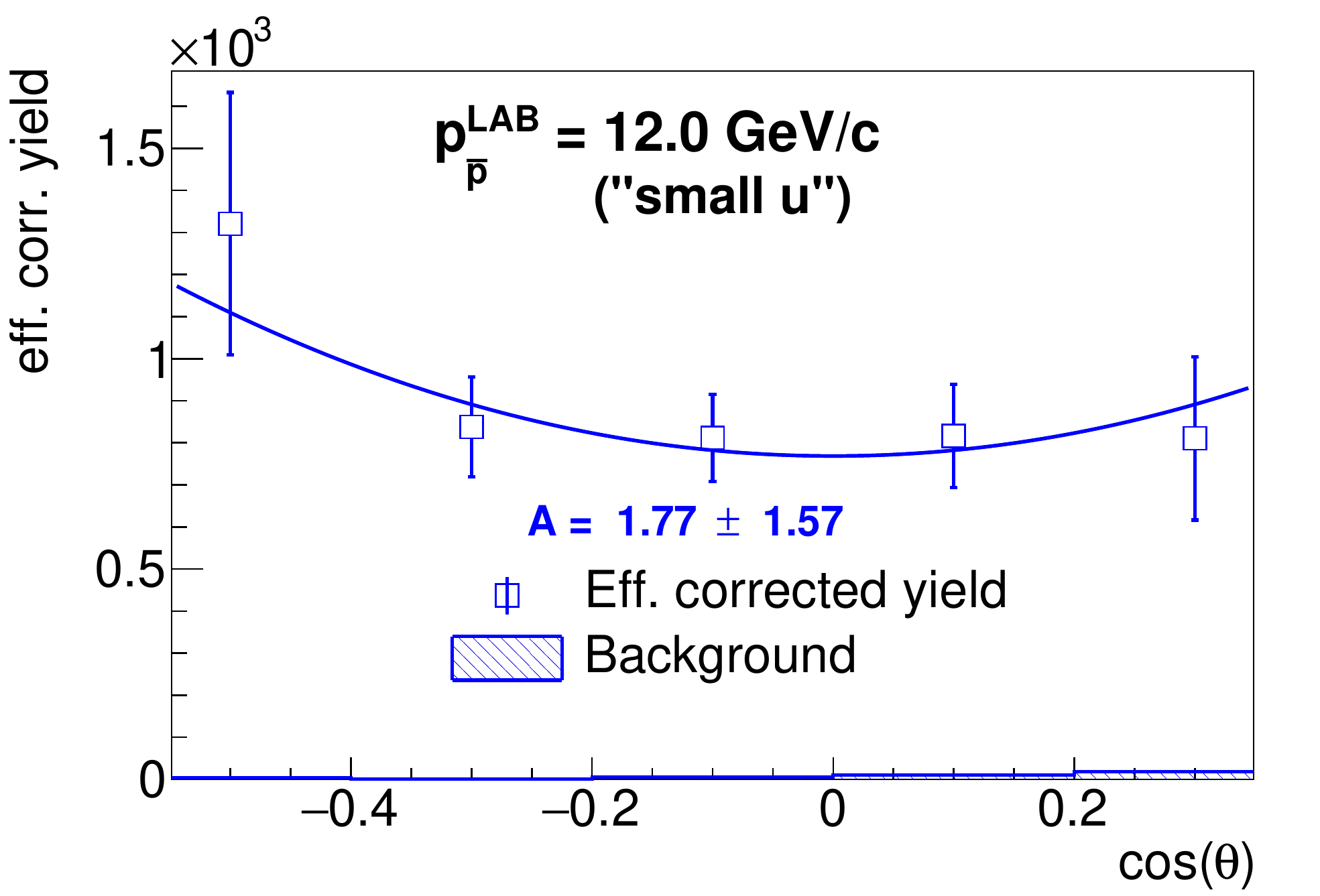}
  \caption{\label{fig:epcth_fit} Efficiency corrected yield of
    $\sigrxn$ \plfm{open markers} and background yield \plfm{shaded
      histogram} as a function of $\epcth$ and the result of the fit
    with the function $B\times(\opcthsq{A})$ \plfm{solid line} at the
    three incident $\pbar$ momenta: \bmomu{0} \plfm{left column},
    \bmomu{1} \plfm{middle column} and \bmomu{2} \plfm{right
      column}. Top row: Near-forward kinematic approximation validity
    range as a function of $t$. Bottom row: Near-backward kinematic
    approximation validity range as a function of $u$.}
\end{figure*}

Following the same analysis procedure as the one used for the
differential cross section plots of Section~\ref{sec:precision}, the
yield of signal events is extracted in bins of reconstructed $\epcth$,
within the small $t$ and small $u$ validity ranges separately at each
energy. The yields are corrected by the efficiency as a function of
$\epcth$ and fitted to the functional form $B\times(\opcthsq{A})$. An
example of the fit is shown in Fig.~\ref{fig:epcth_fit} for events
weighted according to $(\opcthsq{})$. The statistical errors are for
an assumed integrated luminosity of 2~\fbi. There is no sensitivity to
$A$ in the $u$ channel at the highest collision energy simulated due
to low efficiency.

The background contamination is indicated by a shaded histogram in
Fig.~\ref{fig:epcth_fit}, in the same way as for
Fig.~\ref{fig:reco_sig_rate_tu}. At the highest beam momentum, the
contamination is negligible in all bins. At the lowest beam momentum,
the background reaches $\approx$~15\% for some of the bins. About 60\%
of the background is due to the $\pipmpiz$ contribution, which is
subtracted by the sideband analysis, as described in
Section~\ref{sec:sig_cnt}. The remaining contribution is dominated by
the $\jpsipizpiz$ and is $\lesssim$~5\% for all bins. As already
mentioned, this contribution will be measured, allowing for a
subtraction with a residual systematic error below 1\%, which is much
smaller than the statistical errors.

To estimate the expected statistical error on the measurement of
parameter $A$ in a way that does not depend on the particular
statistical fluctuation of the selected simulation sub-sample, the fit
was repeated multiple times, each time using a different set of
simulated events. The root mean square values of the distributions of
these fit results is used to estimate the expected uncertainty on the
measurement. We find that the extraction of the angular distribution
is feasible with \PANDA\ with an integrated luminosity of 2~\fbi,
except in some kinematic zones where the efficiency is too low, e.g.\@
within the backward kinematic approximation validity range of the TDA
model at the largest beam momentum studied. The extraction of the
parameter $A$ will be possible with errors of the order of
$\approx~\pm$0.3 for $A=1$ and about $\approx~\pm$0.2 for $A=0.4$ and
$A=0$. With this level of precision, it will be possible to test the
leading twist approximation employed by the TDA model and potentially
differentiate between models that predict angular distributions that
deviate significantly from $\opcthsq{}$.

\subsection{\label{sec:syst}Systematic Uncertainties}

One of the sources of systematic uncertainty is expected to come from
the determination of beam luminosity. With the LMD, the uncertainty on
the absolute time integrated luminosity will amount to about
5\%~\cite{Panda:LumiTdr}. Another source of systematic uncertainty
will be associated with the signal extraction, namely from the
$\pipmpiz$ background subtraction procedure using the sideband method,
as well as from uncertainties related to the estimation of residual
background from $\jpsipizpiz$ events, which will amount to about 1\%
in the pessimistic scenario assumed in this work. Finally, there will
be a contribution to the systematic uncertainty coming from the
calculation of the efficiency correction, but this can only be
determined by how well the simulation describes the performance of the
fully constructed detector. For the current study, we simulated
$6\times10^6$ events to calculate the efficiency at each beam
momentum; as a result, the statistical uncertainty on the efficiency
is negligible.

\section{\label{sec:concl}Conclusions}

In this study, the feasibility of measuring $\jpsipiz$ production in
$\pbarp$ annihilation reactions with \PANDA\ at the future FAIR
facility was investigated. The study is based on the TDA model for the
description of the signal. A combination of available data and the
established hadronic event generator DPM was used to describe the
background with only pions in the final state. Non-resonant
$\jpsipizpiz$ was also considered, for which a PHSP event generator
was used for the event distributions and conservative estimates were
made for the cross sections given the lack of constraints from
existing data. Generated events were simulated and reconstructed using
the PandaRoot software based on the proposed \PANDA\ setup.

Using events selected to include a $\piz$ and an $\epem$ pair with
invariant mass inside a broad window around the $\jpsi$ mass,
background reactions with no $\jpsi$ in the final state can be reduced
to the level of a few percent. The residual background can be
subtracted using the procedure outlined. The signal efficiency
decreases from 18\% for beam momentum of \bmomu{0} to 9\% for
\bmomu{2}. To reject the $\jpsipizpiz$ background reaction to the 2\%
level, additional selection based on kinematic fits is needed, which
further reduces the signal efficiency by about an additional 30\%. The
severeness of these cuts will however be adjusted depending on the
measured yield for this background channel.

The cross section plots were produced with the assumption of a 2~\fbi\
integrated luminosity which could be accumulated in approximately five
months of data taking at each beam momentum setting with the full
design luminosity of the FAIR accelerator complex. The resulting
uncertainties to measure the differential cross section as a function
of the four momentum transfer in the two validity regions is expected
to be of the order of 5~--~10\%. In addition, the angular distribution
of the leptons in the $\jpsi$ center of mass frame provide important
information to test the leading-twist approximation used in the TDA
model. The distribution of these observables can also serve as a test
of other models, as the recent hadronic model of
Ref.~\cite{Wiele:2013vla}, which will be particularly interesting
outside the validity range of the TDA model at large emission angles.

We conclude that the measurement proposed here can be performed with
\PANDA, even at c.m.\@ energies corresponding to resonances with a
forbidden or weak decay to $\jpsipiz$. Together with $\sigrxnepem$,
the measurement of the $\sigrxnshort$ reaction will enable to test TDA
models and opens exciting perspectives for the study of nucleon and
antinucleon structure.

\begin{acknowledgments}
The success of this work relies critically on the expertise and dedication of the computing organizations that support \PANDA\@.
We acknowledge  financial support from
the Science and Technology Facilities Council (STFC), British funding agency, Great Britain;
the Bhabha Atomic Research Center (BARC) and the Indian Institute of Technology, Mumbai, India;
the Bundesministerium f\"ur Bildung und Forschung (BMBF), Germany;
the Carl-Zeiss-Stiftung 21-0563-2.8/122/1 and 21-0563-2.8/131/1, Mainz, Germany;
the Center for Advanced Radiation Technology (KVI-CART), Gr\"oningen, Netherland;
the CNRS/IN2P3 and the Universit\'e Paris-Sud, France;
the Deutsche Forschungsgemeinschaft (DFG), Germany;
the Deutscher Akademischer Austauschdienst (DAAD), Germany;
the Forschungszentrum J\"ulich GmbH, J\"ulich, Germany;
the FP7 HP3 GA283286, European Commission funding;
the Gesellschaft f\"ur Schwerionenforschung GmbH (GSI), Darmstadt, Germany;
the Helmholtz-Gemeinschaft Deutscher Forschungszentren (HGF), Germany;
the INTAS, European Commission funding;
the Institute of High Energy Physics (IHEP) and the Chinese Academy of Sciences, Beijing, China;
the Istituto Nazionale di Fisica Nucleare (INFN), Italy;
the Ministerio de Educacion y Ciencia (MEC) under grant FPA2006-12120-C03-02;
the Polish Ministry of Science and Higher Education (MNiSW) grant No. 2593/7, PR UE/2012/2, and  the National Science Center (NCN) DEC-2013/09/N/ST2/02180, Poland;
the State Atomic Energy Corporation Rosatom, National Research Center Kurchatov Institute, Russia;
the Schweizerischer Nationalfonds zur Forderung der wissenschaftlichen Forschung (SNF), Swiss;
the Stefan Meyer Institut f\"ur Subatomare Physik and the \"Osterreichische Akademie der Wissenschaften, Wien, Austria;
the Swedish Research Council, Sweden.
\end{acknowledgments}

%\clearpage

\bibliography{tda_jpsi_piz}

\end{document}